\documentclass[10pt]{article}
\usepackage{geometry}
\usepackage{xcolor}
\definecolor{graycolor}{gray}{0.9} 
\usepackage{microtype}
\usepackage{setspace} 
\usepackage[utf8]{inputenc}
\usepackage[english]{babel}
\usepackage{times}
\usepackage{array}
\usepackage{soul}
\usepackage{cite}
\usepackage[numbers,sort&compress]{natbib}
\usepackage{natbib}

\usepackage{titlesec} 
\titleformat {\section} [block] {\raggedright \fontsize{10}{10}\selectfont\bfseries} {\thesection. \space} {0pt} {}
\titlespacing {\section} {0pt} {12pt} {6pt}
\titleformat {\subsection} [block] {\raggedright \fontsize{10}{10}\selectfont\itshape} {\thesubsection .\space} {0pt} {}
\titlespacing {\subsection} {0pt} {12pt} {6pt}
\titleformat {\subsubsection} [block] {\raggedright \fontsize{10}{10}\selectfont} {\thesubsubsection .\space} {0pt} {}
\titlespacing {\subsubsection} {0pt} {12pt} {6pt}
\titleformat {\paragraph} [block] {\raggedright \fontsize{10}{10}\selectfont} {} {0pt} {}
\titlespacing {\paragraph} {0pt} {12pt} {6pt}

\usepackage{array} \newcommand{\PreserveBackslash}[1]{\let\temp=\\#1\let\\=\temp}
\newcolumntype{C}[1]{>{\PreserveBackslash\centering}m{#1}}
\newcolumntype{R}[1]{>{\PreserveBackslash\raggedleft}m{#1}}
\newcolumntype{L}[1]{>{\PreserveBackslash\raggedright}m{#1}}
\usepackage{lineno}
\usepackage{tabularx}
\usepackage{colortbl}
\usepackage{graphicx}
\usepackage{float}
\usepackage[export]{adjustbox}
\usepackage{caption}
\usepackage{fancyhdr} 
\usepackage{bm} 
\usepackage{threeparttable}
\usepackage{lastpage}
\usepackage{rotating}
\usepackage{graphicx}%
\usepackage{multirow}%
\usepackage{amsmath,amssymb,amsfonts}%
\usepackage{amsthm}%
\usepackage{mathrsfs}%
\usepackage[title]{appendix}%
\usepackage{xcolor}%
\usepackage{textcomp}%
\usepackage{manyfoot}%
\usepackage{booktabs}%
\usepackage{algorithm}%
\usepackage{algorithmicx}%
\usepackage{algpseudocode}%
\usepackage{listings}%

\usepackage{makecell}%
\usepackage{threeparttable}%
\usepackage{forest}
\usepackage{tikz}
\usetikzlibrary{shapes.geometric, arrows.meta, positioning}
\usepackage{mathrsfs}

\usepackage{subcaption}
\usepackage{epstopdf}
\usepackage{mathtools}

\usepackage{soul}
\usepackage{xcolor}
\usepackage{mdframed}
\usepackage{soul}

\soulregister{\ref}{1}
\soulregister{\cite}{1}
\soulregister{\label}{1}

\usepackage{doi}  
\graphicspath{{Figure/}}

\theoremstyle{thmstyleone}%
\theoremstyle{thmstyletwo}%

\theoremstyle{thmstylethree}%

\setcitestyle{open={[},close={]},citesep={,\!},numbers}
\usepackage{lastpage}
\usepackage{layout}
\usepackage{setspace} 
\usepackage{enumitem}
\usepackage{booktabs}
\usepackage{arydshln}
\usepackage{multirow}
\usepackage{color}
\usepackage{hyperref} 
\hypersetup{
	colorlinks=true,
	linkcolor=blue,
	filecolor=blue,
	urlcolor=black,
	citecolor=cyan,
}

\renewcommand{\headwidth}{\textwidth}

\fancypagestyle{firstpage}{
    \setlength{\headsep}{2.2cm}
    
    \setlength{\footskip}{1.5cm}
    \fancyhf{}
    \lhead{\begin{table}[H]
        \centering
        \begin{tabular}{L{2.5cm}C{10cm}C{3.1cm}R{2cm}}
            \includegraphics[scale=0.035]{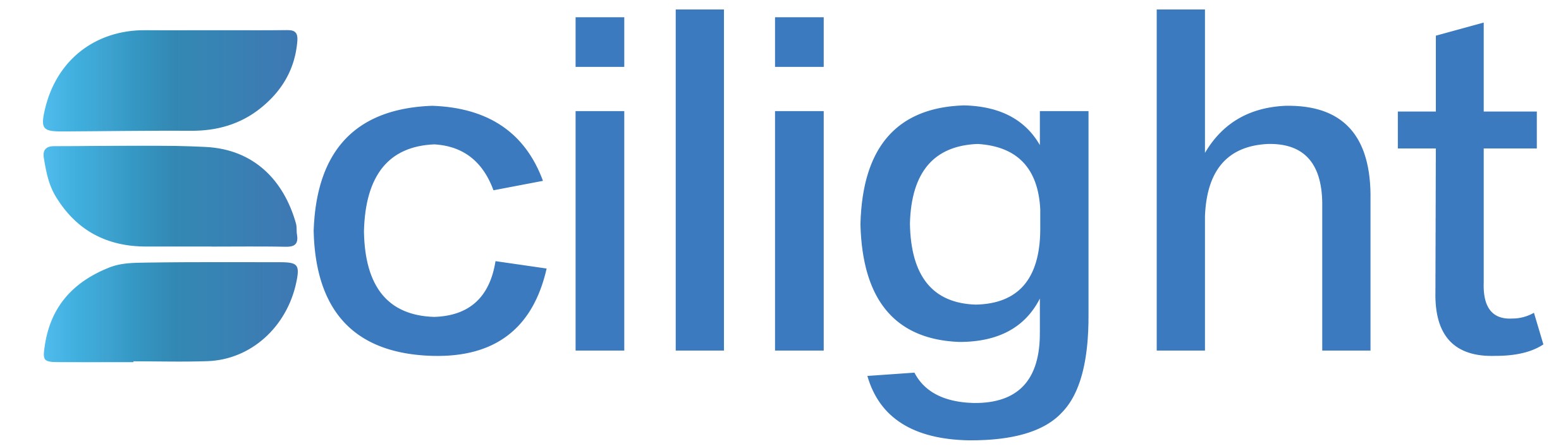} \vspace{-6pt}& \cellcolor{graycolor}\begin{tabular}[c]{@{}c@{}}\textit{Journal of Social Physics}\\ \href{https://www.sciltp.com/journals/jsp}{https://www.sciltp.com/journals/jsp}\end{tabular} & \includegraphics[scale=0.022]{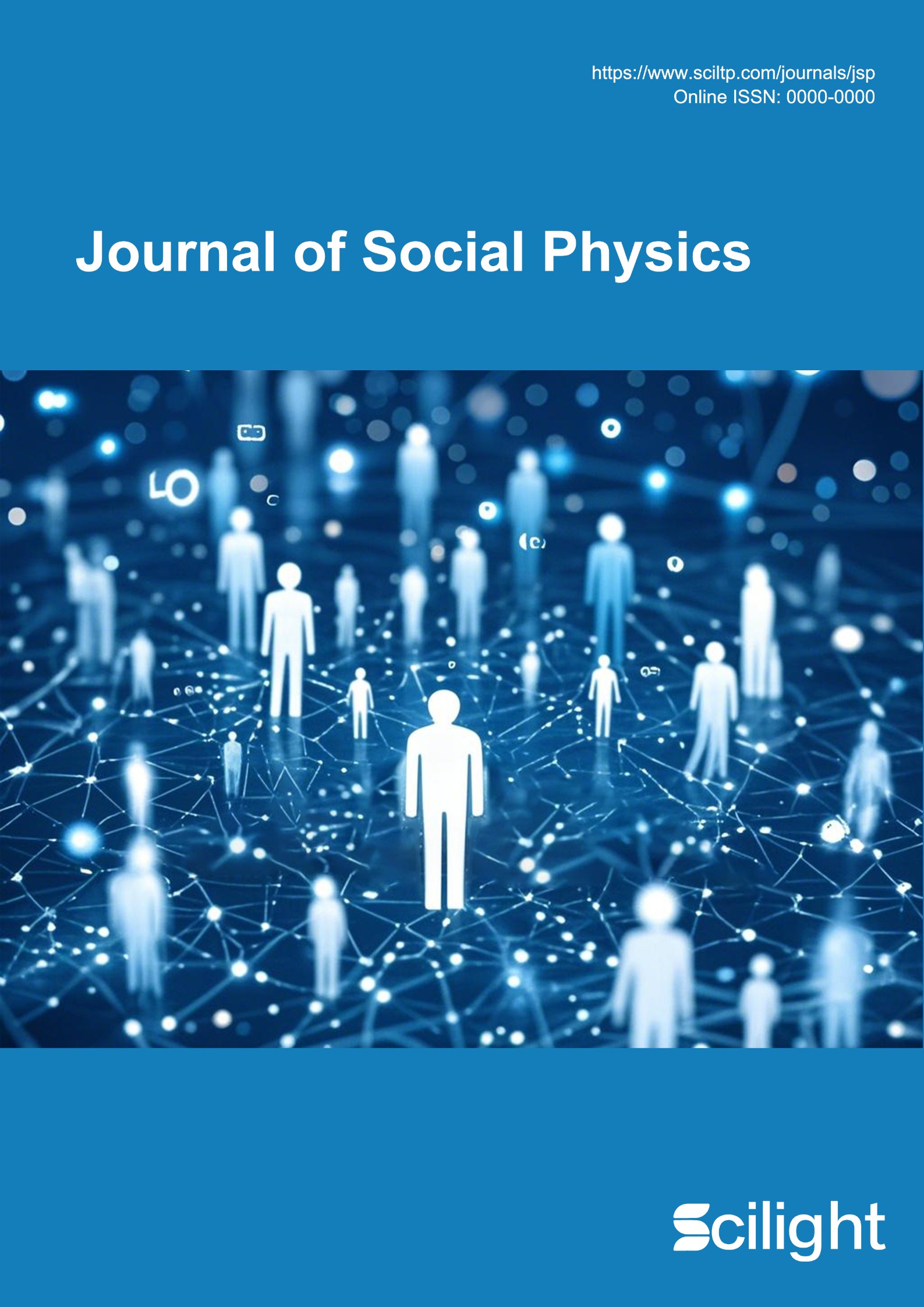} \vspace{-3pt}\\
        \end{tabular}
        \vspace{-22pt}
    \end{table}}
   
    \fancyfoot[C]{
        \vspace{-1.55cm}
        \begin{table}[H]
            \begin{minipage}[c]{0.15\columnwidth}
                \includegraphics[scale=0.5]{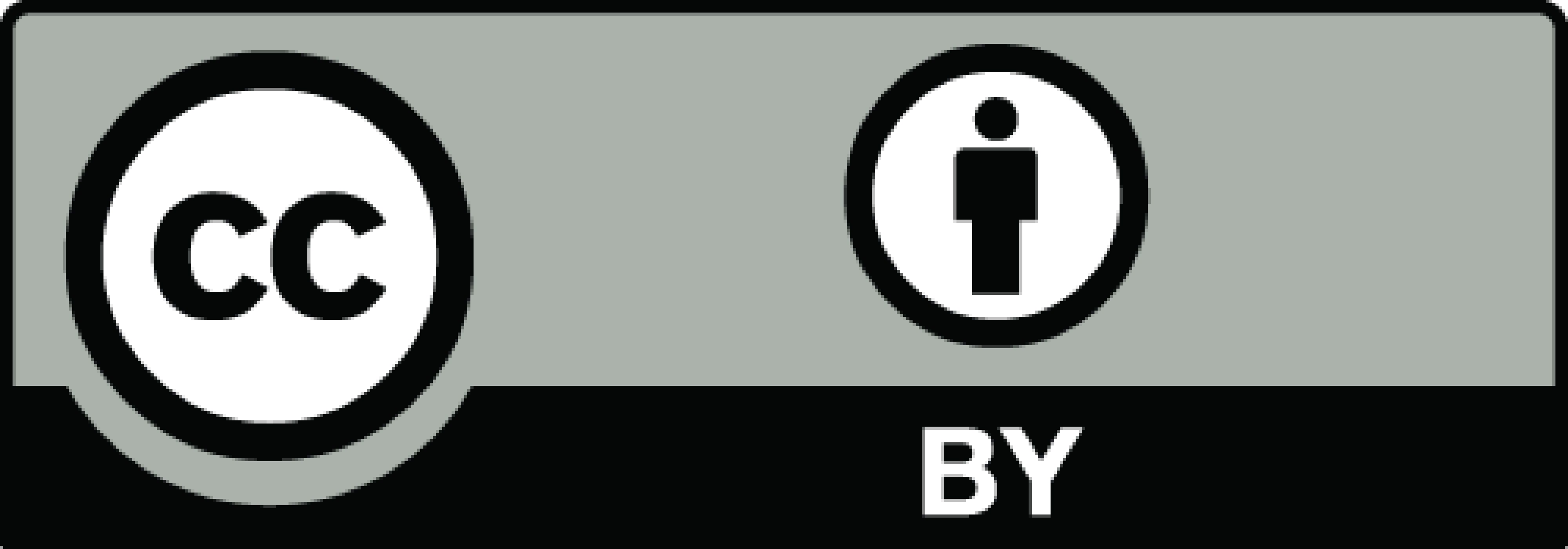} \vspace{1.1pt}
            \end{minipage}
            \hfill
            \begin{minipage}[c]{0.85\columnwidth}
                \scriptsize \textbf{Copyright:} © 2026 by the authors. This is an open access article under the terms and conditions of the Creative Commons Attribution (\mbox{CC BY}) license  (\href{https://creativecommons.org/licenses/by/4.0/}{https://creativecommons.org/licenses/by/4.0/}). \\ \textbf{Publisher’s Note:} Scilight stays neutral with regard to jurisdictional claims in published maps and institutional affiliations.
            \end{minipage}
    \end{table}}
    \vspace{-0.55cm}
}

\begin{document}
\newgeometry{left=2.5cm, right=2.5cm, top=1.8cm, bottom=4cm}
	\thispagestyle{firstpage}
	\nolinenumbers
        {\noindent \textit{{Article}}
} 
\hfill \vspace{3pt}
	\vspace{4pt} \\
	{\fontsize{18pt}{10pt}\textbf{Do Railway Commuters Exhibit Consistent Route Choice ``Rationality" Across Different Contexts and Time? Evidence from Tokyo Metropolitan Commutes}  }
	\vspace{16pt} \\
	{\large {{Yixuan Y. Zheng}}
 \textsuperscript{1}, Hideki Takayasu \textsuperscript{2} and Misako Takayasu \textsuperscript{1,2,*}}
	\vspace{6pt}
	 \begin{spacing}{0.9}
		{\noindent \small
			\textsuperscript{1}	\parbox[t]{0.98\linewidth}{Department of Systems and Control Engineering, Institute of Science Tokyo, Yokohama 2268502, Japan} \\
			\textsuperscript{2} {Department of Computer Science, Institute of Science Tokyo, Yokohama 2268502, Japan}
\\
		    {*}  \parbox[t]{0.98\linewidth}{Correspondence: takayasu@comp.isct.ac.jp} \vspace{6pt}\\
		\footnotesize	\textbf{How To Cite}: {Zheng, Y.Y.}; Takayasu, H.; Takayasu, M. Do Railway Commuters Exhibit Consistent Route Choice ``Rationality" Across Different Contexts and Time? Evidence from Tokyo Metropolitan Commutes. \emph{Journal of Social Physics} \textbf{2026}, \emph{1}(1), 6. \href{}{}}\\
	\end{spacing}

\begin{table}[H]
\noindent\rule[0.15\baselineskip]{\textwidth}{0.5pt} 
\begin{tabular}{lp{12cm}}  
 \small 
  \begin{tabular}[t]{@{}l@{}} 
  \footnotesize  Received: 14 October 2025 \\
  \footnotesize  Revised: 18 May 2026 \\
   \footnotesize Accepted: 10 June 2026 \\
  \footnotesize  Published: 24 June 2026
  \end{tabular} &
  \textbf{Abstract:} In urban railway systems, if every commuter were perfectly rational, concentrated demand on optimal routes would undermine system performance. Understanding the actual degree of collective route choice determinism, and whether it remains stable over time, is essential for transportation policy. Smartphone GPS data offers both the scale and complete trajectory coverage that survey data and smart card data individually lack, yet its coarse spatial accuracy has limited its adoption for railway route identification. To overcome this, we develop a methodology to identify railway commuting routes from one year of GPS data covering over one million daily users in the Tokyo metropolitan area. We apply {a multinomial logit (MNL)} framework, utilizing a standardization convention to extract a comparative measure of collective determinism alongside relative attribute preferences. We find that collective route choices in Tokyo are strongly cost-sensitive but non-deterministic. Both components remain stable across all twelve months of 2023, confirming that this determinism represents a structural consistency of the commuting system. Leveraging the dataset's scale, we further estimate parameters for each {origin--destination (OD)} pair individually, revealing systematic heterogeneity driven by departure time and transport complexity that smaller datasets treat as unobserved. 
  
\\
\\
  & 
\end{tabular}
\noindent\rule[0.15\baselineskip]{\textwidth}{0.5pt} 
\end{table}

	\section{Introduction}%

\label{sect:introduction}

In urban railway systems, a fundamental tension exists between individual cost-minimization and collective system performance. If every commuter were perfectly rational, always selecting the route that minimizes their personal cost, demand would concentrate on a small number of optimal routes, degrading service quality and undermining system resilience. Conversely, if choices were entirely random, infrastructure investments designed around predictable demand patterns would lose their effectiveness. The actual degree of choice determinism, where the collective population falls on this spectrum between perfect predictability and pure randomness, therefore carries direct consequences for capacity planning, congestion management, and disruption response \cite{ardeshiri_lifestyles_2019, wang_free_2021, zhang_analysis_2024, akamatsu_global_2023}. Equally important is whether this behaviour determinism level remains stable over time, or fluctuates with seasons, service changes, and external shocks. A system operating in a consistent state is fundamentally different, from a policy perspective, than one exhibiting volatile behavioral shifts.

Quantifying collective determinism, however, imposes a demanding requirement on data. Because the quantity of interest is a population-level parameter describing the distribution of choices across both {cost-driven and idiosyncratic} commuters, the dataset must be large enough to capture the full spectrum of decision-making behavior. \textls[12]{Small samples risk overrepresenting particular behavioral segments and producing unstable parameter estimates. Moreover, to detect whether choice determinism varies systematically across contexts, such as departure times,}
\restoregeometry   
\noindent commuting distances, or transfer complexity, the dataset must support disaggregate analysis at the level of individual origin-destination (OD) pairs while retaining sufficient observations within each pair for statistical reliability.


\textls[-22]{Traditional route choice research \cite{deng_heterogeneity_2025, deng_unveiling_2025, sakamanee_methods_2020, kaneko_route_2018, li_incorporating_2016, arriagada_incorporating_2025, fosgerau_perturbed_2022, cazor_closed-form_2025, wu_data-driven_2019}, grounded in the Random Utility Maximization (RUM) framework and its extensions such as the Mixed Logit and Latent Class models 
 \cite{mcfadden_conditional_1972, train_discrete_2009, ben_discrete_1985}, has relied primarily on stated or revealed preference survey data involving tens to hundreds of respondents \cite{train_discrete_2009, li_incorporating_2016, deng_heterogeneity_2025, ma_nested_2020, garcia-martinez_transfer_2018,okubo_transportation_2022}. While these models have produced foundational insights into route attribute trade-offs, their sample sizes fundamentally limit the ability to observe population-level phenomena or to estimate parameters for individual OD pairs. Heterogeneity in preferences must be pre-specified as \textls[15]{a parametric distribution \cite{mcfadden_conditional_1972, ben_discrete_1985, train_discrete_2009} (continuous in Mixed Logit \cite{mcfadden_mixed_2000}, discrete in Latent Class \cite{greene_a_latent_2003, kim_finite_2023})} rather than empirically observed from the data \cite{mcfadden_mixed_2000, train_discrete_2009, garcia-martinez_transfer_2018, li_incorporating_2016, ghorbani_enhanced_2025, deng_heterogeneity_2025}. This constraint means that the core question - what is the actual level of collective determinism in a metropolitan commuting system, has remained largely unaddressed.}

\textls[15]{Larger-scale passively collected data sources have begun to relax this constraint. Public transport smart card data (e.g., Suica, Oyster) provides transaction records for millions of trips, but captures only entry and exit stations \cite{tang_mining_2024,kaneko_route_2018, mohammed_origin_2023, yap_public_2025}.} Crucially, transfers between lines operated by the same company do not require re-tapping the card, rendering these transfers invisible to the analyst~\cite{tang_mining_2024, mohammed_origin_2023}. Since transfer behavior~\cite{wu_data-driven_2019, deng_unveiling_2025} is a primary source of route differentiation in dense networks like Tokyo's, this limitation directly undermines route identification accuracy. Smartphone GPS trajectory data offers a complementary advantage: it records complete door-to-door journeys at high temporal resolution, capturing the actual paths taken including all intermediate transfers. The dataset used in this study covers approximately 1.2 million anonymous users per day across the Tokyo metropolitan area (Shutoken) throughout 2023, providing the scale and trajectory completeness that the research question demands. However, GPS data presents its own methodological challenge: the coarse spatial accuracy of smartphone positioning (approximately 10 m on average) has limited its adoption for railway route identification~\cite{cherchi_empirical_2008, sakamanee_methods_2020, sadeghian_stepwise_2022, tang_mining_2024, montini_route_2017}, where distinguishing between routes that share station proximity requires careful spatial analysis~\cite{sakamanee_methods_2020}.

To address this challenge, we develop a comprehensive methodology for extracting commuting trajectories, transportation modes, and transfer stations from large-scale GPS data. Our approach combines rule-based transport mode classification validated against national statistics with a Voronoi tessellation-based spatial matching technique that identifies transfer stations from GPS trajectory interruptions. This pipeline transforms raw location records into structured route choice observations suitable for discrete choice modeling without requiring full path enumeration over the railway network graph.

\textls[-15]{For the modeling framework, we adopt the Multinomial Logit (MNL) 
model~\cite{mcfadden_conditional_1972,ben_discrete_1985,train_discrete_2009}. In standard Random Utility Maximization (RUM) applications,} the scale parameter is typically normalized to one, with its role in reflecting the error term's variance absorbed into the coefficient magnitudes \cite{train_discrete_2009}. 
{In this study, we additionally report the norm and direction of the estimated coefficient vector as descriptive summaries; we apply a standardization convention that constrains the $\ell_2$-norm of the utility weights to unity.} Under this shared scale, we extract a comparative magnitude $\beta$ alongside the unit-length direction $\omega$. This convention serves a specific analytical purpose: $\beta$ captures the overall sensitivity of collective route choices to cost differences, serving as a macroscopic measurement of collective determinism and quantifying how deterministic or random the population's behavior is; while $\boldsymbol{\omega}$ captures the relative importance of each route attribute. Furthermore, because our dataset covers over one million daily users, the sample size is sufficient to estimate $\beta$ and $\boldsymbol{\omega}$ not only at the city-wide level but also for individual OD pairs, allowing systematic heterogeneity to be directly observed from the data rather than pre-specified as a parametric distribution as in the Mixed Logit framework~\cite{mcfadden_mixed_2000}.

In this paper, we focus on quantifying choice determinism
, the degree to which collective behavior is predictable based on observed costs. Although we refer to this as ``rationality'' 
  \cite{helbing_the_physics_2004} in the title, we employ the term strictly as a macroscopic metaphor. It describes the ensemble's tendency to concentrate on optimal routes rather than choosing randomly, without implying any behavioral assumptions about individual commuter intent.

Applying this framework to our dataset, we obtain several findings. First, Tokyo railway commuters' choices exhibit strong but non-deterministic concentration: collective choices strongly but not absolutely favor lower-cost routes, and this level remains stable across all twelve months of 2023, confirming that the commuting system operates in a structural consistency. Second, leveraging the dataset's scale, we estimate parameters for each OD pair individually, revealing systematic heterogeneity driven by departure time and transport complexity. Peak-hour commuters exhibit less deterministic and more strategic, exploratory behavior than off-peak commuters, reflecting behavioral adaptation under capacity constraints. Third, we identify transfers as the dominant source of choice uncertainty. We quantify the transfer-equivalent time penalty as ranging from approximately 7 min for short-to-medium commutes to over 20 min for long-distance suburban routes, demonstrating that these transfer-induced interruptions create bimodal travel time distributions that fundamentally limit model predictability for transfer-intensive routes. These OD-specific analyses are enabled precisely by the unprecedented data scale; with smaller datasets, the per-OD sample sizes would be insufficient for reliable estimation, and this heterogeneity would be absorbed into unobserved random terms.

The remainder of this paper is organized to reflect a continuous analytical pipeline, moving from raw trajectory data to multi-scale behavioral insights, as illustrated in Figure~\ref{fig1}. Section~\ref{sect:data_description_and_processing} establishes the empirical foundation by introducing the large-scale smartphone GPS dataset. To overcome the coarse spatial resolution of this data, Section~\ref{sect:methodology} details our methodology for translating raw coordinates into observable behavior, utilizing transport mode classification and commuting interruption analysis to construct reliable choice sets. Building on this, Section~\ref{sect:model} formulates the theoretical route choice model within a standard Multinomial Logit (MNL) framework, expressing decision-making in terms of collective determinism (quantified by the scale parameter, $\beta$) and relative attribute preferences. Section~\ref{section:results} presents our empirical findings through a top-down approach: first evaluating macro-level, system-wide consistency, and subsequently uncovering micro-level systematic heterogeneity across individual origin-destination pairs. Finally, Section~\ref{sect:conclusion_limitation} synthesizes these structural and behavioral findings, concluding with implications for urban transportation policy.

\begin{figure}[H]
    \centering
 \resizebox{0.7\textwidth}{!}{
\begin{tikzpicture}[
    node distance=1.1cm and 0.5cm,
    every node/.style={font=\sffamily\normalsize, align=center},
    data/.style={rectangle, rounded corners=5pt, draw=blue!80, fill=blue!5, thick, inner sep=8pt, minimum width=4cm},
    process/.style={rectangle, draw=gray!80, fill=white, thick, inner sep=6pt, minimum width=5cm},
    formula/.style={rectangle, draw=black, fill=gray!5, thick, inner sep=8pt, font=\small, dashed},
    header/.style={font=\sffamily\bfseries\color{gray!90}, inner sep=2pt},
    rq_box/.style={rectangle, rounded corners=3pt, thick, inner sep=5pt, minimum width=2cm, minimum height=1.5cm},
    arrow/.style={-{Stealth[scale=1.2]}, thick, draw=gray!60}
]

    \node[data] (raw) {Raw GPS Trajectories};
    
    \node[header, below=0.4cm of raw] (meth_h) {Methodology (Section 3)};
    
    \node[process, below=0.4cm of meth_h] (extract) {Status $\rightarrow$ Mode $\rightarrow$ Transfer $\rightarrow$ Route Extraction};
    
    \node[data, below=of extract] (Xnj) {Route Choice Sets \& Attributes};

    \node[header, below=0.4cm of Xnj] (model_h) {Model (Section 4)};
    
    \node[process, below=0.4cm of model_h] (mnl) {MNL with Scale-Direction Decomposition};
    
    \node[formula, below=0.6cm of mnl] (math) {
        $P_{nj} = \dfrac{\exp(-\beta \omega^T X_{nj})}{\sum_{k \in C_n} \exp(-\beta \omega^T X_{nk})}$ \\[6pt]
        s.t. $||\omega||_2 = 1$
    };

    \node[rq_box, draw=orange!80, fill=orange!5, below left=1cm and -0.5cm of math] (Model 1) {
        \textbf{Model 1} \\ city-wide parameters};
    
    \node[rq_box, draw=teal!80, fill=teal!5, below right=1cm and -0.5cm of math] (Model 2) {
        \textbf{Model 2} \\ per OD pair parameters};

       \node[header, below=3cm of math] (RQ) {Research Questions (Section 5)};

    \node[rq_box, draw=orange!60, fill=orange!2, below=1.2cm of Model 1, xshift=-1cm] (rq1) {
        \textbf{[RQ1]} \\ How ``rational"? \\ Stable over time? \\ \scriptsize ($\beta$)
    };
    \node[rq_box, draw=orange!60, fill=orange!2, right=0.3cm of rq1] (rq2) {
        \textbf{[RQ2]} \\ What attributes matter? \\ Stable over time? \\ \scriptsize ($\mathbf{\omega}$)
    };
    
    \node[rq_box, draw=teal!60, fill=teal!2, below=1.2cm of Model 2, xshift=-1cm] (rq3) {
        \textbf{[RQ3]} \\ Systematic \\ heterogeneity \\ across OD pairs? \\ \scriptsize ($\beta$ and $\mathbf{\omega}$, distribution)
    };
    \node[rq_box, draw=teal!60, fill=teal!2, right=0.3cm of rq3] (rq4) {
        \textbf{[RQ4]} \\ Heterogeneity \\ drivers \\ \scriptsize (transfer penalty)
    };

    \draw[arrow] (raw) -- (extract);
    \draw[arrow] (extract) -- (Xnj);
    \draw[arrow] (Xnj) -- (mnl);
    \draw[arrow] (mnl) -- (math);
    
    \draw[arrow] (math.south) -- ++(0,-0.4) -| (Model 1.north);
    \draw[arrow] (math.south) -- ++(0,-0.4) -| (Model 2.north);
    
    \draw[arrow] (Model 1.south) -- ++(0,-0.4) -| (rq1.north);
    \draw[arrow] (Model 1.south) -- ++(0,-0.4) -| (rq2.north);
    \draw[arrow] (Model 2.south) -- ++(0,-0.4) -| (rq3.north);
    \draw[arrow] (Model 2.south) -- ++(0,-0.4) -| (rq4.north);

\end{tikzpicture}}
    \caption{Conceptual framework of this study. \textls[-15]{The analytical pipeline proceeds from raw GPS trajectories through methodology (Section~\ref{sect:methodology}) to the MNL route choice model with scale-direction decomposition (Section~\ref{sect:model}). The model is estimated in two variants: Model 1 (city-wide pooled estimation) and Model 2 (OD-specific estimation), which together address four research questions on collective determinism, temporal stability, and systematic heterogeneity (Section~\ref{section:results}).}}
    \label{fig1}
\end{figure}
\vspace{-24pt}

\subsection*{Comparing with Previous Research}
\label{subsect: comparing_with_previous_research}

Table~\ref{tab:previous_study} provides a structured comparison 
with representative route choice studies. Three dimensions 
distinguish the current study. First, in data scale: previous 
GPS-based studies~\cite{tang_mining_2024, montini_route_2017} 
tracked 35--156 users, while survey-based 
studies~\cite{garcia-martinez_transfer_2018, ma_nested_2020} 
covered at most several thousand respondents. Our dataset 
comprises approximately 1.2 million daily users, providing 
sufficient observations for OD-pair-level estimation. Second, 
in route identification: rather than generating synthetic 
alternatives via network algorithms (e.g., BFS-LE) or 
hypothetical survey designs, we construct choice sets directly 
from observed GPS traces by identifying transfer station 
sequences. Third, in variable derivation: transfer penalties 
and peak-hour effects are estimated from revealed behavior 
rather than stated preference scenarios.

\newpage
\paperwidth=\pdfpageheight
\paperheight=\pdfpagewidth
\pdfpageheight=\paperheight
\pdfpagewidth=\paperwidth
\newgeometry{layoutwidth=297mm,layoutheight=210 mm, left=2.5cm,right=2.5cm,top=0.7cm,bottom=1.5cm, includehead,includefoot}
\fancyheadoffset[LO,RE]{0cm}
\fancyheadoffset[RO,LE]{0cm}
\vspace*{-24pt}
\begin{table}
\centering
\caption{Comparison of route choice models in previous studies.}%
\label{tab:previous_study}
\scriptsize
\begin{tabular*}{\textwidth}{@{\extracolsep{\fill}}
    >{\centering\arraybackslash}m{0.1\textwidth}
    >{\centering\arraybackslash}m{0.15\textwidth}
    >{\centering\arraybackslash}m{0.15\textwidth}
    >{\centering\arraybackslash}m{0.15\textwidth}
    >{\centering\arraybackslash}m{0.15\textwidth}
    >{\centering\arraybackslash}m{0.2\textwidth}
}
\toprule
\textbf{Item} 
& \textbf{Tang et al. \cite{tang_mining_2024}}
& \textbf{Ma et al. \cite{ma_nested_2020}}
& \textbf{Montini et al. \cite{montini_route_2017}}
& \textbf{Garcia-Martinez et al. \cite{garcia-martinez_transfer_2018}}
& \textbf{This Study} \\
\midrule

Data Type

& GPS and GIS data; survey data 
& 2011 Household Travel Survey 
& GPS tracking, OpenStreetMap, elevation, and survey data 
& RP and SP survey data 
& GPS trajectory data with land use and \par station-level attributes \\

   \midrule
Sample Size
& 35 individuals (109 trips) 
& 2967 samples from 29,670 trips by 9100 participants 
& 156 valid users 
& 78 respondents; 295 valid answers 
& 1.2 million users, per user per path choice \\

 \midrule
Route Choice Condition
& Fixed ODs; observed GPS over multiple commuting days 
& Mode-time combinations \par (not spatial routes),\par based on survey 
& Observed GPS; ODs vary; \par includes multiple trip purposes 
& Routes defined by survey design 
& Fixed ODs; commuter inferred \par from status algorithm \\

   \midrule
Model Used
& Neural Network (ANN) 
& Nested Logit (time-mode hierarchy) 
& MNL and Path Size Logit \par (joint mode-route) 
& Error Component Logit (ECL) 
& MNL with GPS-derived transfer penalties \\ 

   \midrule
Alternative Routes
& 10 route options per commuter generated using an extended BFS-LE algorithm (includes observed, shortest, and 8 synthetic routes based on network and traffic data) 
& Predefined combinations of mode and travel time bands; \par not actual physical routes 
& BFS-LE used for car/bike/walk; via-point and stop-based methods for public transport 
& Hypothetical public transport trips with 0, 1, or 2 transfers; generated via Ngene software in SP survey 
& Observed choice sets from GPS traces; \par alternatives identified by transfer patterns \\

   \midrule
Variables Used
& Traveler classification used demographics, job type, and household information. Route choice model included route length, detour rate, freeway access, number of intersections, and left turns 
& Mode-time choice modeled using demographics (age, income, job), household vehicle/bike ownership, and commute distance 
& Route utility based on travel time, road type ratio, elevation, number of turns/signals, trip purpose, and environmental attitudes 
& Utility modeled using transfer count, walking/waiting time, crowding, stairs, and user characteristics \par  (e.g., age, gender, habits) 
& Transfer count, commuting time, peak hour dummy, station capacity; variables derived from real movement data \\

   \midrule
{Results}
& ANN classification accuracy: 79\%; route choice prediction 
\par   accuracy: 67.3\% 
& Model fit: $R^{2} = 0.428/0.111$; \par  hit rate: 86.3\% 
& PT model: $R^{2} = 0.541$ \par (Basic CSG), $0.595$ (Via CSG); 
 \par joint model: $R^{2} = 0.326$ 
& No fit index reported; perceived transfer penalty estimated 
  at approximately 15--18 \par in-vehicle minutes 
& $R^2 = 0.232$; hit rate $\approx 55.8\%$; Spearman $\rho = 0.72$ (city-wide), $0.86$ (OD-specific); parameters stable across 12 months \\
\bottomrule
\end{tabular*}
\end{table}

\newpage
\restoregeometry
\paperwidth=\pdfpageheight
\paperheight=\pdfpagewidth
\pdfpageheight=\paperheight
\pdfpagewidth=\paperwidth
\headwidth=\textwidth

\section{Data Description and Preprocessing}%
\label{sect:data_description_and_processing}

This and next sections outline our approach to acquiring, processing, and refining large-scale GPS trajectory data for commuting pattern analysis. In this section  we describe the primary dataset (Section~\ref{subsect:data_description}), which provides high-resolution spatiotemporal mobility information from smartphone  across Japan's Shutoken metropolitan area. We then detail our first-stage data processing (Section~\ref{subsect:data_preprocessing}), including filtering criteria and population renormalization techniques to address sampling biases. 

\subsection{Data Description}%
\label{subsect:data_description}

This study utilizes a large-scale mobility GPS dataset provided by a private company, Agoop Corporation, containing anonymized location data from approximately 1.2 million smartphones per day across Japan, with each user having an average of 125 trajectory points per weekday \cite{ozaki_direct_2022, shida_potential_2022}. The GPS data provide timestamped latitude-longitude coordinates with high temporal resolution, typically captured at one-minute intervals, and an average spatial accuracy of approximately 10 m. Each data point includes a randomized user ID (reset nightly to protect privacy), timestamp, geographical coordinates, and associated home and work city codes. Additionally, we utilize railway network Geographic Information System (GIS) data from the Ministry of Land, Infrastructure, Transport and Tourism~\cite{mlit_railway_data}, which contains station locations (coordinates and attributes), railway line geometries (polyline representations), and transfer station indicators for transportation mode identification and route matching.

To ensure robust analysis of regular commuting patterns, we focused exclusively on weekday data throughout 2023 for the Shutoken area of Japan (Tokyo, Kanagawa, Chiba, Saitama), excluding weekends and holidays. This study period and area allowed us to concentrate on typical workday route choice patterns that represent routine commuting behavior.
 
We should note that the dataset has certain unavoidable limitations. First, as it relies on smartphone applications issued by a private company, there is a demographic bias in representation. According to the Ministry of Internal Affairs and Communications data, elderly individuals and children under 13 are underrepresented due to lower smartphone adoption rates in these groups \cite{ozaki_direct_2022}. Second, privacy protections result in certain data limitations—user IDs are randomized each night, preventing multi-day trajectory analysis for a single individual, and precise residential locations are blurred to central points within grid areas.

\subsection{Data Preprocessing}%
\label{subsect:data_preprocessing}

\textls[-15]{To prepare this extensive dataset for commuting pattern analysis, we implemented several data preprocessing steps to ensure quality and relevance. Several filtering criteria were applied to focus specifically on commuting behavior:}

\begin{itemize}[leftmargin=*, labelsep=6.2mm, itemsep=0pt, topsep=3pt, partopsep=0pt, parsep=0pt]
  \item Only  users with more than 100 location points per day were included to ensure sufficient trajectory information.
  \item Selected IDs where both home city code and work city code were within the study area.
  \item Applied population renormalization to align the GPS user sample with actual population figures based on official statistics, where the real population is consistently around 1.7 times the recorded monthly number of our dataset, as shown in {Supplementary Figure S4.}
\end{itemize}


\section{Methodology}%
\label{sect:methodology}

The model formulated in Section~\ref{sect:model} requires four quantities per recorded user in the GPS data: whether the user is a railway commuter, the location information of their origin and destination (OD pair), the sequence of transfer stations defining their chosen route, and the attributes characterizing each route alternative. This section describes how each quantity is extracted from raw GPS trajectories.

We first classify every GPS point into an activity status---\textit{home, work, move, stroll, stay}
---which provides the building blocks for everything that follows (Section~\ref{subsect:methodology_status_identification}). These status labels allow us to locate each user's home and work to define the OD pair, extract the morning commute segment between them, and distinguish short pauses (\textit{{stroll}}) from prolonged stops (\textit{{stay}}).

Next, we classify each commuter's primary transport mode---railway, car or bus, bicycle, walking
---and restrict the subsequent analysis to railway users (Section~\ref{subsect:methodology_transportation_mode}).

Finally, from each railway commuter's trajectory we extract \textit{{interruptions}
}, defined as \textit{{stroll}} states bounded by \textit{{move}} states on both sides. An interruption occurring within a station-area grid cell is identified as a transfer (Section~\ref{subsect:methodology_commuting_motif}).

\subsection{Status Identification}%
\label{subsect:methodology_status_identification}

Based on previous research \cite{su_pattern_2020, shida_potential_2022, ozaki_direct_2022}, we further developed an algorithm to classify each point within the trajectory data to detailed movement states (\textit{home, work, move, stroll, stay}), according to Table~\ref{tab:movement_states}. Let $S(t)$ represent the activity status of a commuter at time $t$, defined as
\begin{equation}
  S(t) \in \{home, work, move, stroll, stay\}
\end{equation}

The complete trajectory can therefore be expressed as an ordered sequence:
\begin{equation}
  \{S(t_1), S(t_2), \ldots, S(t_n)\} \quad \text{where} \quad t_1 < t_2 < \cdots < t_n
\end{equation}

This classification forms the foundation for our subsequent analysis of commuting patterns. The time-based status distribution is shown in {Supplementary Figure S9.}

\vspace{-12pt}
\begin{table}[H]
\begin{threeparttable}
\caption{Criteria for movement state classification.}%
\label{tab:movement_states}
\begin{tabular*}{\textwidth}{@{\extracolsep{\fill}}
    >{\centering\arraybackslash}m{0.15\textwidth}
    >{\raggedright\arraybackslash}m{0.8\textwidth} 
}
\toprule
\textbf{Status} & \multicolumn{1}{>{\centering\arraybackslash}m{0.8\textwidth}}{\textbf{Classification Criteria}}  \\
\midrule
Home & Located within the identified 100m residential grid cell;
       stay duration exceeding 4 h; \par
       first observation after 5:00 AM \\ \midrule
Work & Located within the identified 100m workplace grid cell;
       stay duration exceeding 5 h; \par
       not coinciding with home location \\   \midrule
Move & Speed exceeding 8 km/h \footnotemark[1];
       or stopping time less than 4 min \footnotemark[2] within two consecutive \par mesh cells \\ \midrule
Stroll & Speed below 8 km/h;
         stopping time exceeding 4 min \footnotemark[2];
         continuous duration in a 1 km grid cell for less than 30 min \\  \midrule
Stay & Speed below 8 km/h;
       continuous duration in a 1 km grid cell exceeding 30 min \\
\bottomrule%
\end{tabular*}
\begin{tablenotes}[flushleft]
\fontsize{9}{10.8}\selectfont
\item[1] {The 8~km/h threshold (approximately 2.2~m/s) exceeds typical walking speed (~1.3~m/s~\cite{zhang_activity_2024}) and approximates adult running speed, ensuring that the \textit{move} state captures active transport (by any mode) rather than pedestrian movement.} \textsuperscript{2} {Pauses shorter than 4 min are treated as part of the \textit{move} status because they typically reflect incidental halts within an ongoing trip rather than a distinct visit. Pauses exceeding this threshold are more likely to 
correspond to structurally meaningful interruptions; the 4-min cutoff is calibrated against the average interruption duration reported in~\cite{mlit_transfer_survey_2016}.}
\end{tablenotes}
\end{threeparttable}
\end{table}

Then we represent a user's complete daily trajectory $\mathcal{T}'$ as a sequence of tuples $(p_i, t_i, S(t_i))$ of $i$-th GPS location $p_i$, timestamp $t_i$, and user's status $S(t_i)$. We then extract the morning commute segment as all trajectory points between home departure and work arrival, by identifying the departure time $t_{\text{departure}}$ (time of the first non-home status after being at home) and arrival time $t_{\text{arrival}}$ (time of the first work status after leaving home), and then calculate the door-to-door commuting time $T_{\text{commute}}$ and commuting distance $D_{\text{commute}}$:
\begin{subequations}
  \begin{align}
    T_{\text{commute}} &\coloneqq t_{\text{arrival}} - t_{\text{departure}} \\
    D_{\text{commute}} &\coloneqq \sum_{i=1}^{M-1} d(p_i, p_{i+1}), \quad \text{for all } p_i \text{ where } t_{\text{departure}} \leq t_i \leq t_{\text{arrival}}
  \end{align}
  \label{eq:commuting_time_and_distance_for_trajectory}%
\end{subequations}
where $M$ is the number of GPS points recorded during the commute period, and $d(p_i, p_{i+1})$ is the geographic distance between consecutive points calculated using the Haversine formula to account for Earth's curvature.

For visualization purposes, Supplementary Figure S5 illustrates the daily trajectory of a typical commuter, showcasing the complete movement pattern: departure from the \textit{home} location in the morning, extended stay at the \textit{work} location during daytime hours, and return journey to the \textit{home} location in the evening.

Each of the five states plays a distinct role in the downstream pipeline. home and work locate the OD pair and bound the morning commute segment (this section); move enables transport mode classification through GIS proximity (Section \ref{subsect:methodology_transportation_mode}); also move, stroll and stay are the raw input to interruption and transfer detection (Section~\ref{subsect:methodology_commuting_motif}) and excludes non-commuting trajectories.

\subsection{Transportation Mode Identification}%
\label{subsect:methodology_transportation_mode}

The commuting motifs provide valuable insight into the structural patterns of daily journeys; however, understanding transportation mode choice is essential for interpreting these motifs within the broader context of urban mobility systems \cite{montini_route_2017, takahashi_transportation_2019, sadeghian_review_2021, markos_unsupervised_2020, montazeri-gh_traffic_2011}. During a commuting process, users may employ multiple combinations of transportation modes when selecting their routes \cite{rasmussen_improved_2015, sadeghian_review_2021}. To address this complexity, we developed a comprehensive approach to identify each user's primary mode of transport, defined as the mode accounting for the largest proportion of total travel distance among all trip segments.

We then constructed an algorithm to classify commuting trajectories into four transportation modes, railway, walking, cycling, bus or car, by extracting sequence-based features of velocity, acceleration, and spatial patterns from GPS trajectories, partially following the methodologies of \cite{sadeghian_stepwise_2022, zheng_learning_2008}. Since cars and buses share the same road network and distinguishing them requires rather high-resolution data \cite{dabiri_inferring_2018, sadeghian_review_2021, che_identifying_2016}, they are combined into a single category in this study. The classification procedure follows a rule-based framework, as detailed in Table~\ref{tab:transportation_modes}.

\vspace{-12pt}
\begin{table}[H]
\small
\begin{threeparttable}
\caption{Criteria for transportation mode classification.}%
\label{tab:transportation_modes}
\begin{tabular*}{\textwidth}{@{\extracolsep{\fill}}
    >{\centering\arraybackslash}m{0.2\textwidth}
    >{\raggedright\arraybackslash}m{0.75\textwidth} 
}
\toprule
\textbf{Mode} &  \multicolumn{1}{>{\centering\arraybackslash}m{0.75\textwidth}}{\textbf{Classification Criteria}} 
 \\
\midrule
Walking & Mean speed below 6 km/h;  \par
          Maximum speed below 12 km/h; 
          Total travel distance less than 4 km \\ \midrule
Cycling & Mean speed below 18 km/h;  \par
          Maximum speed below 30 km/h; 
          Total travel distance less than 10 km \\ \midrule
Railway & Travel distance greater than 1 km; \par
          Average moving points' distance to railway network less than 90 m \footnotemark[1] \\ \midrule
Bus or Car & All remaining trajectories after other mode classification \\
\bottomrule%
\end{tabular*}
\begin{tablenotes}[flushleft]
\fontsize{9}{10.8}\selectfont
\item[1] {Railway classification incorporates GIS-based proximity analysis.}
\end{tablenotes}
\end{threeparttable}
\end{table}
\vspace{-3pt}

\begin{figure}[H]
\centering
\includegraphics[width=0.85\textwidth]{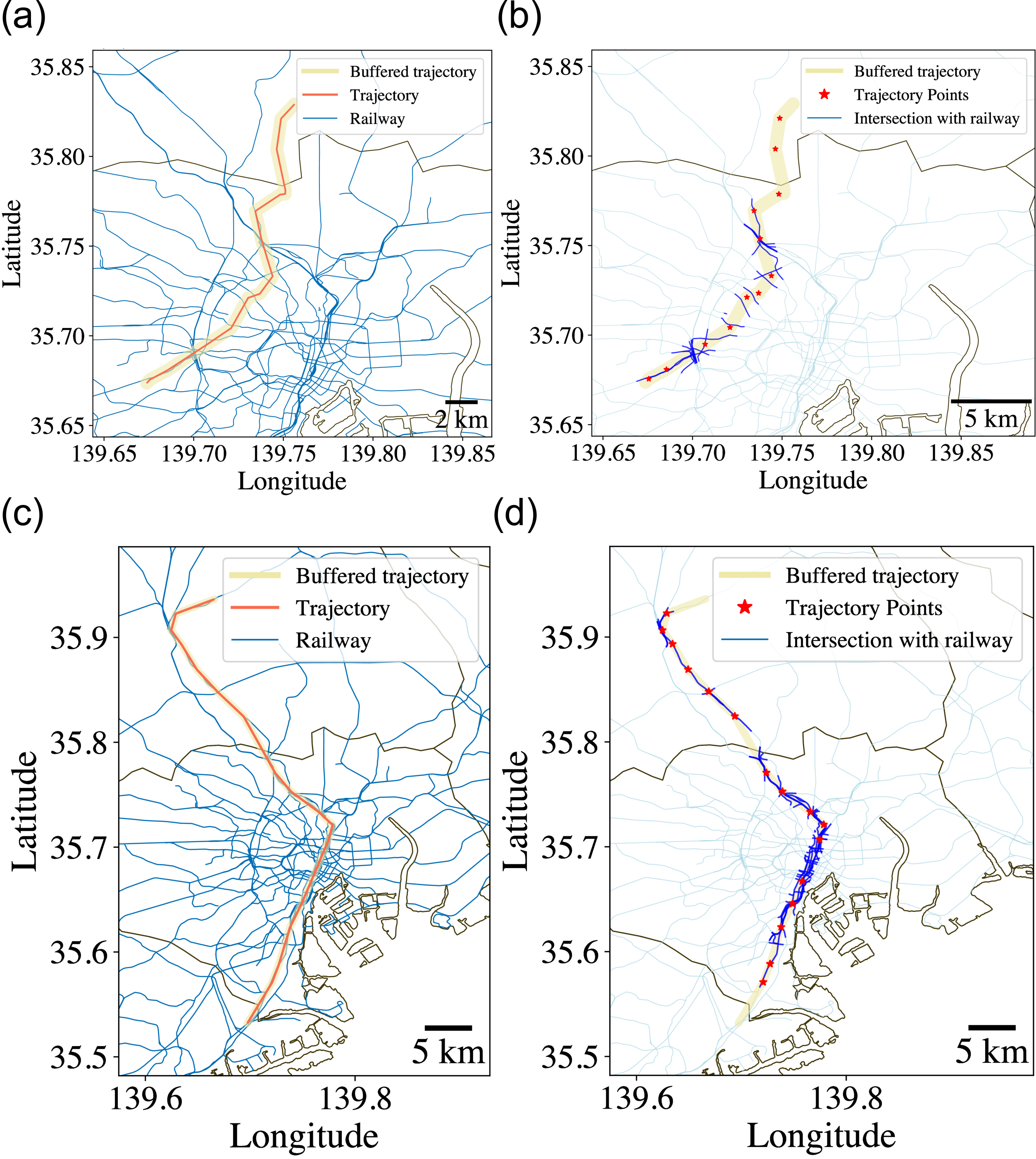}
\caption{Transportation mode identification process for railway and car commuters. Raw GPS trajectories (orange lines) are overlaid on the railway network (light blue lines) with 100-meter buffered zones (yellow). Dark blue segments indicate intersections between buffered zones and railway lines. Red stars mark identified trajectory points within these intersection areas. (\textbf{a},\textbf{b}) Car commuter example: large distances between \textit{move} status GPS points and railway network result in car mode classification. (\textbf{c},\textbf{d}) Railway commuter example: short distances between  \textit{move} status GPS points and railway network result in railway mode classification.}
\label{fig3}
\end{figure}

Among these classification methods, railway identification required a particularly sophisticated approach due to the unique characteristics of rail transport. For railway identification specifically, we employed a hybrid approach combining numerical analysis and GIS methods. As shown in Figure~\ref{fig3} we extracted railway networks from geospatial data and calculated the proximity of GPS trajectory points to these networks. To improve classification accuracy, we created a 100-meter buffer around railway lines to account for GPS measurement accuracy, and calculated the average distance from each trajectory point with  \textit{move} status to the nearest railway segment. Additionally, a trajectory is classified as a railway commuter only if all trajectory points in moving status have their nearest railway line within 5 candidate lines, which prevents misclassification in areas where road and rail networks overlap densely.

\begin{figure}[H]
\centering
\includegraphics[width=0.85\textwidth]{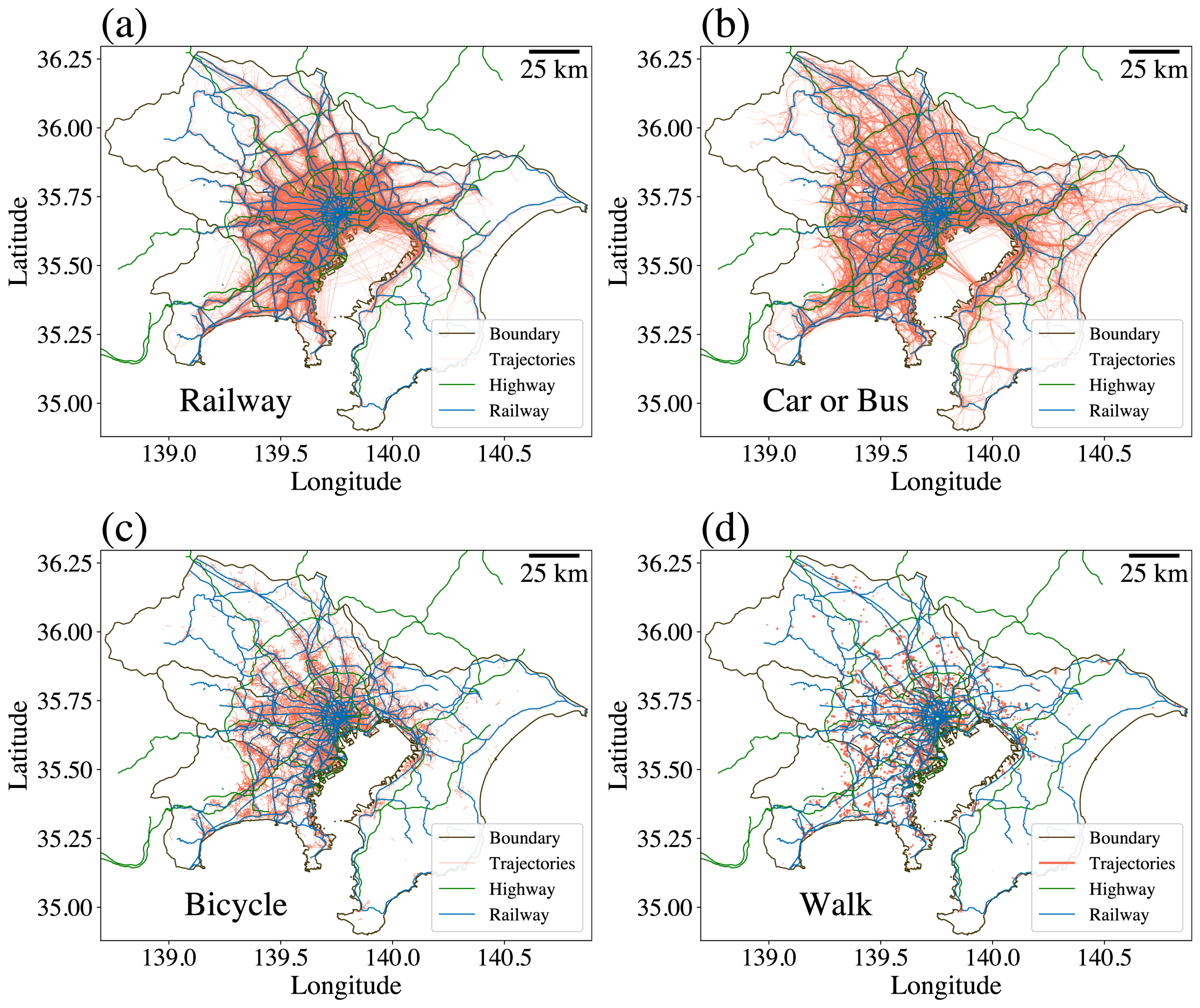}
\caption{Aggregated trajectories for four recognized transportation modes. 
 (\textbf{a}) Railway trajectories (red lines) closely follow the railway network (blue lines), demonstrating high spatial correlation with rail infrastructure. While some trajectories, such as those across the sea (Tokyo Bay), may appear discontinuous due to signal loss from tunnels or urban canyons, these instances are infrequent and do not compromise the overall dataset integrity for aggregate analysis. \textls[-15]{(\textbf{b}) Car or bus trajectories closely align the highway network (green lines) and urban road systems, instead of the railway network. Notably, the trajectories across Tokyo Bay align with the shape of the Tokyo Bay Aqua-Line. (\textbf{c}) Bicycle trajectories exhibit intermediate-range mobility patterns with moderate spatial coverage. (\textbf{d}) Walking trajectories display short-distance, localized movement patterns concentrated in urban areas. The distinct spatial characteristics of each mode validate the effectiveness of the transportation mode classification algorithm in this study}.}
\label{fig4}
\end{figure}

\textls[-22]{This multi-modal classification approach enables detailed analysis of transportation behavior across different user segments. Figure~\ref{fig4} illustrates the aggregated trajectory linestrings for each identified transportation mode. Railway users predominantly travel along railway lines, cars and buses follow more complex road systems, while bicycles and pedestrians exhibit notably shorter path lengths. We validated our classification results by examining Figure~\ref{fig4} and comparing the transportation mode distribution (Supplementary Figure S1) with Japan's national transportation mode choice report \cite{tokyo_public_transport_status}. Our data shows a higher proportion of railway users, likely reflecting sampling bias in smartphone data that potentially excludes individuals under 13 and over 65 years of age, and non-commuting users. After implementing this classification framework across our dataset, we validated our approach through comparison with official statistics.}

\subsection{Transfer Identification and Validation}%
\label{subsect:methodology_commuting_motif}

\textls[-15]{The route choice model requires two quantities from each railway commuter's trajectory: the number of burdensome transfers and the sequence of transfer stations defining each route alternative. We extract both by first reducing the high-resolution status sequence into a compact representation capturing only structural stops, the \textit{{commuting motif}}, following network-science conventions \cite{schneider_unravelling_2013, jiang_activity-based_2017, cao_characterizing_2019}, and then identifying which stops correspond to transfers.}



The motif representation serves two analytical purposes beyond providing model input. It validates the transfer-identification pipeline at this stage of the refinement chain. We compare the observed $n \geq 3$ share (approximately 10\%) against the independent estimate for Tokyo reported by \cite{at_home_trend_2018}, which uses a survey-based transfer definition that, like the motif-level count here, includes both burdensome and strategic (express-switching) transfers. Matching the survey's definition is what makes the comparison meaningful; the model-specific refinement to {burdensome} transfers is applied later, in Section~\ref{subsect:methodology_route_alternatives}.

\vspace{-18pt}
The extraction proceeds in two steps:

\begin{enumerate}[leftmargin=2.2em, labelsep=4mm, label=(\arabic*), itemsep=0pt, topsep=3pt, parsep=0pt]

     \item Status sequence simplification: 
     Converting the complete status sequence $\{S(t_i)\}_{i=1}^M$ (with $M$ time points, Section~\ref{subsect:methodology_status_identification}) into a simplified sequence $\{S'(t_c)\}_{c=1}^m$ (with $m < M$ time points) by retaining only status changes. Each consecutive status must differ from the previous one: $S'(t_c) \neq S'(t_{c-1})$ for all $c>1$. For example, if the original sequence is \textit{(home--home--home--move--move--work)}, the simplified sequence becomes \textit{(home--move--work)}.

    \item Interruption identification: \textls[-22]{Detecting interruptions where $S'(t_k)=\text{stroll}$ occurs between two movement periods:}
    \begin{equation}
    \label{eq:stroll_between_move}
      S'(t_{k-1})=\text{move} \quad \text{and} \quad S'(t_{k+1})=\text{move}
    \end{equation}
    
\end{enumerate}

Through our analysis, we identified several characteristic commuting motifs that represent different degrees of commuting smoothness in Table~\ref{tab:motif}. At this stage, the definitions apply to all transportation modes; for railway commuters specifically, we impose additional spatial constraints below (and summarize the full refinement pipeline in Table
~\ref{tab:interruption_evolution}).

\vspace{-36pt}
\begin{table}[H]
\small
\centering
\caption{Characteristic commuting motifs that represent different degrees of commuting smoothness.}%
\label{tab:motif}

\begin{tabular*}{\textwidth}{@{\extracolsep{\fill}}
    >{\centering\arraybackslash}m{0.13\textwidth}
    >{\centering\arraybackslash}m{0.1\textwidth} 
  >{\centering\arraybackslash}m{0.2\textwidth} 
  >{\centering\arraybackslash}m{0.5\textwidth} 
}
\toprule
\textbf{Stop Count} & \textbf{Motif} & \textbf{Description}&\textbf{Route} \\
\midrule

$n = 0$ &
\begin{tikzpicture}
  \node[circle, draw, fill=black, inner sep=1.5pt] (home) at (0,0) {};
  \node[circle, draw, fill=black, inner sep=1.5pt] (work) at (1,0) {};
  \draw[->, thick] (home) -- (work);
\end{tikzpicture}
&
Smooth
& Home$\to$move$\to$work \\
\midrule

$n = 1$ &
\begin{tikzpicture}
  \node[circle, draw, fill=black, inner sep=1.5pt] (home) at (0,0) {};
  \node[circle, draw, fill=black, inner sep=1.5pt] (stop) at (0.5,0.5) {};
  \node[circle, draw, fill=black, inner sep=1.5pt] (work) at (1,0) {};
  \draw[->, thick] (home) -- (stop) -- (work);
\end{tikzpicture}
&
Interrupt once & Home$\to$move$\to$\textbf{{stroll}}$\to$move$\to$work \\
\midrule

$n = 2$ &
\begin{tikzpicture}
  \node[circle, draw, fill=black, inner sep=1.5pt] (home) at (0,0) {};
  \node[circle, draw, fill=black, inner sep=1.5pt] (stop1) at (0.25,0.5) {};
  \node[circle, draw, fill=black, inner sep=1.5pt] (stop2) at (0.75,0.5) {};
  \node[circle, draw, fill=black, inner sep=1.5pt] (work) at (1,0) {};
  \draw[->, thick] (home) -- (stop1) -- (stop2) -- (work);
\end{tikzpicture}
&
Interrupt twice & Home$\to$move$\to$\textbf{{stroll}}$\to$move$\to$\textbf{{stroll}}$\to$move$\to$work \\
\midrule

$n\geq3$ &
\begin{tikzpicture}
  \node[circle, draw, fill=black, inner sep=1.5pt] (home) at (0,0) {};
  \node[circle, draw, fill=black, inner sep=1.5pt] (stop1) at (0.25,0.5) {};
   \node[circle, draw, fill=black, inner sep=1.5pt] (stop2) at (0.5,0.25) {};
  \node[circle, draw, fill=black, inner sep=1.5pt] (stop3) at (0.75,0.5) {};
  \node[circle, draw, fill=black, inner sep=1.5pt] (work) at (1,0) {};
  \draw[->, thick] (home) -- (stop1) -- (stop2) -- (stop3) -- (work);
\end{tikzpicture}
&
Interrupt multiple & Home $\to \ldots \to$ work \\
\midrule

$\text{Other}$ &
\begin{tikzpicture}
  \node[circle, draw, fill=black, inner sep=1.5pt] (home) at (0,0) {};
  \node[rectangle, draw, fill=red, minimum width=5pt, minimum height=5pt] (visit) at (0.5,0.5) {};
  \node[circle, draw, fill=black, inner sep=1.5pt] (work) at (1,0) {};
  \draw[->, thick] (home) -- (visit) -- (work);
\end{tikzpicture}
&
Visited over 30 min & Home$\to$\textbf{{stay}}$\to$work \\
\bottomrule


\end{tabular*}
\end{table}

A pattern \textit{home-stay-work} indicating a prolonged stationary period between home and work locations. These behaviors do not correspond to typical transfer behavior and are excluded from the route-choice model; their inclusion in Table~\ref{tab:motif} is for pattern display only.

For railway commuters, we additionally require that the \textit{stroll} points fall within a 1~km grid cell containing a railway station, ensuring that only station-area pauses, predominantly transfers are counted.

The 4-min lower bound on the \textit{stroll} duration (Table~\ref{tab:movement_states}) is chosen to match the average transfer time in Tokyo reported by the MLIT transfer survey \cite{mlit_transfer_survey_2016}: 4.3~min at peak and 3.4~min off-peak. This threshold captures the majority of genuine transfers while excluding brief operational pauses such as signal adjustments or short platform holds. The 30-min upper bound further excludes non-transfer deviations such as dropping off children or mid-commute meetings, which are assigned to the \textit{stay} status rather than counted as interruptions.

\vspace{-12pt}
\begin{table}[H]
\small
\begin{threeparttable}
\caption{Progressive refinement of the interruption concept through the analytical pipeline.}%
\label{tab:interruption_evolution}
\begin{tabular*}{\textwidth}{@{\extracolsep{\fill}}
    >{\centering\arraybackslash}m{0.2\textwidth}
    >{\centering\arraybackslash}m{0.15\textwidth} 
 >{\raggedright\arraybackslash}m{0.6\textwidth} 
}
\toprule
\textbf{Stage} & \textbf{Section} & \multicolumn{1}{>{\centering\arraybackslash}m{0.55\textwidth}}{\textbf{Definition and Scope}} \\
\midrule
Stroll status & Section~\ref{subsect:methodology_status_identification} & All modes
: Speed $<$ 8\,km/h; duration from 4 to 30 min in a 1\,km grid cell. \\ \midrule
Interruption & Section~\ref{subsect:methodology_commuting_motif} & All modes: A stroll between two move states (Equation~(\ref{eq:stroll_between_move})).\\ \midrule
Transfer & Section~\ref{subsect:methodology_commuting_motif} & Railway users: An interruption whose stroll points fall within a station-area grid cell \footnotemark[1]. \\ \midrule
Express transfer & Section~\ref{subsect:methodology_route_alternatives} & Railway users: A transfer where the transfer route is faster than the fastest direct route for the same OD.\\ \midrule
Model variable $\mathrm{NT}_{nj}$ & Section~\ref{subsect:model_variable_specification} & Railway users only: Number of pure burdensome transfers, excluding express transfers (Equation~(\ref{eq:x_spec})).\\
\bottomrule%
\end{tabular*}
\begin{tablenotes}[flushleft]
\fontsize{9}{10.8}\selectfont 
 \item \textsuperscript{1} {A 1\,km grid cell containing a railway station; threshold based on MLIT transfer survey \cite{mlit_transfer_survey_2016}.}
\end{tablenotes}
\end{threeparttable}
\end{table}

\begin{figure}[H]
\centering
\includegraphics[width=0.5\textwidth]{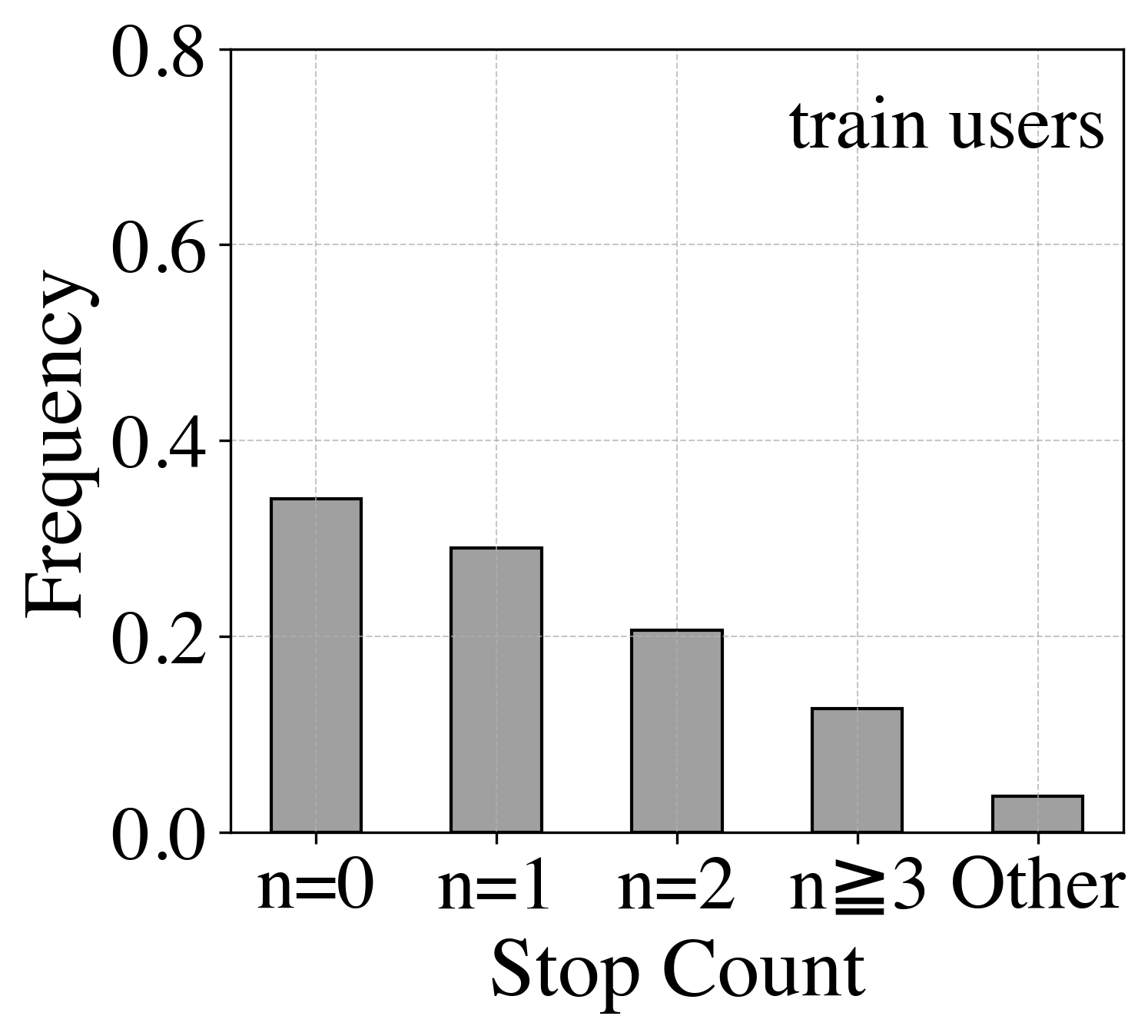}
\caption{Commuting motifs distribution for railway commuters. 
Motifs are classified by the number of interruptions (stop count) between home and work, where each interruption occurring within a 1~km grid cell containing a railway station is identified as a transfer. Direct routes (n = 0) account for approximately 35\%, with single-transfer (n = 1) and double-transfer (n = 2) routes comprising a substantial share, reflecting the structural complexity of Tokyo's railway network. The \textit{Other} category captures trips containing a \textit{stay} state (pause exceeding 30~min); these are excluded from the route-choice model.}
\label{fig2}
\end{figure}

The distribution in Figure~\ref{fig2} confirms that transfer behavior is the dominant source of route variation in Tokyo's railway network, justifying our focus on transfer counts as a primary attribute in the route-choice model. As a validation, Ref. \cite{at_home_trend_2018} reports that commutes involving three or more transfers represent approximately 10\% of the Tokyo commuting population, closely matching the 10\% observed in our $n \geq 3$ bin. The corresponding distribution for non-railway commuters, together with the cross-mode comparison, is provided in Supplementary Figure~S13.

\vspace{-26pt}

\subsection{Route Alternatives and Choice Set Construction}%
\label{subsect:methodology_route_alternatives}

The route-choice analysis is restricted to railway commuters. Each choice situation is defined by a fixed Origin--Destination (OD) pair, spatially anchored to 1~km grid cells containing the commuter's home and work locations. We use 1~km grids rather than station-based Voronoi meshes~\cite{sakamanee_methods_2020} because the high station density in Tokyo's Central Business District (CBD) produces extremely small, irregular Voronoi cells. These meshes are sensitive to minor GPS errors and frequently misclassify origins and destinations; a visual comparison is provided in Figure~\ref{fig6}a. The 1~km grid provides a robust, standardized spatial unit that mitigates GPS noise and guarantees sufficient sample sizes per OD pair.

We distinguish two concepts used throughout the remainder of the paper. A \textit{route alternative} is an abstraction: the commuting strategy shared by all realized commutes within the same OD pair that share the same departure time window and transfer-station sequence. We also refer to the \textit{physical route} as the transfer-station sequence alone, independent of peak-hour status; a single physical route may correspond to two route alternatives (one peak, one off-peak) if it is taken by commuters at both times of day. Travel time and distance of a route alternative are defined as the median over the realized commutes that instantiate it. The choice set faced by each commuter consists of the route alternatives available within their OD pair, and the route-choice model describes which alternative each commuter's trajectory is assumed to represent.

\subsubsection{Per-Trajectory Extraction}
For each raw trajectory we record:
\begin{enumerate}[leftmargin=2.2em, labelsep=4mm, label=(\arabic*), itemsep=0pt, topsep=3pt, parsep=0pt]
    \item Origin and destination. 
The 1~km grid cells containing the first and last GPS points, identified via the \textit{home} and \textit{work} statuses from Section~\ref{subsect:methodology_status_identification}.
    \item Departure time and peak-hour indicator. The timestamp of the first non-\textit{home} point after morning departure. The trajectory is tagged \textit{peak} if this timestamp falls between 07:00 and 10:00, and \textit{off-peak} otherwise.
    \item Transfer count and transfer stations. The number of transfers (defined in Section~\ref{subsect:methodology_commuting_motif}) and the station at which each occurred, encoded by its Voronoi mesh identifier (Figure~\ref{fig6}a). 
\end{enumerate}

\subsubsection{Route Tagging and Aggregation}
Each trajectory is assigned a \textit{route tag} combining its peak-hour status, transfer pattern, and transfer-station identifiers. Two trajectories within the same OD pair and with identical tags represent the same route alternative; their travel times and distances are aggregated by the median. Table~\ref{tab:route_tag_example} shows the resulting records for a representative OD pair.
\vspace{-10pt}

\begin{table}[htbp]
\begin{threeparttable}
\centering
\caption{Sample of route classification results and feature encoding}%
\label{tab:route_tag_example}
\scriptsize
\newcolumntype{c}{>{\centering\arraybackslash}X}
\begin{tabularx}{\textwidth}{llcccccc} 
\toprule
\textbf{OD} & \textbf{Route Tag} & $\textbf{Time}$ & $\textbf{Distance}$ & $\textbf{Departure}$ & $\textbf{No. of Transfer}$ & $\bm{\ldots}$ & $\textbf{Station}$ \\
\midrule
OD$_1$ & peak\_direct\_-1 & 65.0 & 60.03 & 7:10 & 0 & $\ldots$ &---\\
OD$_1$ & peak\_direct\_$Vor_{station}$ & 55.0 & 53.85 & 10:10 & \textbf{0} & $\ldots$ & $Vor_{station}$ \\
OD$_1$ & offpeak\_transfer\_$Vor_{station}$ & 59.0 & 54.85 & 10:10 & 1 & $\ldots$ & $Vor_{station}$ \\
\bottomrule
\end{tabularx}

\begin{tablenotes}[flushleft]
\item \textbf{Note.} The route alternatives for each origin–destination pair together form the choice set. Each row represents a route alternative; its Time and Distance values are the medians over all realized commutes sharing the alternative's tag. The Transfer Station column refers to the Voronoi mesh ID $Vor_{\text{station}}$ of the transfer location; when two transfers occur, two separate columns are used. The Voronoi meshes are visualized in Figure~\ref{fig6}a. For OD$_1$, the first row is a genuine non-transfer commute (Transfer Station $= -1$), while the second row captures a local-to-express switch at $Vor_{\text{station}}$. Although express switching does not increment No. of Transfer, the recorded transfer station produces a distinct route tag, preserving the slow-train vs. express-train distinction.


\end{tablenotes}
\end{threeparttable}
\end{table}
\vspace{-8pt}

\begin{figure}[H]
\centering
\includegraphics[width=0.95\linewidth]{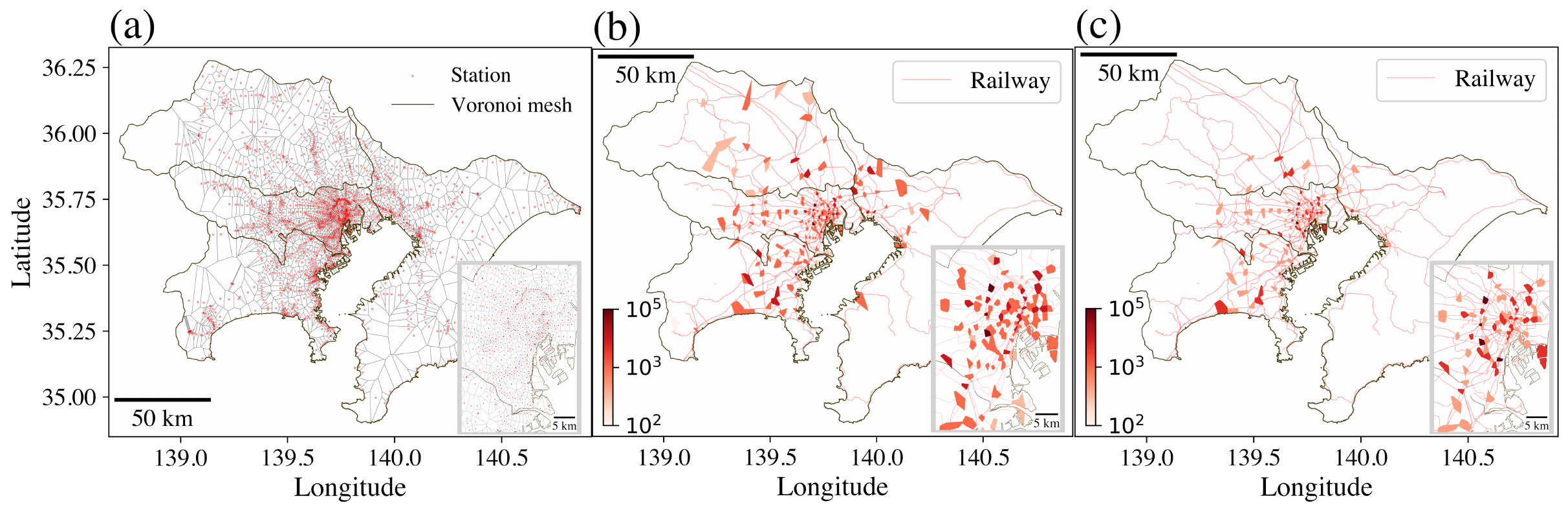}\\
\caption{Voronoi tessellation of railway stations and geographical distribution of transfer stations in the study area. 
(\textbf{a}) Voronoi cells constructed based on station locations, with red points indicating the nearest station for each cell. (\textbf{b}) Geographical distribution of first transfer stations; the inset shows the zoomed Tokyo Central Business District (CBD). The color scale represents the number of commuters making their first (or only) transfer at each station. (\textbf{c}) Geographical distribution of second transfer stations. High-usage stations are more concentrated toward the CBD compared to first transfer stations.}
\label{fig6}
\end{figure}
\vspace{-16pt}

Two filters are applied before estimation. First, OD pairs with fewer than 20 realized commutes are excluded, and the sensitivity test to test whether 20 is statistically sufficient for the model is provided in Supplementary Section~10 (Robustness Check). Second, OD pairs with fewer than 2 or more than 6 physical routes are excluded, following~\cite{hibino_alternative_2001}: single-route OD pairs contain no choice to model, while OD pairs with very large choice sets typically reflect non-routine travel or GPS noise rather than stable commuting behavior.

\subsubsection{Express-Transition Adjustment}
\label{para:express}

Tokyo's railway network runs local and express services on many lines, so a small number of transfers are strategic rather than burdensome: a commuter accepts a transfer in order to board a faster express train. Because such transfers save time rather than add cost, we reclassify them before finalizing the transfer count:
\begin{enumerate}[leftmargin=2.2em, labelsep=4mm, label=(\arabic*), itemsep=0pt, topsep=3pt, parsep=0pt]
    \item For each OD pair, find the travel time of the fastest route alternative whose tag indicates no transfer.
    \item Any route alternative with one or more transfers but a median travel time lower than this benchmark is reclassified: its transfer count is set to zero and its transfer-status tag component is reset to \textit{direct}, while the transfer-station identifiers are retained. The stations are kept in the tag to preserve the distinction between local and express services sharing the same endpoints.
    \item After this adjustment, all transfers remaining in the dataset are burdensome, i.e.\ time-adding.
\end{enumerate}

As illustrated in the second row of Table~\ref{tab:route_tag_example}, an express transition produces a record with zero transfers but a non-empty transfer-station field.

Because transfers are identified through successive filtering stages, we summarize the progressive refinement in Table~\ref{tab:interruption_evolution} to clarify how each stage narrows the concept.

\section{Model}%
\label{sect:model}

\textls[-30]{Our modeling framework employs a discrete choice approach based on the Multinomial Logit (MNL) model 
\cite{mcfadden_conditional_1972, train_discrete_2009},} utilizing 
the route and feature information constructed in 
Section~\ref{sect:methodology}. Commuters are assumed to select 
routes that minimize a generalized cost incorporating travel time, 
transfers, and crowding. 
Section~\ref{subsect:model_canonical} introduces the MNL framework with an explicit scale--direction decomposition of the estimated parameter vector. Section~\ref{subsect:model_variable_specification} 
specifies the route attributes entering the model and the rationale 
for their selection. Section~\ref{subsect:model_estimation} describes parameter estimation, the {two estimation schemes (Model 1 and Model 2)}, and the metrics used for model evaluation.

\subsection{Route Choice Model}%
\label{subsect:model_canonical}

We model route choice using a Multinomial Logit (MNL) 
framework~\cite{mcfadden_conditional_1972,train_discrete_2009} in which commuters select routes that minimize a generalized energy cost. Let $C_n$ denote the set of feasible routes available to commuter $n$. The cost of route $j \in C_n$ is modeled as a linear 
combination of its attributes:

\begin{equation} 
\label{eq:energy} 
E_{nj} = \mathbf{X}_{nj}^{\top}\boldsymbol{\omega}
\end{equation} 
where $\mathbf{X}_{nj}$ and $\boldsymbol{\omega}$ are vectors of 
the same dimension, determined by the number of route attributes. 
$\mathbf{X}_{nj}$ contains the attribute values of route $j$ for 
commuter $n$ (e.g., travel time, number of transfers), and 
$\boldsymbol{\omega}$ is the weight vector quantifying each 
attribute's contribution to the total cost, a normalized direction vector. A positive element of $\boldsymbol{\omega}$ indicates that the corresponding attribute increases cost and thus reduces choice probability; a negative element indicates the opposite.

The probability that commuter $n$ chooses route $j$ follows the 
MNL form:

\begin{equation}
  \label{eq:prob}
  P_{nj} = \frac{\exp(-\beta E_{nj})}{\sum_{k \in C_n} \exp(-\beta E_{nk})}
\end{equation}

Equation~(\ref{eq:prob}) can be derived from the standard random 
utility framework~\cite{train_discrete_2009, ben_discrete_1985}, 
which makes the role of $\beta$ explicit. Each commuter $n$ is 
assumed to associate each route $j$ with a utility

\begin{equation}
\label{eq:utility_with_error}
U_{nj} = -E_{nj} + \varepsilon_{nj}
\end{equation}
\textls[-15]{where $E_{nj}$ is the deterministic cost defined in Equation~(\ref{eq:energy}) and $\varepsilon_{nj}$ is an unobserved error term capturing idiosyncratic factors. When $\varepsilon_{nj}$ is i.i.d. and Gumbel-distributed with scale parameter $\mu$, integrating over the error yields:}

\begin{equation}
\label{eq:rum_prob}
P_{nj} = \frac{\exp(-E_{nj}/\mu)}{\sum_{k \in C_n} \exp(-E_{nk}/\mu)}
\end{equation}
which recovers Equation~(\ref{eq:prob}) with $\beta = 1/\mu$. In this 
view, $\beta$ is the inverse of the Gumbel error scale, 
controlling how sharply the deterministic cost dominates the 
unobserved stochastic component.

In standard applications, the error scale is typically absorbed into the utility coefficients, preventing separate identification from choice data \cite{train_discrete_2009}. Rather than claiming structural identification, we apply a standardization convention by constraining the $\ell_2$-norm of the weight vector ($\lVert \boldsymbol{\omega} \rVert_2 = 1$). In practice, after estimating an unconstrained vector $\boldsymbol{\omega}_{\text{raw}}$ via maximum likelihood, we define:

\begin{equation}
  \label{eq:decomp}
  \beta = \lVert \boldsymbol{\omega}_{\text{raw}} \rVert_2, 
  \quad 
  \boldsymbol{\omega} = \frac{\boldsymbol{\omega}_{\text{raw}}}{\lVert \boldsymbol{\omega}_{\text{raw}} \rVert_2}
\end{equation}

Fixing $\|\omega\|_2 = 1$ is strictly an identifying convention. It places $\beta$ on a standardized scale, enabling direct comparisons of \textit{choice determinism} across OD pairs, months, and studies that adopt the same $[0,1]$ feature scaling. Under this shared metric, a larger $\beta$ indicates that collective choices respond more sharply to cost differences, while a smaller $\beta$ approaches random selection.

Under this convention, $\beta$ inherits the interpretation from 
Equation~(\ref{eq:rum_prob}) as the inverse Gumbel error scale: it 
measures how sharply the population's collective choices respond 
to cost differences, with larger $\beta$ producing more 
deterministic choices and smaller $\beta$ approaching random 
selection. Meanwhile, $\boldsymbol{\omega}$ captures the relative 
importance of each route attribute.

The MNL specification in Equation~(\ref{eq:prob}) carries the Independence of Irrelevant Alternatives (IIA) assumption: the relative odds of any two alternatives depend only on those two alternatives' attributes, not on the presence or attributes of a third~\cite{train_discrete_2009}. IIA becomes problematic when alternatives share substantial unobserved features, such as overlapping physical segments in a route-choice context. Two modeling choices in this study mitigate the assumption's consequences. First, we define alternatives at the category level (peak status, transfer-station sequence), which aggregates partially overlapping physical paths into structurally distinct options and reduces link-level redundancy. Second, the OD-specific estimation variant (Section~\ref{subsubsect:model_two_model_variants}) absorbs correlation effects that are local to each network topology, which a pooled model would have to treat as IIA violations. A full path-size correction 
~\cite{dixit_perception_2023} remains a direction for future work (Section~\ref{sect:conclusion_limitation}).

The equivalence between this MNL formulation and the canonical ensemble \cite{anas_discrete_1983, helbing_the_physics_2004} of statistical physics is discussed in Section~\ref{sect:conclusion_limitation}; the full Maximum Entropy derivation is provided in Appendix~\ref{appendix:maxent}.

\subsection{Feature Vector Specification}%
\label{subsect:model_variable_specification}

\subsubsection{Variables Included}
For commuter $n$ choosing route $j$, the deterministic attribute vector is
\begin{equation}
  \label{eq:x_spec}
  \mathbf{X}_{nj} = \left( \text{CT}_{nj},\ \text{CD}_{nj},\ \text{PK}_{nj},\ \text{NT}_{nj},\ \text{CAP}_{nj1},\ \text{CAP}_{nj2} \right),
\end{equation}
with each variable defined as follows:
\begin{itemize}[leftmargin=*, labelsep=6.2mm, itemsep=0pt, topsep=3pt, partopsep=0pt, parsep=0pt]
  \item $\text{CT}_{nj}$: door-to-door commuting time (minutes).
  \item $\text{CD}_{nj}$: commuting distance (kilometers).
  \item $\text{PK}_{nj}$: peak-hour indicator (1 if departure is between 07:00--10:00, 0 otherwise).
  \item $\text{NT}_{nj}$: number of burdensome transfers, as refined in Table~\ref{tab:interruption_evolution}.
  \item $\text{CAP}_{nj1},\ \text{CAP}_{nj2}$: capacity at the first and second transfer stations (defined below).
\end{itemize}

These six variables capture the two dimensions commuters trade off in route choice: \textit{efficiency} (time, distance) and \textit{inconvenience} (transfer count, peak-hour demand pressure, and station capacity as a proxy for crowding). The remainder of this subsection gives the rationale for each variable, the candidate variables we considered and excluded, and the final specification selected by empirical testing.

\subsubsection{Why These Variables}
Travel time $\text{CT}$ and distance $\text{CD}$ are standard components of generalized travel cost in transportation research~\cite{train_discrete_2009}. We retain both rather than collapsing them into a single composite because time and distance separately carry the non-monetary and monetary dimensions of route disutility in the Tokyo context: fares in Japan are largely distance-determined~\cite{yoo_2025_railway}, so $\text{CD}$ effectively proxies the monetary cost component without requiring fare data at the route level.

Transfer count $\text{NT}$ enters because transfers impose a well-documented psychological and physical penalty beyond the added travel time~\cite{garcia-martinez_transfer_2018}. Our refined definition (Table~\ref{tab:interruption_evolution}) counts only burdensome transfers, ensuring $\text{NT}$ reflects the kind of transfer behavior the model is meant to explain.

Peak-hour status $\text{PK}$ is included because departure time correlates systematically with crowding, service frequency, and headway variability, all of which affect perceived route disutility independently of the other variables.

Station capacity $\text{CAP}_k$ addresses the role of crowding in route choice while avoiding the simultaneity bias inherent in real-time crowding measures. Because observed crowding is itself the outcome of collective route choices, using real-time counts would make the explanatory variable endogenous to the choice being modeled. Instead we define station capacity as the maximum hourly population observed at station $s$ during the study year:
\begin{equation}
  \label{eq:capacity}
  \text{CAP}_{s} = \max_{t \in \text{study year 2023}} Pop_{st},
\end{equation}
\textls[-15]{where $Pop_{st}$ is the hourly passenger count at station $s$ during hour $t$. This measure is predetermined with respect to any individual commuter's choice and therefore exogenous, while still capturing the infrastructural crowding pressure commuters experience at major hubs. We enter $\text{CAP}_1$ and $\text{CAP}_2$ as separate variables rather than a single composite because first and second transfer stations play asymmetric roles in Tokyo's hierarchical network (see Section~\ref{subsubsect:result_general_route_attribute_preference}).}

A note on congestion. Because Tokyo's railway network operates on grade-separated infrastructure independent of the road system, road-traffic congestion does not enter the model: railway travel time and route choice are decoupled from road conditions. Congestion in this paper therefore refers to station-level density and transfer complexity, not roadway flow.

\subsubsection{Variables Considered and Excluded}
Two further candidate variables were considered and excluded:

{Monetary cost at the route level.} 
 For a fixed OD pair, fares are typically uniform across lines operated by the same company, and most employers reimburse commuting passes (\textit{ts\={u}kin teikiken}), so out-of-pocket cost exerts limited influence on daily decisions. The remaining distance-driven variation in cost is already captured by $\text{CD}$.

{Higher-order transfer capacity ($\text{CAP}_{3}$, $\text{CAP}_{4}$, etc.).} Only about 10\% of observed commutes involve more than two transfers. Industry survey data~\cite{at_home_trend_2018} confirms that regular commutes with three or more transfers are rare; properties requiring such commutes are generally recommended only for full-time remote workers. Most of the 10\% in our data likely represent non-routine trips (e.g., daytime client meetings) rather than regular commuting, so adding $\text{CAP}_{3}$ or $\text{CAP}_{4}$ would fit noise rather than structure.

\subsubsection{Final Specification}

Beyond rejecting these individual candidates, we systematically compared ten alternative functional specifications of the retained variables, including linear-vs-logarithmic transformations of $\text{CD}$ and $\text{CAP}$, aggregating $\text{CAP}_1$ and $\text{CAP}_2$ into a single maximum, and including higher-order capacity terms. Details are provided in Supplementary Table~S2.

The specification test selects logarithmic transformations of distance and capacity, yielding the estimated form

\begin{equation}
  \label{eq:x_spec_final}
  \mathbf{X}_{nj} = \left( \text{CT}_{nj},\ \log\text{CD}_{nj},\ \text{PK}_{nj},\ \text{NT}_{nj},\ \log\text{CAP}_{nj1},\ \log\text{CAP}_{nj2} \right),
\end{equation}

Multicollinearity diagnostics confirm that the retained variables are non-redundant (all variance inflation factors below 5, the conventional threshold~\cite{neter_applied_1996, hair_multivariate_2009}; see Supplementary Table~S1).

\subsection{Model Estimation and Evaluation}%
\label{subsect:model_estimation}

We estimate model parameters by maximizing the following sample log-likelihood
\begin{equation}
\ell(\boldsymbol{\omega}_{\text{raw}}) = \sum_{n=1}^{N} \sum_{j \in C_n} y_{nj} \log P_{nj}
\end{equation}
where $y_{nj} = 1$ if commuter $n$ chooses route $j$ and 0 otherwise. The optimization is performed on the raw, unconstrained weight vector $\boldsymbol{\omega}_{\text{raw}}$ without any direct constraints. Then we obtain $\beta$ and $\boldsymbol{\omega}$ by Equation \eqref{eq:decomp}.

We compute the standard error of the estimated parameter $\boldsymbol{\omega}_{\text{raw}}$ using the inverse of the Hessian matrix of the log-likelihood according to the asymptotic theory of the maximum likelihood estimator. Then we use the delta method \cite{casella_statistical_2002, greene_econometric_2018} \textls[-22]{to compute the standard error of the normalized parameters $\beta$ and $\boldsymbol{\omega}$ (see Appendix for detail procedure).}

\subsubsection{Two Estimation Schemes}%
\label{subsubsect:model_two_model_variants}

We estimate the model in two variants:

\begin{itemize}[leftmargin=*, labelsep=6.2mm, itemsep=0pt, topsep=3pt, partopsep=0pt, parsep=0pt]
  \item {Common Parameter Over all ODs (Model 1). }
A single parameter vector $\boldsymbol{\omega}_{\text{raw}}$ estimated using all routes pooled across all OD pairs. One estimate per month.
  \item {OD-Specific Parameters (Model 2). }A separate parameter vector $(\boldsymbol{\omega}_{\text{raw}})_{\text{OD}}$ estimated for each OD pair. One estimate per OD pair.
\end{itemize}

The two variants share the same mathematical form (Equation~(\ref{eq:prob})) but are estimated at different scopes, which lets them answer different questions. {Model 1} measures whether system-wide collective determinism is stable over time: by pooling all OD pairs each month and comparing the resulting twelve monthly estimates, it speaks to the temporal dimension of our research question (Section~\ref{subsect:result_CPOD}). {Model 2} measures whether determinism varies with commuting context: by estimating parameters per OD pair, it exposes systematic heterogeneity that a pooled model absorbs into its unobserved random term (Section~\ref{subsubsect:result_temperal_dimension_peak_offpeak}, Section~\ref{subsubsect:result_explaining_counterintuitive}). Neither variant subsumes the other---they estimate the same object at different granularities, and the Model 1 estimate serves as the aggregate benchmark against which the mode of the Model 2 distribution is validated.

\paragraph{Normalization and the $\beta$ Baseline}
Prior to estimation, each route attribute is Min--Max normalized to the interval $[0, 1]$. For Model 1, the minimum and maximum are computed across all routes in the study area; for Model 2, within each OD pair. Both variants therefore share the same $[0, 1]$ scale, which gives $\beta$ a concrete interpretation: it measures choice sensitivity when the worst-to-best cost gap is unity. This normalization convention allows direct comparison of $\beta$ estimates across OD pairs within this study, across months, and across different metropolitan networks that adopt the same convention. The city-wide Model 1 estimate $\beta_{\text{city}}$ thus serves as an empirical baseline against which each OD-specific $\beta^{\text{OD}}$ can be judged (Section~\ref{subsect:result_ODSP}): $\beta^{\text{OD}} > \beta_{\text{city}}$ indicates that a given corridor's commuters make more deterministic choices than the metropolitan average, while $\beta^{\text{OD}} < \beta_{\text{city}}$ indicates greater stochasticity.

\subsubsection{Model Evaluation}%
\label{subsubsect:model_evaluation}

To assess the performance of our estimated model, we employ two complementary metrics:
\begin{enumerate}[leftmargin=2.2em, labelsep=4mm, label=(\arabic*), itemsep=0pt, topsep=3pt, parsep=0pt]
  \item {McFadden's pseudo-$R^2$}
, calculated as
  \[R^{2}_{\text{McF}} = 1 - \frac{\ell_{\text{full}}}{\ell_{\text{null}}}\]
  where $\ell_{\text{full}}$ is the log-likelihood of the estimated model and $\ell_{\text{null}}$ is the log-likelihood of a null model with equal alternative probabilities \cite{mcfadden_conditional_1972}. 
  
  \item {Hit rate}, measuring the proportion of correctly predicted choices:
  \[\text{Hit Rate} = \frac{\text{Number of Correct Predictions}}{\text{Number of Choice Situations}}\]
  This metric directly reflects the model's ability to reproduce individual choices \cite{ben_discrete_1985, train_discrete_2009, ma_nested_2020}.
\end{enumerate}

Together, these metrics provide a comprehensive evaluation of how well our model captures railway commuters' path choice behavior in the complex Tokyo transit network.

\section{Results}
\label{section:results}

This section presents the results of the railway commuting route choice model, revealing that Tokyo commuters exhibit highly structured, yet fundamentally non-deterministic, preferences that remains stable throughout the year. The following subsections establish this result and then examine its variation across OD pairs. We begin with an aggregated-level model that pools all Origin-Destination (OD) pairs (Section~\ref{subsect:result_CPOD}). This initial analysis addresses three key questions regarding overall commuter behavior: (1) Overall determinism: How deterministic are commuters in the aggregate? (Section~\ref{subsubsect:result_commuter_rationality_and_choice_determinism});
(2) Temporal Consistency: How consistent is overall choice behavior over time? (Section~\ref{subsubsect:result_rationality_consistency_of_choice_behavior});
(3) Attribute Preference: Which route attributes are regarded attractive by commuters? (Section~\ref{subsubsect:result_general_route_attribute_preference})  with a case analysis (Section~\ref{subsubsect:result_case_analysis}).

Following the aggregate analysis, we apply the same model structure to individual OD pairs to examine heterogeneity and address: (1) Existence of Heterogeneity: Can OD pairs be grouped by similar choice patterns? (Section~\ref{subsect:result_ODSP}); (2) Behavioral Drivers of Heterogeneity: Which OD pair characteristics account for different choice behaviors, including the time-equivalent cost of transfers across specific OD pairs? (Sections~\ref{subsubsect:result_temperal_dimension_peak_offpeak}, \ref{subsubsect:result_explaining_counterintuitive})

Due to computational constraints in analyzing the massive 12-month trajectory dataset, a representative month was selected for most analyses. Except for the temporal consistency analysis (Section~\ref{subsubsect:result_rationality_consistency_of_choice_behavior}), which uses all 12 months to examine stability over time, all other results are based on April 2023 data. April was chosen as the most representative month due to its largest number of observed users, maximizing statistical reliability of OD-specific estimates. The complete set of estimated parameters (same format as Table~\ref{tab:para}) for all 12 months is provided in Supplementary Section 4, confirming month-to-month stability of core findings.


\vspace{-12pt}
\begin{table}[H]
\small
\centering
\begin{threeparttable}
\caption{Estimation results of railway commuting route choice model.}
\label{tab:para}
\begin{tabular*}{\textwidth}{@{\extracolsep{\fill}}
    >{\centering\arraybackslash}m{0.35\textwidth}
    >{\centering\arraybackslash}m{0.2\textwidth}
>{\centering\arraybackslash}m{0.2\textwidth}
>{\centering\arraybackslash}m{0.2\textwidth}
}
\toprule
\textbf{Parameter} & \textbf{Estimate} & \textbf{Std. Error} & \textbf{\textit{t}-Value} \\
\midrule
Scale parameter ($\beta$) & 1.658 & 0.116 & 285.841 \\

No. of Transfers (NT$_{nj}$) & $0.569\thinspace^{***}$ & $0.014$ & $156.309$ \\

Commuting Time (CT$_{nj}$, min) & $0.087\thinspace^{***}$ & $0.022$ & $80.378$ \\

Commuting Distance ($\log(\mathrm{CD_{nj}})$, km) & $0.579\thinspace^{***}$ & $0.034$ & $133.157$ \\

Peak Departure (PK$_{nj}$, dummy) & $-0.117\thinspace^{***}$ & $0.005$ & $-209.778$ \\

Capacity at 1st transfer ($\log(\mathrm{CAP_{nj1}})$) & $0.523\thinspace^{*}$ & $0.017$ & $2.543$ \\

Capacity at 2nd transfer ($\log(\mathrm{CAP_{nj2}})$) & $-0.340\thinspace^{***}$ & $0.013$ & $-210.615$ \\
\midrule
\multicolumn{2}{c}{Number of users} & \multicolumn{2}{c}{125,368} \\
\multicolumn{2}{c}{Number of routes} & \multicolumn{2}{c}{23,887} \\
\multicolumn{2}{c}{Number of ODs} & \multicolumn{2}{c}{5,908} \\
\multicolumn{2}{c}{Initial log-likelihood $\mathcal{L}(\hat{\beta})$} & \multicolumn{2}{c}{$-170,016.797$} \\
\multicolumn{2}{c}{Final log-likelihood $\mathcal{L}(\beta_0)$} & \multicolumn{2}{c}{$-130,530.778$} \\
\multicolumn{2}{c}{AIC} & \multicolumn{2}{c}{261,075.555} \\
\multicolumn{2}{c}{McFadden's $R^2$} & \multicolumn{2}{c}{0.232} \\
\multicolumn{2}{c}{Hit rate} & \multicolumn{2}{c}{$55.8\%$} \\
\multicolumn{2}{c}{Spearman $\rho$} & \multicolumn{2}{c}{0.724} \\
\bottomrule
\end{tabular*}

\begin{tablenotes}[flushleft]
\item {Note:} $^{***}$ $p < 0.001$, $^{**}$ $p < 0.01$, $^{*}$ $p < 0.05$


\end{tablenotes}
\end{threeparttable}
\end{table}

\subsection{Common Parameters for All ODs: Model Validation and Performance}
\label{subsect:result_CPOD}

As mentioned in Section~\ref{subsubsect:model_two_model_variants}, the aggregate model demonstrates strong predictive performance and statistical significance (Table~\ref{tab:para}), with reliability established through multiple validation metrics.

All explanatory variables are statistically significant according to their t-statistics and p-values (Table~\ref{tab:para}). The model correctly predicts the chosen route in 55.8\% of observations, substantially exceeding the 16.7\% random benchmark (based on a max of six alternatives per OD pair). The McFadden's pseudo $R^2$ of 0.232 is acceptable for empirical social science modeling \cite{ozili_acceptable_2023}.

\begin{figure}[H]
\centering
\includegraphics[width=0.76\linewidth]{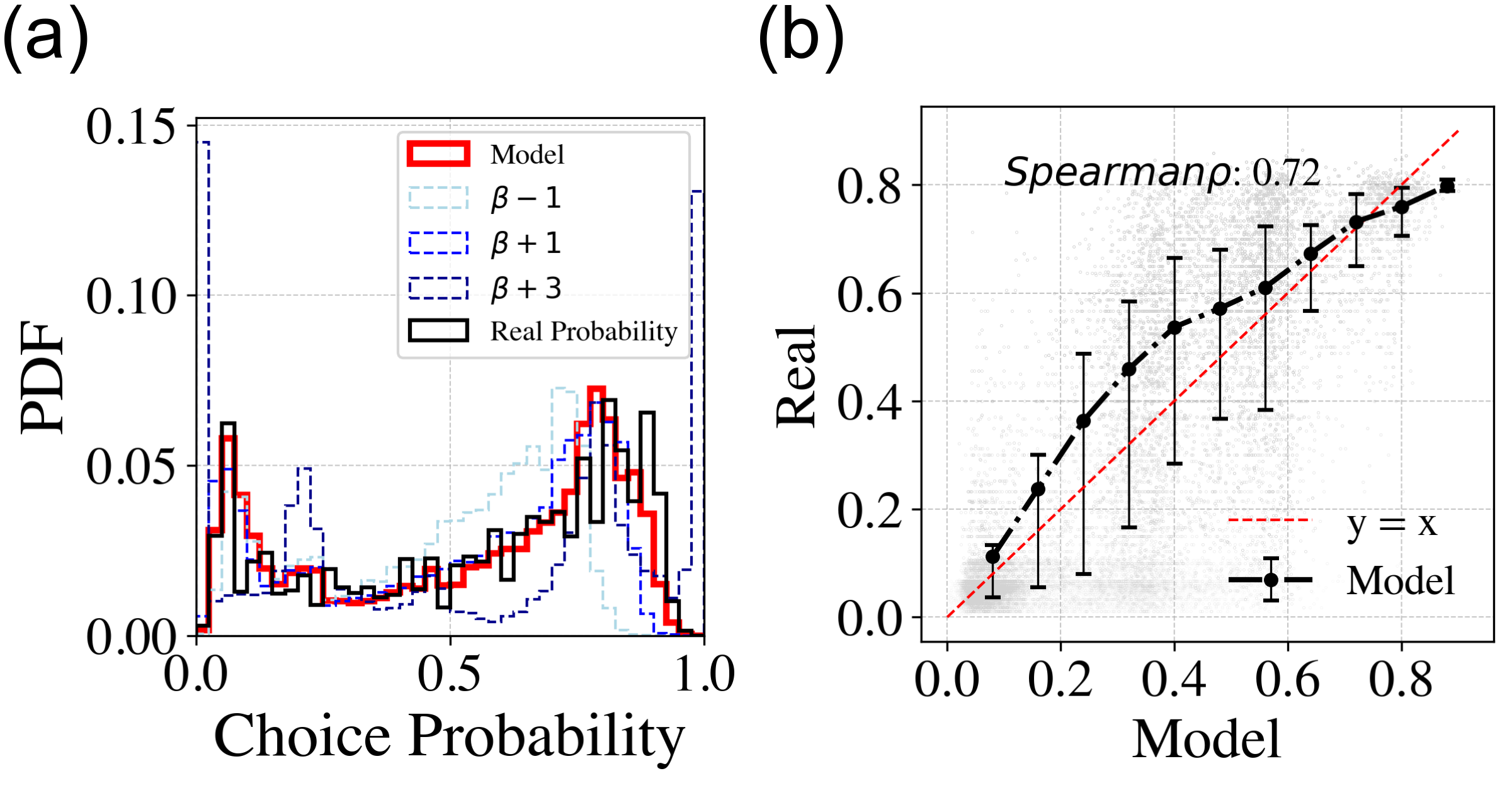}
\caption{Model validation and sensitivity to scale parameter $\beta$. 
(\textbf{a}) Predicted route choice probability distribution and its relationship with scale parameter $\beta$. The red line shows model predictions with the estimated $\beta$ value; the light blue line shows the scenario with $\beta$ decreased by 1; darker blue lines show $\beta$ increased by 1 and 3, respectively. The dashed black line represents the observed probability distribution. (\textbf{b}) Model validation showing predicted versus observed route choice probabilities. Black dots represent aggregated predictions with error bars indicating the 25th and 75th percentiles; grey dots show individual observations. The red dashed line indicates perfect prediction ($y=x$). Spearman $\rho = 0.72$.}
\label{fig5}
\end{figure}

Most importantly for a discrete choice model, the model achieves a high Spearman correlation coefficient ($\rho$) of 0.72 between predicted and observed route choice probabilities (Figure~\ref{fig5}b). This metric confirms that the model accurately captures the relative ranking of alternatives, assigning higher probabilities to routes that are chosen more frequently, the fundamental objective in understanding choice behavior.

\vspace{-16pt}

The moderate McFadden's pseudo $R^2$ can be attributed to three factors: (1) subtle differences between route alternatives in Tokyo's dense railway network; (2) spatial aggregation causing within-group heterogeneity in route variables; and (3) inherent limitations in GPS data precision.

\subsubsection{Commuter Choice Determinism}
\label{subsubsect:result_commuter_rationality_and_choice_determinism}

Under the standard Random Utility Maximization framework, $\beta$ is proportional to the inverse of the Gumbel scale parameter, quantifying the variance of unobserved idiosyncratic factors ($\epsilon_{nj}$ in Equation~(\ref{eq:utility_with_error})) relative to deterministic route costs. Behaviorally, $\beta$ measures \textit{choice determinism}: the extent to which the commuter population relies on observed cost attributes rather than unmeasured preferences or habits.

{High}  $\beta$: 
The system is highly deterministic. Even small cost differences produce large probability differences, meaning commuters almost always select the lowest-cost route.

{Low} $\beta$: The system is highly stochastic (random). Cost differences have minimal impact on choices, meaning commuters are nearly equally likely to select high-cost and low-cost routes.

To make the interpretive scale of $\beta$ concrete, we examine a standardized scenario: a two-route OD pair where the alternatives differ by exactly one generalized energy unit ($\Delta E = 1$). Under the MNL form (Equation (\ref{eq:prob})), the odds ratio of choosing the better alternative over the worse one ($P_{\text{better}} / P_{\text{worse}}$) is exactly $e^{-\beta(-\Delta E)} = e^{\beta}$, meaning the better route is $e^{\beta}$ times more likely to be chosen. Consequently, since there are only two routes, the probability of selecting the better route evaluates to $P_{\text{better}} = e^{\beta} / (1 + e^{\beta}) = 1 / (1 + e^{-\beta})$. Evaluating this probability at the empirical Tokyo value, and at its $\pm 1$ and $+3$ perturbations visualized in Figure~\ref{fig5}a, yields the following correspondence between $\beta$ and collective route choice probability (Table~\ref{tab:beta_interpretation}):

\vspace{-12pt}
\begin{table}[H]
\begin{threeparttable}
\small
\centering
\caption{The correspondence between $\beta$ and choice preference.} 
\label{tab:beta_interpretation}
    \begin{tabular*}{\textwidth}{@{\extracolsep{\fill}}
    >{\centering\arraybackslash}m{0.15\textwidth}
    >{\centering\arraybackslash}m{0.15\textwidth}
 >{\centering\arraybackslash}m{0.15\textwidth}
  >{\raggedright\arraybackslash}m{0.6\textwidth} }
\toprule
$\bm{\beta}$ & $\bm{e^{\,\beta}}$ & $\bm{P_}{\textbf{Better}}$ & \multicolumn{1}{>{\centering\arraybackslash}m{0.5\textwidth}}{\textbf{Interpretation}} \\
\midrule
$0$              & $1$              & $50\%$    & Fully random (all routes equally weighted) \\
$0.658~(\beta-1)$ & $1.93$           & $\approx 66\%$ & Near-random; modest preference for better route \\
{$\mathbf{1.658}$}
 & {$\mathbf{5.25}$}  & {$\mathbf{\approx 84\%}$} & \textbf{{This study (Tokyo, city-wide)}} \\
$2.658~(\beta+1)$ & $14.3$           & $\approx 93\%$ & Substantially more deterministic \\
$4.658~(\beta+3)$ & $105$            & $\approx 99\%$ & Near-deterministic \\
$\to\infty$      & $\to\infty$      & $\to 100\%$    & Perfectly deterministic \\
\bottomrule
\end{tabular*}
\begin{tablenotes}[flushleft]
\fontsize{9}{10.8}\selectfont 
\item {Note: {Bold row indicates the empirical estimate from this study (Tokyo, city-wide, $\beta = 1.658$).} At the Tokyo empirical estimate, a 1-unit advantage in generalized energy means that the better route is 5.25 times more likely to be chosen than its alternative.}
\end{tablenotes}
\end{threeparttable}
\end{table}

\vspace{-6pt}
Therefore, $\beta$ serves as an aggregate measure of choice determinism, the degree to which commuters consistently select routes in alignment with the calculated cost function.

As shown in Figure~\ref{fig5}a, both the model predictions (red line) and observed probabilities (black line) exhibit a distinctive two-peak pattern: a smaller peak near zero probability and a larger peak near one. This bimodal distribution reveals the underlying choice structure, as most routes either have low choice probabilities (unlikely alternatives) or relatively high probabilities (preferred alternatives), with few routes falling in the intermediate range. This pattern indicates that commuters make relatively clear distinctions between desirable and undesirable routes based on their attributes, rather than distributing choices uniformly across alternatives. \textls[-15]{At the same time, the peak near probability 1 does not saturate at 1, and the peak near 0 does not collapse to 0, a non-negligible mass remains across the full probability range. Tokyo commuters thus exhibit a strong but non-deterministic preference: clearly biased toward low-cost routes, yet far from the strict cost minimization that a purely deterministic system would produce.}

The close alignment between the red and black lines demonstrates that the estimated $\beta$ value (1.658) accurately captures this choice behavior. Figure~\ref{fig5}a further illustrates how variations in $\beta$ affect the probability distribution. As $\beta$ increases to 3 (darker blue lines), the system becomes near-deterministic: low-cost routes are assigned probabilities approaching 1, while high-cost routes approach zero, resulting in more pronounced peaks at the extremes. Conversely, decreasing $\beta$ (light blue line) shifts the distribution toward less deterministic choices, with more probability mass concentrated around 0.5, indicating greater randomness in route selection. The observed distribution sits clearly between these two extremes, visually confirming that Tokyo's choice determinism level is substantial but not absolute.

The next section illustrates if the current $\beta$ value of 1.658 positions the system at a balanced state or not.

\subsubsection{Consistency of Choice Behavior}
\label{subsubsect:result_rationality_consistency_of_choice_behavior}

 This section evaluates the temporal consistency of commuters' overall determinism and route choice strategy by examining the monthly stability of the estimated model parameters.
 
\begin{figure}[H]
 \centering \includegraphics[width=0.99\linewidth]{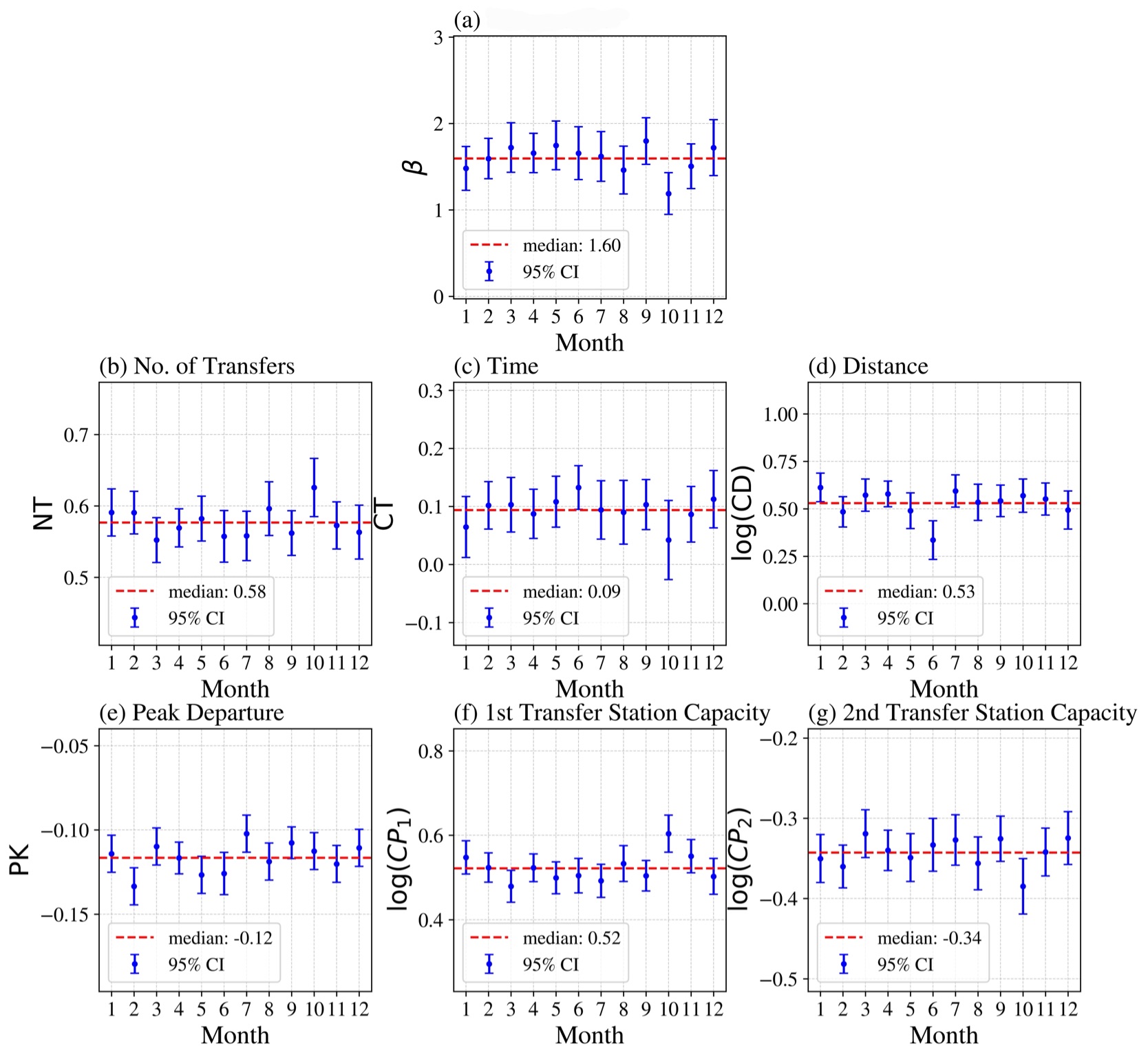}\\
 \caption{Monthly estimates and 95\% confidence intervals (CI) (calculated by delta method) for the normalized weights, $\boldsymbol{\omega}$, across twelve months.
 (\textbf{a}) Inverse of the unobserved error scale($\beta$) ranges from 1.189 (October) to 1.797 (September) with median 1.60; confidence intervals overlap across all months. 
 (\textbf{b}) Number of transfers coefficient (NT) ranges from 0.552 (March) to 0.626 (October) with median 0.58; confidence intervals generally overlap throughout the year. 
 (\textbf{c}) Time coefficient ranges from 0.042 (October) to 0.133 (June) with median 0.09; confidence intervals overlap across months. 
 (\textbf{d}) Distance coefficient (log(CD)) ranges from 0.336 (June) to 0.613 (January) with median 0.53; confidence intervals overlap for all months. 
 (\textbf{e}) Peak departure coefficient (PK) ranges from $-$0.134 (February) to $-$0.102 (July) with median $-$0.12; confidence intervals overlap throughout the study period. 
 (\textbf{f}) First transfer station capacity (log(CP1)) ranges from 0.479 (March) to 0.604 (October) with median 0.52; confidence intervals overlap across all months. 
 (\textbf{g}) Second transfer station capacity (log(CP2)) ranges from $-$0.385 (October) to $-$0.319 (March) with median $-$0.34; confidence intervals overlap for all months. 
 The dashed red line in each subplot indicates the annual median value.}
 \label{fig7}
 \end{figure}

The scale parameter $\beta$ demonstrates remarkable consistency throughout the year, with monthly values clustering around the annual mean of 1.60 (Figure~\ref{fig7}a). This temporal consistency indicates that the aggregate level of choice determinism and the sensitivity to route costs remain relatively constant across seasons.

\vspace{-12pt}
Similarly, the normalized coefficients ($\omega$) for all other route attributes (Figure \ref{fig7}b--g) exhibit minor fluctuations, confirming the stability of the relative importance of factors like travel time, transfer frequency, and station capacity. The narrow and overlapping 95\% confidence intervals (CIs)(See Method Section~\ref{subsect:model_estimation}) across all twelve months provide strong statistical evidence that the identified behavioral patterns represent stable structural features of commuter decision-making, rather than being artifacts of specific time periods or data collection conditions.

Despite this overall stability, we observe modest seasonal fluctuations. The lowest $\beta$ value occurs in October (approximately 1.20), suggesting slightly increased randomness in route choices, likely attributable to Japan's momiji (autumn foliage) season when tourist activity may disrupt normal transportation patterns. Conversely, the highest $\beta$ value appears in September (approximately 1.80), indicating more deterministic choice behavior, which correlates with the resumption of highly structured commuting routines following the summer period.

To assess the practical significance of these seasonal variations in commuter route choice determinism, Figure~\ref{fig5}a examines how changes in $\beta$ affect the predicted probability distribution. The observed monthly fluctuations (approximately $\pm0.3$ around the median $\beta$ value) produce minimal changes in the predicted choice probabilities, demonstrating model robustness. These limited effects indicate that seasonal variations have negligible practical impact on route choice predictions. 

This temporal consistency confirms that the identified route choice patterns represent stable structural features of commuter behavior rather than temporary phenomena.

\subsubsection{General Route Attribute Preferences}
\label{subsubsect:result_general_route_attribute_preference}

The estimation results, summarized in Table~\ref{tab:para}, reveal system-wide commuter behavioral preferences and directly identify which route attributes constitute a higher or lower \textit{energy cost} in the choice decision. Consistent with a model seeking to minimize route cost, the signs of the coefficients reveal the general attractiveness or deterrent effect of each attribute.

\paragraph{Route Deterrents (Positive Coefficients)} 

Three fundamental travel energy costs consistently deter route selection: The positive coefficients for number of transfers ($NT_{nj}$), commuting time ($CT_{nj}$), and log commuting distance ($\log(\mathrm{CD_{nj}}$)) confirm that commuters system-wide prefer routes that minimize transfers, travel time, and distance.

Notably, the positive coefficient for $\log(\mathrm{CAP_{nj1}})$ (Capacity at 1st transfer or single transfer) reveals that transferring at high-capacity stations imposes higher perceived costs. This counterintuitive result likely reflects the compounding negative effects experienced at major transfer hubs: difficulty boarding crowded trains, low probability of securing a seat, longer walking distances between platforms, and greater uncertainty during service disruptions. These factors collectively outweigh any potential benefits of higher service frequency at large stations during the initial transfer.

\paragraph{Route Attractions (Negative Coefficients)} 

Two attributes significantly reduce perceived route costs: The negative coefficient for second transfer station capacity ($\log(\mathrm{CAP_{nj2}}$)) indicates that high capacity at the second transfer point reduces perceived travel burden. This asymmetric effect compared to first transfers reflects the distinct spatial and functional roles of transfer locations. As shown in Figure \ref{fig6}, second transfers predominantly occur at major CBD hubs (e.g., Shinjuku, Shibuya) where high capacity is structurally coupled with superior service frequency and better infrastructure. This suggests that commuters have adapted their expectations, perceiving high capacity at these specific, centralized locations as an expected feature that facilitates the final leg of the journey, rather than a deterrent.

The negative coefficient for peak departure ($PK_{nj}$) reflects a structural constraint of commuting demand. The majority of trips are temporally fixed to peak hours due to work schedules, effectively making peak-hour travel a \textit{default} and therefore less costly (or necessary) choice relative to the perceived cost of missing the required arrival time. While peak-hour departure is often constrained by work schedules, the rise of telework and flexible shifting in post-pandemic Tokyo has introduced greater temporal choice, making the $PK_{nj}$ variable a relevant indicator of shifted commuting constraints. The negative and significant coefficient suggests that temporal constraints often override comfort considerations in route selection for the commuters.

Approximately 10\% of commuters in our sample make more than two transfers. We did not include third or fourth transfer capacity variables for two reasons. First, the sample size for these cases is too small for reliable statistical estimation. Second, trips with multiple transfers often represent non-routine activities (e.g., client meetings, special errands) rather than regular commuting patterns. Our model focuses on typical commuting behavior to maintain generalizability and interpretability.

\subsubsection{Case Analysis}
\label{subsubsect:result_case_analysis}

To validate the model's predictive capability and capture nuanced commuter behavior, we analyzed a specific origin-destination (OD) pair: Musashi-Shinj\={o} to Shinjuku. We selected this case based on three criteria: 
(i) Sufficient sample size for stable monthly probability estimates (typically 60+ commuters; Table~\ref{tab:route_info}); 
(ii) A choice set of at least four physically distinct alternatives, making the asymmetric roles of transfer station capacity observable within a single case.
This OD pair provides a representative lens into the complex trade-offs commuters make between travel time, distance, and transfers.

\begin{figure}[H]
\centering
\includegraphics[width=0.99\linewidth]{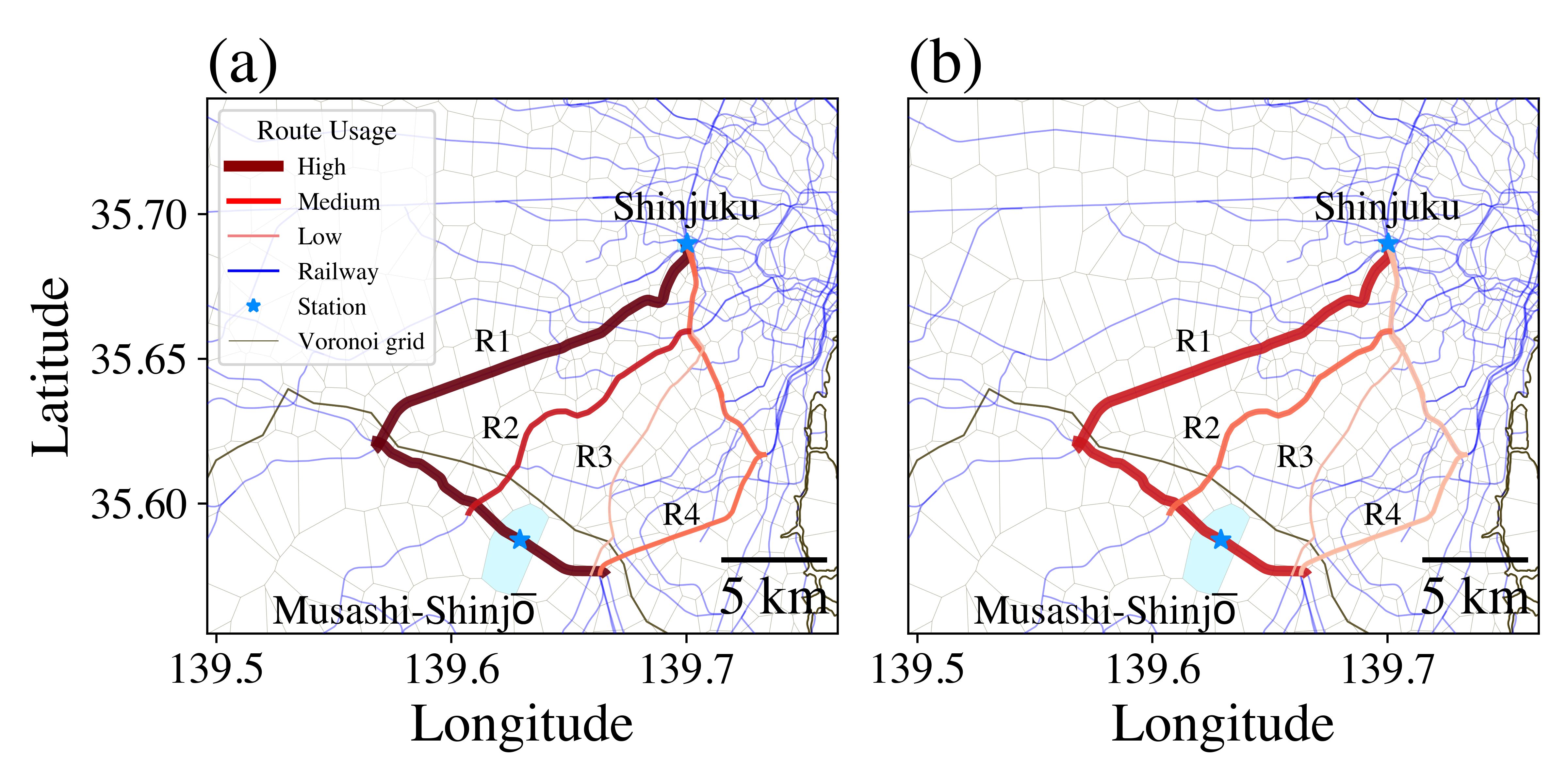}\\
\caption{Real and model-predicted route choice maps for the Musashi--Shinj\={o} to Shinjuku OD pair. 
(\textbf{a}) Actual route usage based on GPS data showing high, medium, and low-usage routes. (\textbf{b}) Route usage predicted by the Model 1 (city-wide pooled estimation) model variants. Line thickness and darkness indicate usage intensity. Four primary route alternatives are examined (detailed in Table~\ref{tab:route_info}). The model successfully captures the most and least utilized routes, with predicted patterns closely resembling observed usage.}
\label{fig8}
\end{figure}

As shown in Figure ~\ref{fig8}, the model's predicted route usage (Figure~\ref{fig8}b) closely aligns with the actual route usage derived from GPS data (Figure~\ref{fig8}a). The most utilized route, R1, which corresponds to the Odakyu Line, exhibits high predicted usage consistent with its superior attributes, including the shortest travel time and a single transfer. Conversely, the least utilized route, R3 (T\={o}yoko Line), which involves two transfers and a longer travel time, is correctly assigned a low probability by our model.

\vspace{-34pt}

\begin{table}[H]
\footnotesize
\begin{threeparttable}
\centering
\caption{Route options from left to right visualized in Figure~\ref{fig8} for Musashi-Shinj\={o} to Shinjuku OD pair (at 7:00 am).}
\label{tab:route_info}
\begin{tabular*}{\textwidth}{@{\extracolsep{\fill}}
    >{\centering\arraybackslash}m{0.1\textwidth}
    >{\centering\arraybackslash}m{0.2\textwidth}
>{\centering\arraybackslash}m{0.05\textwidth} 
>{\centering\arraybackslash}m{0.05\textwidth} 
>{\centering\arraybackslash}m{0.12\textwidth} 
>{\centering\arraybackslash}m{0.15\textwidth} 
>{\centering\arraybackslash}m{0.15\textwidth}  
}
\toprule
\textbf{Route ID} & \textbf{Railway Lines} & \textbf{NT (Times)} & \textbf{CT (min)} & \textbf{CD \bm{$\log(\mathrm{CD})$} (km)} & \textbf{CAP$_1$ \bm{$\log(\mathrm{CAP_{1}})$} (Persons)} & \textbf{CAP$_2$ \bm{$\log(\mathrm{CAP_{2}})$} (Persons)}\\
\midrule
R1 & Nambu$\rightarrow$  Odakyu & 1 & 33 & 22  (3.09) & 110,466  (11.61) & - \\
\midrule
R2 & Nambu$\rightarrow$  Den-en-toshi$\rightarrow$  Yamanote & 2 & 39 & 19.6  (2.98) & 169,774  (12.04) & 245,291  (12.41) \\
\midrule
R3 & Nambu$\rightarrow$  T\={o}yoko$\rightarrow$  Yamanote & 2 & 41 & 18.2  (2.90) & 169,898  (12.04) & 245,291  (12.41) \\
\midrule
R4 & Nambu$\rightarrow$  Sh\={o}nan-Shinjuku & 1 & 34 & 23.6  (3.16) & 169,898  (12.04) & - \\
\bottomrule
\end{tabular*}
\begin{tablenotes}[flushleft]
\item {Note:} Transfer stations and their approximate daily passenger capacity rankings are as follows: (1) Shibuya (2nd transfer for R2, R3); (2) Musashi-Kosugi (1st transfer for R3, R4); (3) Musashi-Mizonokuchi (1st transfer for R2); (4) Noborito (1st transfer for R1). 
\end{tablenotes}
\end{threeparttable}
\end{table}

Table~\ref{tab:monthly_prob} further quantifies these results by showing the monthly observed route choices for the four primary alternatives detailed in Table~\ref{tab:route_info}. As visually depicted in Figure ~\ref{fig8}, R1 is the most frequently chosen option, while R2 and R4 are selected at a roughly equal, moderate rate, and R3 is the least preferred. The \textit{Model Predicted} row, calculated using the route features from Table~\ref{tab:route_info} and the parameter estimates from the model shown in Table~\ref{tab:para}, successfully captures this relative popularity, correctly identifying the order of preference as R1 $>$ R2 $\simeq$ R4 $>$ R3.

A more subtle and insightful observation arises from the intermediate route choices. The Den-en-toshi Line is also commonly used despite an extra transfer, with usage comparable to the Sh\={o}nan-Shinjuku Line's route with a single transfer. This occurs because the Den-en-toshi Line's transfer happens at Shibuya, a major station with large capacity and numerous accessible lines. Our model's positive coefficient for $\log(\mathrm{CP}_2)$ correctly captures the positive effect of high-capacity second transfer station. 

The choice to exclude monetary cost as a feature warrants specific justification. As detailed in Section~\ref{sect:methodology}, the widespread use of employer-reimbursed commuting passes in Japan, combined with the typically uniform fare structures across major rail networks, suggests that monetary cost is often not the primary decision variable for the average commuter. Empirical data from specific OD pairs, such as Musashi-Shinj\={o} to Shinjuku, supports this structural assumption. Within this pair, routes R2 (540 JPY) and R3 (560 JPY) are frequently chosen by commuters over the less expensive routes R1 (450 JPY) and R4 (410 JPY). This behavior, where a substantial number of commuters consistently select a route that is approximately 100 JPY more expensive, provides strong revealed preference evidence. This difference in price is clearly outweighed by the non-monetary costs captured in our model, such as travel time, transfers, and station capacity. This observation validates our model's focus on non-monetary costs as the dominant drivers of route choice for the average commuter in this system, ensuring the model's parameters accurately reflect the key trade-offs they prioritize.

\vspace{-12pt}
\begin{table}[H]
\begin{threeparttable}
\centering
\caption{Monthly observed and model calculated route choice probabilities and total sample sizes for Musashi-Shinj\={o} to Shinjuku OD pair.}
\label{tab:monthly_prob}
\begin{tabular*}{1\textwidth}{@{\extracolsep{\fill}}c c c c c c}
\toprule
\textbf{Month} & \textbf{R1} & \textbf{R2} & \textbf{R3} & \textbf{R4} & \textbf{Sample Size} \\
\midrule
1  & 0.56 & 0.21 & 0.10 & 0.13 & 30 \\
2  & 0.57 & 0.20 & 0.12 & 0.11 & 60 \\
3  & 0.34 & 0.28 & 0.12 & 0.26 & 57 \\
4  & 0.53 & 0.18 & 0.13 & 0.16 & 60 \\
5  & 0.42 & 0.15 & 0.13 & 0.30 & 58 \\
6  & 0.39 & 0.22 & 0.14 & 0.25 & 57 \\
7  & 0.50 & 0.16 & 0.13 & 0.21 & 73 \\
8  & 0.31 & 0.27 & 0.16 & 0.26 & 62 \\
9  & 0.44 & 0.18 & 0.15 & 0.23 & 65 \\
10 & 0.32 & 0.30 & 0.24 & 0.14 & 64 \\
11 & 0.35 & 0.22 & 0.16 & 0.27 & 60 \\
12 & 0.56 & 0.17 & 0.10 & 0.17 & 55 \\
\midrule
Model Predicted (Model 1 variants) 
& 0.32 & 0.27 & 0.18 & 0.23 &   \\
\bottomrule
\end{tabular*}
\begin{tablenotes}[flushleft]
\fontsize{9}{10.8}\selectfont 
\item {Note}: To minimize the influence of individual commuting strategies, this origin-destination (OD) pair was selected because commuters almost exclusively travel during peak hours, ensuring the peak dummy variable (PK) is consistently equal to 1. 
\end{tablenotes}
\end{threeparttable}
\end{table}
\vspace{-6pt}

Additional cases are shown in the Supplement Figures S2 and S3, including scenarios where people choose less crowded transfer stations despite slightly longer travel times, and cases where people tolerate extra transfers for less crowded lines.

Consequently, the model's predicted route usage is more evenly distributed across all options compared to the highly concentrated usage observed in raw GPS data, where minor behavioral factors not captured by the model lead to more dispersed choices.

Despite this moderate uncertainty, the model successfully captures the fundamental ranking of alternatives, as evidenced by the high Spearman correlation. This indicates that while we may not perfectly predict the exact probability for each route, the model reliably identifies which routes are more or less likely to be chosen: the essential information for transportation planning and network optimization.

\subsection{OD-Specific Parameter Estimation: Systematic Heterogeneity}
\label{subsect:result_ODSP}

\textls[-15]{Following the aggregate analysis, we apply the same model structure to individual OD pairs to examine heterogeneity and address: (1) whether OD pairs can be grouped by distinct choice patterns (Section~\ref{subsect:result_ODSP}); and (2) which OD-pair characteristics drive these differences, including the time-equivalent cost of transfers (Sections~\ref{subsubsect:result_temperal_dimension_peak_offpeak} and \ref{subsubsect:result_explaining_counterintuitive}).}

\begin{figure}[H]
\centering
\includegraphics[width=0.95\linewidth]{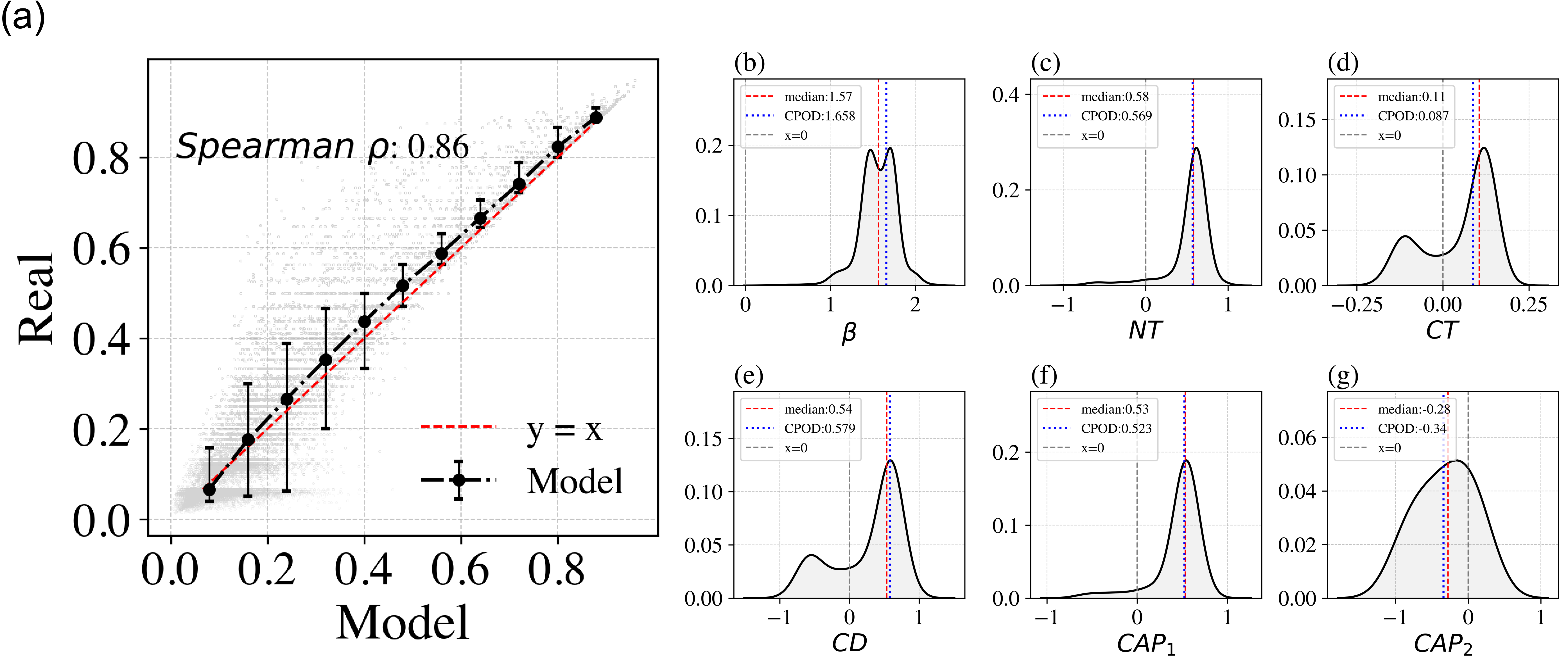}\\
\caption{OD-specific model performance and parameter heterogeneity. (\textbf{a}) Model validation for OD-specific estimations showing predicted versus observed route choice probabilities. Error bars represent 25th-75th percentiles; the red dashed line indicates perfect prediction ($y=x$); Spearman $\rho = 0.86$. (\textbf{b}--\textbf{g}) Probability density functions of estimated parameters across all OD pairs: (\textbf{b}) inverse of the unobserved error scale parameter $\beta$; (\textbf{c}) number of transfers (NT) coefficient; (\textbf{d}) commuting time coefficient (CT); (\textbf{e}) commuting distance coefficient (CD); (\textbf{f}) first transfer capacity coefficient (CAP$_1$); and (\textbf{g}) second transfer capacity coefficient (CAP$_2$). Blue dotted lines indicate the aggregate Model 1 (city-wide pooled estimation) estimates from Table~\ref{tab:para}, confirming that the city-wide parameters align with the primary mode of the OD-specific distributions, red vertical lines indicate median values; grey vertical lines indicate zero. The distributions reveal systematic heterogeneity in route choice preferences across different origin-destination pairs.}
\label{fig9}
\end{figure}

Table~\ref{tab:odsp_summary} summarizes the Model 2(OD-specific estimation) estimation results alongside the Model 1(city-wide pooled estimation) aggregate estimates from Table~\ref{tab:para}, providing a unified comparison of both model variants. The probability density functions of the OD-specific parameters (Figure~\ref{fig9}b--g) reveal that preference variation across OD pairs is systematic rather than random. Most parameters exhibit unimodal distributions whose primary modes align closely with the Model 1 aggregate estimates (indicated by blue dotted lines in Figure~\ref{fig9}), confirming that the city-wide parameters represent the central tendency of OD-level behavior. Because both Model 1 and Model 2 estimates share the same $[0,1]$ normalization (Section~\ref{subsubsect:model_two_model_variants}), $\beta^{OD}$ values are directly comparable to the city-wide benchmark $\beta_{\mathrm{city}}=1.658$: $\beta^{OD}>\beta_{\mathrm{city}}$ indicates more deterministic choices than the metropolitan average, $\beta^{OD}<\beta_{\mathrm{city}}$ indicates greater stochasticity (Figure~\ref{fig9}). The same convention enables standardized cross-city comparison once comparable datasets become available.

\vspace{-36pt}
\begin{table}[H]
\centering
\begin{threeparttable}
\caption{Comparison of Model 1 and Model 2 estimation results.}
\label{tab:odsp_summary}
\begin{tabular*}{\textwidth}{@{\extracolsep{\fill}}
    >{\raggedright\arraybackslash}m{0.4\textwidth}
    >{\centering\arraybackslash}m{0.12\textwidth} 
 >{\centering\arraybackslash}m{0.2\textwidth} 
 >{\centering\arraybackslash}m{0.2\textwidth} 
}
\toprule
\multirow{2}{*}{\textbf{Parameter}} 
& \multirow{2}{*}{\begin{tabular}{@{}c@{}}\textbf{Model 1} \\ \textbf{Estimate}\end{tabular}} 
& \multicolumn{2}{c}{\textbf{Model 2}} \\
\cmidrule(l{0.5em}r{0.5em}){3-4}
& & \textbf{Mean} & \textbf{Median} \\
\midrule
Scale parameter ($\beta$) & 1.658 & 1.561 & 1.567 \\

No. of Transfers (NT$_{nj}$) & $0.569\thinspace^{***}$ & 0.536 & 0.580 \\

Commuting Time (CT$_{nj}$, min) & $0.087\thinspace^{***}$ & 0.048 & 0.105 \\

Commuting Distance ($\log(\mathrm{CD_{nj}})$, km) & $0.579\thinspace^{***}$ & 0.262 & 0.535 \\

Peak Departure (PK$_{nj}$, dummy) & $-0.117\thinspace^{***}$ & --- & --- \\

Capacity at 1st transfer ($\log(\mathrm{CAP_{nj1}})$) & $0.523\thinspace^{*}$ & 0.465 & 0.533 \\

Capacity at 2nd transfer ($\log(\mathrm{CAP_{nj2}})$) & $-0.340\thinspace^{***}$ & $-0.312$ & $-0.277$ \\
\midrule
\multicolumn{2}{l}{} & \textbf{Model 1} & \textbf{Model 2 (median)} \\
\midrule
\multicolumn{2}{@{}l@{}}{Number of OD pairs} & 5,908 & 5,908 \\
\multicolumn{2}{@{}l@{}}{McFadden's $R^2$} & 0.232 & 0.383 \\
\multicolumn{2}{@{}l@{}}{Hit rate} & $55.8\%$ & $71.0\%$ \\
\multicolumn{2}{@{}l@{}}{Spearman $\rho$} & 0.724 & 0.860 \\
\multicolumn{2}{@{}l@{}}{\% of ODs with $R^2 > 0.232$} & --- & $74.8\%$ \\
\multicolumn{2}{@{}l@{}}{\% of ODs with hit rate $> 55.8\%$} & --- & $73.8\%$ \\
\bottomrule
\end{tabular*}

\begin{tablenotes}[flushleft]
\fontsize{9}{10.8}\selectfont 
\item {Note:} {$^{***}$ $p < 0.001$, {$^{**}$ $p < 0.01$}, $^{*}$ $p < 0.05$. 
The Peak Departure variable (PK) is not estimated in the Model 2 model because the Model 2 analysis separates peak and off-peak commuters into distinct sub-samples.}

\end{tablenotes}
\end{threeparttable}
\end{table}
\vspace{-6pt}

However, the distributions for the inverse of the unobserved error scale parameter ($\beta$) and the coefficients for Commuting Time (CT) and Commuting Distance (CD) display distinct bimodal patterns (Figure~\ref{fig9}b,d,e). This bimodality is a key finding: it reveals the existence of structurally distinct behavioral groups within the network, rather than a continuum of idiosyncratic variation. 

Specifically, while the main peak for CT and CD coefficients is positive (aligning with the aggregate result, where greater time or distance increases perceived cost), a substantial subset of OD pairs exhibits negative coefficients. In the context of our energy-cost model, a negative cost coefficient for time or distance indicates that, for these specific corridors, longer times or distances are associated with lower perceived energy cost. This counter-intuitive result immediately establishes that two structurally distinct OD groups coexist in how commuting time and distance enter the realized choice distribution; their behavioral interpretation will be examined in Section~\ref{subsubsect:result_explaining_counterintuitive}. The heterogeneity in $\beta$, meanwhile, will be explored in Section~\ref{subsubsect:result_temperal_dimension_peak_offpeak} to account for temporal differences between peak and off-peak departures. Here, systematic heterogeneity refers strictly to the variation in commuter preferences for a single route attribute across different origin-destination pairs.

To ensure that these bimodal patterns are not artifacts of limited observations in certain OD pairs, we verified the statistical adequacy of our observation-to-parameter ratio and conducted a robustness analysis restricting the sample to OD pairs with at least 40 commuters. As detailed in the Supplementary Material (Robustness Check, Figure~S8), the bimodal structures remained consistent, confirming the robustness of the finding.

 The Model 2 (OD-specific estimation) model's predictive performance confirms the reliability of these OD-level estimates: as shown in Table~\ref{tab:odsp_summary}, the median McFadden's $R^2$ is 0.383 and the median hit rate is 71.0\%, with 74.8\% and 73.8\% of OD pairs exceeding the corresponding Model 1(city-wide pooled estimation) benchmarks, respectively (see Supplementary Material Figure~S12 for distributions).

\subsubsection{Temporal Dimension: Peak versus Off-Peak Dynamics}
\label{subsubsect:result_temperal_dimension_peak_offpeak}

Analysis of the OD-specific parameters separated by departure time (peak versus off-peak) reveals distinct behavioral dynamics, which concurrently explains the bimodal distribution observed for the inverse of the unobserved error scale parameter ($\beta$) in Figure~\ref{fig9}b.

\begin{figure}[H]
\centering
\includegraphics[width=0.75\linewidth]{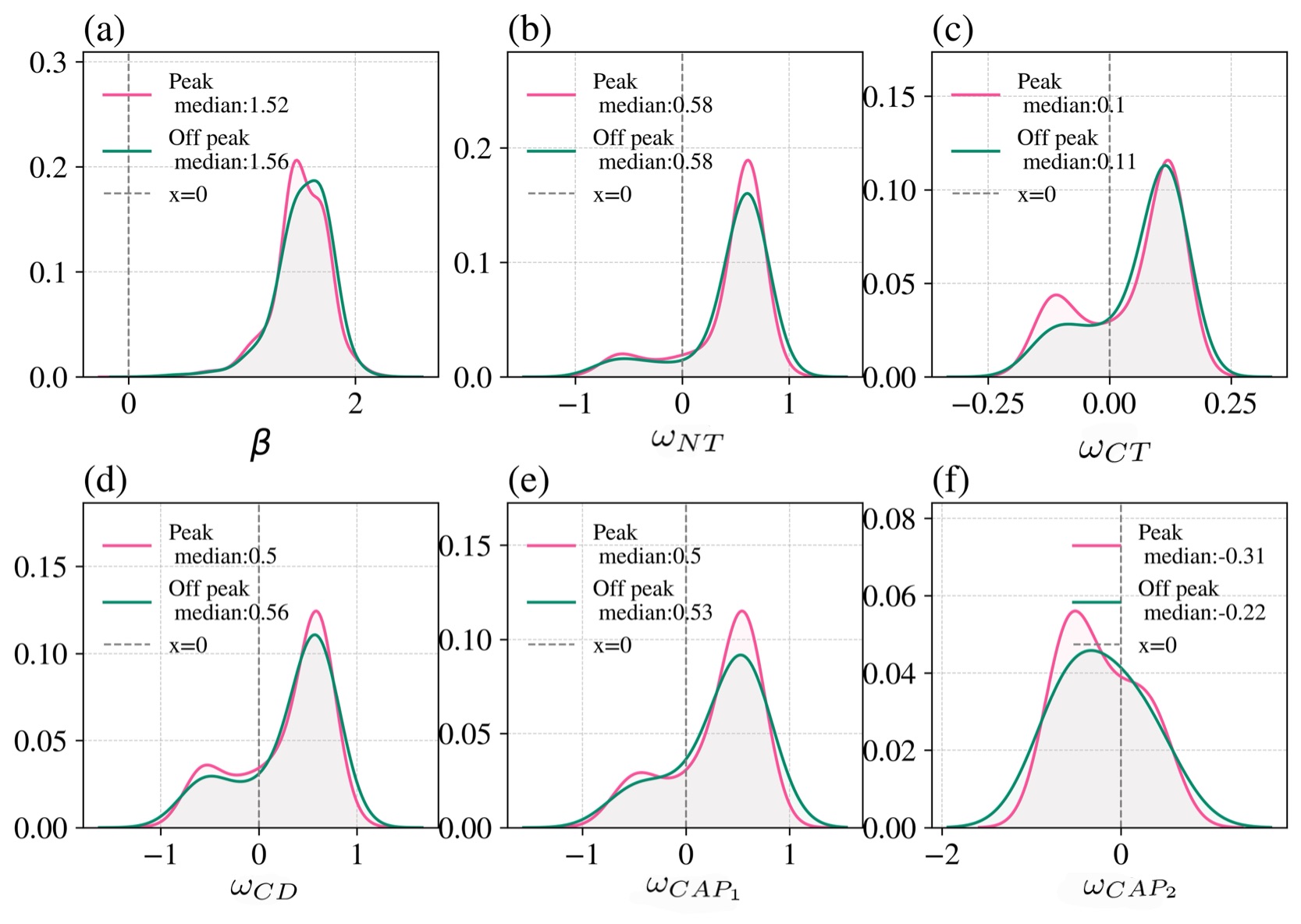}\\
\caption{Comparison of parameter distributions between peak and off-peak periods. 
Probability density functions of estimated parameters across OD pairs, separated by departure time: pink lines show peak-hour distributions; green lines show off-peak distributions. (\textbf{a}) Scale parameter $\beta$; (\textbf{b}) number of transfers coefficient (NT); (\textbf{c}) commuting time coefficient (CT); (\textbf{d}) commuting distance coefficient (CD); (\textbf{e}) first transfer capacity coefficient (CAP$_1$); (\textbf{f}) second transfer capacity coefficient (CAP$_2$). Grey dashed lines indicate zero reference.}
\label{fig10}
\end{figure}

Figure~\ref{fig10}a shows a clear distinction in the inverse of the unobserved error scale parameter ($\beta$). Peak-hour commuters exhibit lower $\beta$ values (median $\simeq$ 1.52), indicating less deterministic and more exploratory choice behavior. This reduction in determinism can be attributed to two main factors: (1) strategic Exploration: Commuters actively deviate from minimal-cost routes to navigate severe crowding and capacity constraints, often prioritizing securing a seat or avoiding specific congested points; (2) peak-hour operation is more susceptible to service disruptions and delays, which introduce higher variance in actual commuting times. As shown in Figure \ref{fig10}c, peak-hour commuting time exhibits more negative parameters compared to off-peak hours, indicating that for more OD pairs, longer routes are taken more frequently during peak periods. This counterintuitive pattern provides evidence that unexpected factors (such as crowding and delays) make peak-hour commuting less predictable. The underlying reasons for these negative time parameters are discussed further in the following section.

\vspace{-12pt}

Off-Peak Commuters: Exhibit higher $\beta$ values (median $\simeq$1.56), reflecting more deterministic behavior. With reduced crowding and greater capacity, the need for strategic deviation is minimized. Furthermore, the train operation schedule is generally more stable during off-peak hours, leading to lower time variance and allowing commuters to more consistently select routes based on minimized cost or habit.

The remaining route attributes (Figure  \ref{fig10}b--f) exhibit similar central tendencies between peak and off-peak periods. This finding suggests that fundamental cost, such as the inherent aversion to transfers, time, and distance, remain consistent regardless of departure time. Consequently, for the subsequent detailed analysis of the bimodal heterogeneity observed in the Commuting Time and Commuting Distance coefficients(Figure  \ref{fig9}c,d), we proceed without separating peak and off-peak commuters, focusing instead on other contextual factors that drive these systematic differences.

\subsubsection{Explaining Counterintuitive Parameters: OD Pair Characteristics}
\label{subsubsect:result_explaining_counterintuitive}

The systematic heterogeneity revealed in the OD-specific model, particularly the bimodal distributions of the Commuting Time (CT) and Commuting Distance (CD) coefficients, presents a crucial puzzle: why do some OD pairs exhibit negative coefficients, suggesting commuters prefer routes with longer times or distances?

\begin{figure}[H]
\centering
\includegraphics[width=0.88\linewidth]{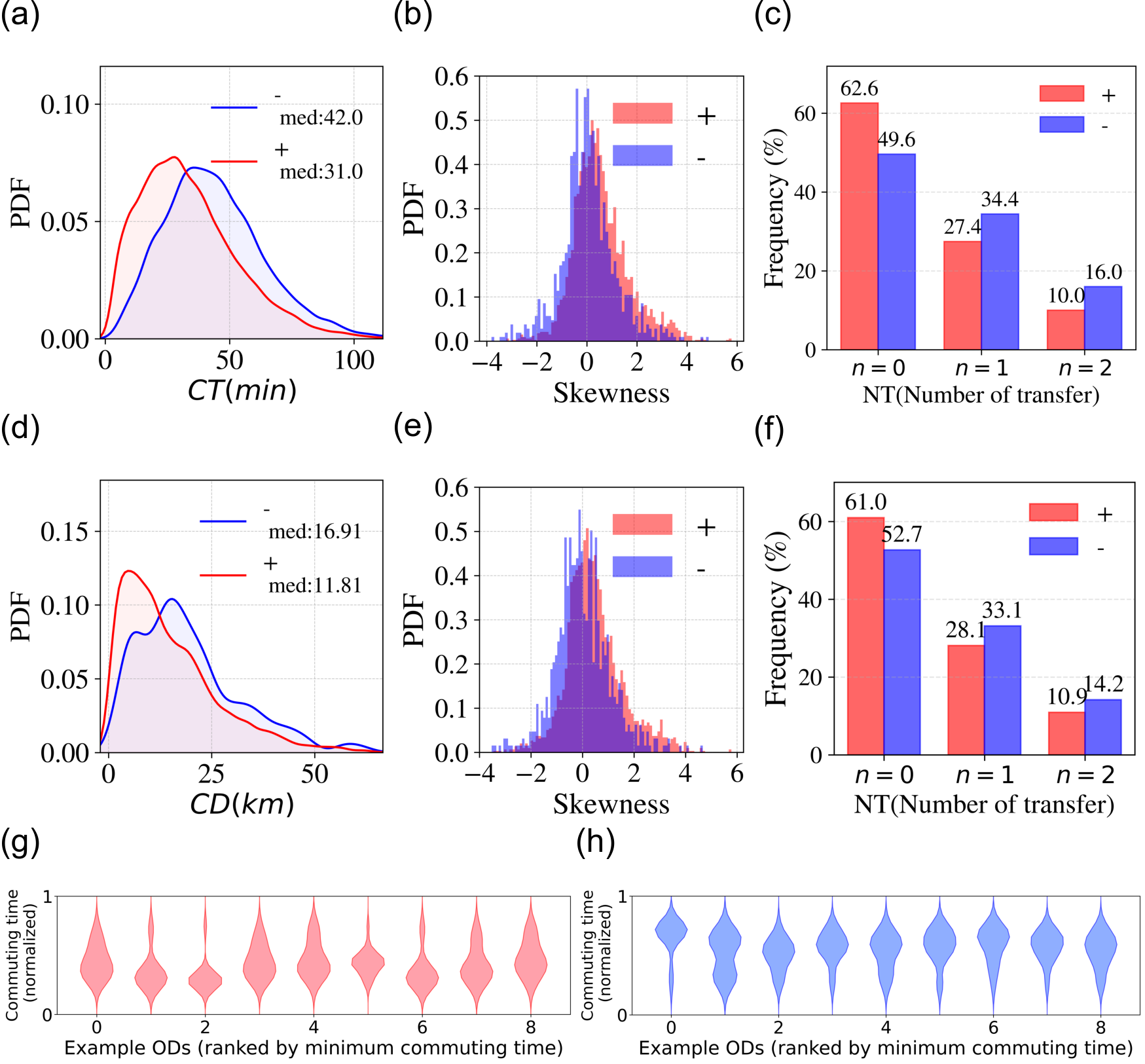}\\
\caption{Characterization of OD pairs with positive versus negative commuting time and distance coefficients. 
(\textbf{a}) Probability density distribution of commuting time (CT) for OD pairs with positive (red, "+") versus negative (blue, "-") coefficients, with median values shown in legends. (\textbf{b}) Distribution of skewness values for within-OD commuting time distributions, separated by coefficient sign (red "+" and blue "-"). (\textbf{c}) Frequency distribution of number of transfers (NT) for CT-positive versus CT-negative OD pairs, with percentages indicating the proportion of ODs in each category. (\textbf{d}) Probability density distribution of commuting distance (CD) for OD pairs with positive versus negative coefficients, with median values shown in legends. (\textbf{e}) Distribution of skewness values for within-OD commuting distance distributions. (\textbf{f}) Frequency distribution of number of transfers for CD-positive versus CD-negative OD pairs. (\textbf{g},\textbf{h}) Violin plots showing commuting time distributions for route alternatives within nine representative OD pairs, with commuting time rescaled to the unit interval [0,1] via min-max normalization (0 represents the minimum commuting time, 1 represents the maximum commuting time within each OD pair), ordered from left to right by minimum commuting time: (\textbf{g}) OD pairs with positive CT coefficients exhibit concentrated distributions near lower normalized values, while (\textbf{h}) OD pairs with negative CT coefficients show distributions with peaks shifted toward longer travel times.}
\label{fig11}
\end{figure}

To investigate this phenomenon, we segregated OD pairs based on the sign of their CT and CD coefficients and analyzed three key characteristics: (1) the central tendency of route attribute values (time/distance); (2) the distributional shape through skewness of within-OD attribute distributions; and (3) the frequency of transfers (Table~\ref{tab:motif}). Figure~\ref{fig11} presents comparative analyses ((a)--(f)) and representative examples through violin plots ((g)--(h)). A sensitivity analysis (Supplementary Section Sensitivity Analysis) confirms that this heterogeneity persists even when restricting the dataset to OD pairs with 40 or more commuters, ruling out small-sample variance as the primary cause.

\paragraph{Commuting Time Heterogeneity}

Why do commuters appear to ``prefer'' longer routes? Railway network service constraints and crowd-avoidance behavior drive this pattern. The negative commuting time (CT) coefficients observed in some OD pairs do not imply a genuine behavioral preference for longer travel times. Instead, this apparent anomaly is a statistical artifact driven by the underlying travel-time distributions. When unobserved utility trade-offs (e.g., choosing slower local trains to secure seating) or structural network constraints (e.g., fewer headways, longer transfer waiting times) cause a dominant subset of commuters to cluster at higher time values, the MNL regression mechanically yields a negative CT coefficient.

We establish this mechanical reversal through three interconnected observations. First, negative-coefficient ODs systematically involve more complex route profiles with longer journeys (median 42.0 vs.\ 31.0 min) and a higher reliance on transfers, compared to positive-coefficient ODs (Figure~\ref{fig11}a,c), indicating greater reliance on interruption-heavy network structures. Second, realized commuting times in these ODs are frequently bimodal. Rather than concentrating near the minimum possible time, the probability mass shifts toward the upper travel-time bound, reflecting distinct route choice clusters within the same OD pair (Figure~\ref{fig11}b,g,h). Finally, this structural bimodality induces a mechanical sign reversal. Individual commuters are still maximizing their overall utility by incorporating unobserved comforts or avoiding transfer uncertainties. However, because their chosen commutes cluster at the longer-time mode of the bimodal distribution, the standard MNL regression mechanically assigns a negative parameter to commuting time.

\paragraph{Commuting Distance Heterogeneity}

Distance coefficients exhibit parallel patterns. Negative-coefficient ODs correspond to longer journeys (median: 16.9 km vs. 11.8 km, Figure~\ref{fig11}d) with lower skewness (Figure~\ref{fig11}e), mirroring the time analysis. Transfer complexity follows the same pattern (Figure~\ref{fig11}f): fewer direct travelers, more multi-transfer users. 

Importantly, commuters may occasionally prefer geometrically longer routes if those specific alternatives happen to offer express services, higher service frequency, or more reliable connections compared to the more direct paths. Network topology constrains certain OD pairs into choice sets where preferred (lower perceived-cost) alternatives are not necessarily the shortest.

\paragraph{Mechanisms Underlying Bimodal Distributions} 

Bimodal patterns in both time and distance arise from two structural mechanisms, complemented by a behavioral explanation. First, transfers function as trip interruptions analogous to traffic signals. Previous research~\cite{ji_estimation_2015} showed traffic signals induce bimodal travel time distributions; our analysis reveals this extends to railway commuting. A possible explanation is that service frequency shapes these distributions: with long headways, commuters face binary outcomes—those arriving just before departure board immediately, while those just missing a train wait prolonged periods. This waiting-time clustering introduces significant delays and variance (Figure~\ref{fig11}c,f), explaining why high-transfer ODs develop bimodal characteristics concentrated at longer times and distances. 

Second, even without transfers ($n=0$), bimodality emerges from the coexistence of local and express train services. For longer-distance OD pairs, the time and distance differentials between local trains (all stops) and express trains (limited stops) increase proportionally with journey length, creating two distinct travel-time clusters that intensify with distance. Beyond these structural mechanisms, travel comfort offers a complementary behavioral explanation: commuters on long-distance routes may prefer slower direct services that offer greater seating availability over faster but less predictable transfer-dependent alternatives, further reinforcing the separation between the two modes.
\vspace{6pt}

\vspace{-24pt}

\paragraph{Quantifying Transfer Penalty}
\label{subsubsect:result_quantifying_tansfer_penalty}

By analyzing the ratio of model coefficients, we can quantify the implicit trade-offs in commuters' route selection. Specifically, the transfer penalty expressed in time-equivalent units can be calculated as:

\begin{equation}
\text{Transfer Penalty (minutes)} = \frac{\omega_{NT}}{\omega_{CT}},
\label{eq:transfer_penalty}
\end{equation}
where $\omega_{NT}$ is the coefficient for number of transfers and $\omega_{CT}$ is the coefficient for commuting time. This ratio quantifies the energy cost of transfers in travel time equivalents. Dimensionally, dividing the transfer coefficient (energy cost per transfer) by the time coefficient (energy cost per minute) cancels the implicit energy unit, translating the transfer burden directly into an equivalent value of commuting minutes.

Using the city-wide coefficients in Section~\ref{subsect:result_CPOD}, Table~\ref{tab:para} ($\omega_{NT} = 0.569$, $\omega_{CT} = 0.087$), Equation~(\eqref{eq:transfer_penalty}) yields an average transfer penalty of approximately 7 min. In other words, a typical Tokyo commuter treats one transfer as imposing a mental and physical cost equivalent to 7 extra minutes of in-vehicle time, and will only switch from a direct route to a one-transfer alternative when the latter saves at least 7 min.

\begin{figure}[H]
\centering
\includegraphics[width=0.7\textwidth]{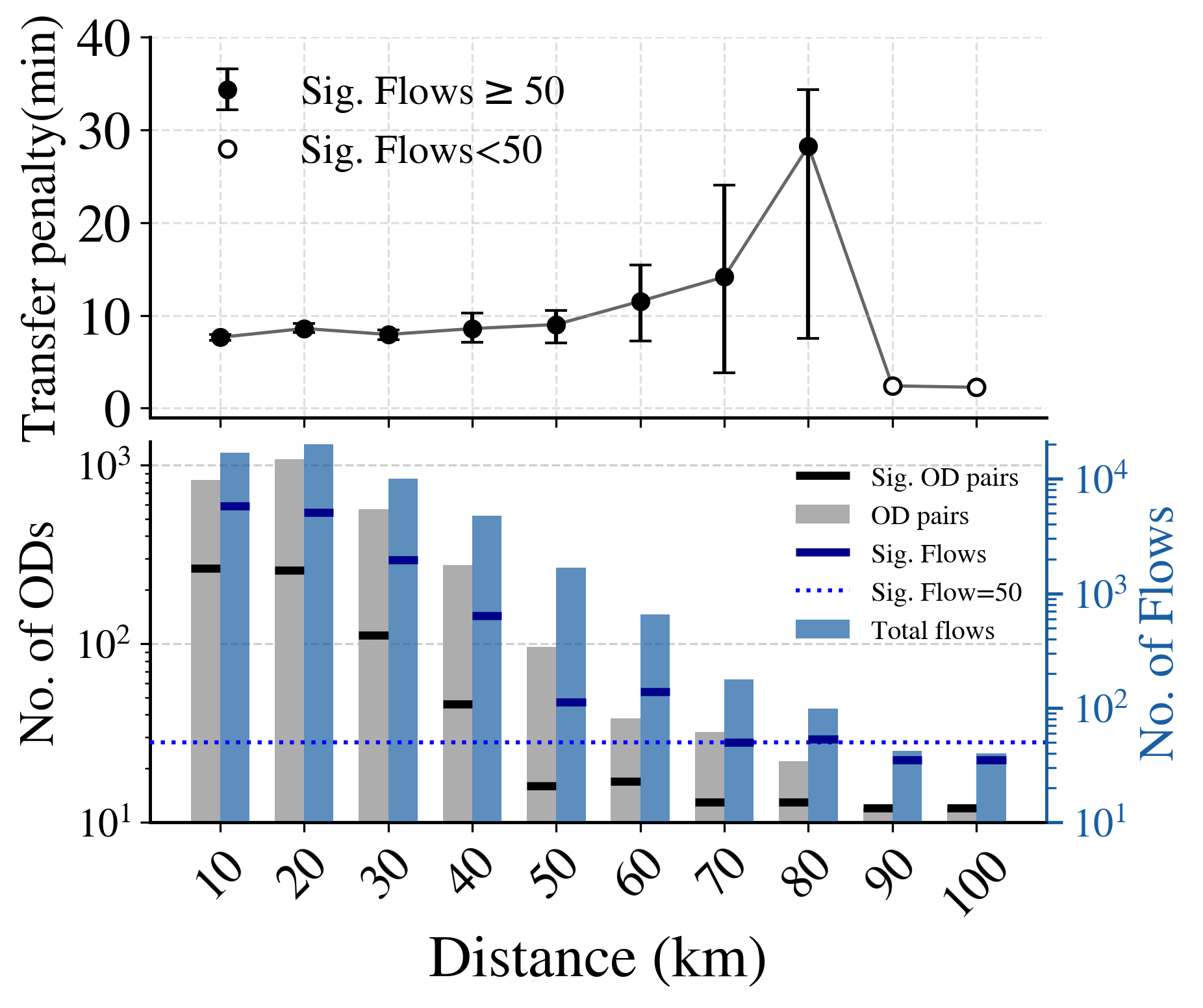}\\
\caption{Transfer penalty as a function of commuting distance.  
Based on the 2557 jointly significant OD pairs identified in Table~\ref{tab:stats}. Top: Per-OD transfer penalty $\omega_{NT}/\omega_{CT}$ summarized by the bin-wise median; error bars are 95\% percentile bootstrap confidence intervals of the median (10{,}000 resamples). Filled markers indicate bins with at least 50 significant flows; open markers indicate bins below this threshold, which should be interpreted with caution. Bottom: Number of OD pairs (grey, left axis) and total trajectories (blue, right axis) per distance bin. The blue dotted line marks the 50-flow threshold separating filled and open markers in the top panel.}
\label{fig12}
\end{figure}

To examine spatial variations, we apply Equation~(\ref{eq:transfer_penalty}) at the OD level. As detailed in Table~\ref{tab:stats}, this calculation is strictly performed on the final subset of 2{,}557 OD pairs that admit transfer alternatives and exhibit statistically significant positive coefficients for both time and transfers. Aggregating these OD-specific penalties by commuting distance (Figure~\ref{fig12}) reveals that the 7-min city-wide average conceals substantial distance dependence. For short- to medium-distance commutes ($<$50 km), penalties cluster below 10 min, close to the Model 1(city-wide pooled estimation) baseline. Between 50 and 70 km, the penalty steadily rises to $\sim$10--15 min. While the 80 km bin exhibits a sharp peak near 30 min, its wide confidence intervals suggest high volatility. Beyond this point, the penalty declines for the longest trips, where data becomes sparse. This escalation toward the periphery likely reflects the heightened inconvenience of transfers for suburban commuters, who often face longer headways and less seat availability at peripheral stations; this increase aligns with the spatial heterogeneity observed in \cite{kumagai_visualizing_2025}.

\vspace{-18pt}

Our estimated transfer penalty (approximately 7 min) is substantially lower than the 15--18 min reported by Garcia-Martinez et al. (2018) \cite{garcia-martinez_transfer_2018} for Madrid.  There are two potential reasons for this discrepancy.

First, in our state-determination procedure, each 1 km grid cell is treated as a single state. Consequently, the intra-cell walking time, specifically the duration from entering the grid to reaching the exact physical destination, is inherently omitted. Estimating the travel time coefficient $\omega_{CT}$ using this undercounted travel time introduces an upward bias in $\omega_{CT}$, thereby compressing the penalty ratio $\omega_{NT}/\omega_{CT}$.

Second, stated-preference (SP) and revealed-preference (RP) estimates capture distinct constructs. The survey-based 15--18 min penalty reflects subjective psychological resistance (e.g., \textit{anxiety}). However, as Garcia-Martinez et al.~\cite{garcia-martinez_transfer_2018} themselves note, habitual transferring confers a positive utility that offsets this resistance. Our GPS-based revealed-preference estimates capture habituated commuters who have already internalized these psychological costs, yielding a lower, behaviorally realized penalty.

\vspace{-12pt}
\begin{table}[H]
\begin{threeparttable}
\centering
\caption{Identification quality of OD-level estimates and parameter regimes}
\label{tab:stats}
\begin{tabular*}{\textwidth}{@{\extracolsep{\fill}}
    >{\raggedright\arraybackslash}m{0.5\textwidth}
    >{\centering\arraybackslash}m{0.1\textwidth} 
>{\centering\arraybackslash}m{0.3\textwidth} 
}
\toprule
\multicolumn{1}{>{\centering\arraybackslash}m{0.5\textwidth}}{\textbf{Criterion}} & \bm{$N$} & \textbf{\% of Total OD Pairs 5908} \\
\midrule
All ODs with Model 2 estimates & 5908 & 100.0 \\
\quad Without transfer alternatives & 1457 & 24.7 \\
\quad With transfer alternatives & 4451 & 75.3 \\
\midrule
\addlinespace
Criterion & $N$ & \% of 4451 \\
\midrule
\multicolumn{3}{l}{For ODs with transfer alternatives} 
\\
\quad Positive regime ($\omega_{NT} > 0$ and $\omega_{CT} > 0$) 
& 3653 & 82.1 \\
\quad Negative regime (at least one $\leq 0$) & 798 & 17.9 \\
\midrule
\addlinespace
Criterion & $N$ & \% of 3653 \\
\midrule
\multicolumn{3}{l}{Statistical significance (positive regime only for calculating transfer penalty):} \\
\quad $\omega_{CT}$ significantly $> 0$ ($p < 0.05$, one-sided) & 2559 & 70.0  \\
\quad $\omega_{NT}$ significantly $> 0$ ($p < 0.05$, one-sided) &  3364 & 92.1 \\
\quad Both significantly $> 0$ & 2557 & 70.0 \\
\bottomrule%
\end{tabular*}
\begin{tablenotes}[flushleft]
\fontsize{9}{10.8}\selectfont 
\item Note: This table illustrates the sequential filtering 
of OD pairs. Percentages in each section are calculated relative 
to the subset defined in the preceding section, as indicated by 
the column headers. The ``positive regime'' is defined by point 
estimates where both transfer ($\omega_{NT}$) and time 
($\omega_{CT}$) coefficients are strictly positive. Statistical 
significance is evaluated via the asymptotic Wald test applied 
separately to each coefficient, with one-sided $p$-values 
computed under $H_0: \omega \leq 0$ against $H_1: \omega > 0$.

\end{tablenotes}
\end{threeparttable}
\end{table}
\vspace{-12pt}

\section{Conclusions and Limitations}
\label{sect:conclusion_limitation}

This study addresses the central research question: Do commuters exhibit consistent route choice ``rationality" across different contexts and time in Tokyo's morning commute? The analysis reveals that Tokyo railway commuters exhibit strong but non-deterministic preference, with route choices clustering into highly preferred or highly unlikely alternatives with few intermediate options. These choices are driven primarily by factors including travel time, distance, and transfers. Additionally, commuters tend to avoid high-capacity stations for their first (or only) transfer, but choose high-capacity stations for their second transfer. Importantly, the results demonstrate nuanced consistency: the aggregate level of choice determinism and route attribute preferences remain temporally stable across seasons, showing structural consistency in decision-making.

Furthermore, the OD-pair-based heterogeneity analysis reveals significant contextual variation. Peak-hour commuters are less deterministic (indicated by lower $\beta$), exhibiting more exploratory behavior as they strategically deviate from minimal-cost routes to secure seats and ensure punctual arrival under capacity constraints. In contrast, off-peak commuters show more deterministic choices. Critically, we identify transfer interruptions as a key factor introducing behavioral heterogeneity and choice uncertainty. The disaggregate analysis confirms that this variation is driven primarily by network structure: OD pairs with high transfer complexity display counterintuitive preferences (choosing longer routes). Transfers create bimodal travel time distributions, where some commuters arrive significantly earlier while others arrive later \cite{ji_estimation_2015}, generating psychological pressure of~{\textit{I could have arrived much earlier if I had caught the train leaving 1 minute before.}} 
This structural uncertainty, amplified by the high transfer penalty (7-min equivalent), fundamentally limits model predictability for transfer-intensive routes \cite{kumagai_visualizing_2025}. These findings emphasize that policymakers must exercise greater caution in transfer route design, as transfer complexity drives both commuter stress and behavioral unpredictability in dense urban railway networks.

Methodologically, we demonstrate the successful application of this framework to a large-scale, passively collected GPS dataset. We addressed the technical challenges of processing raw smartphone GPS data, including its variable spatial resolution and noisy trajectories, by developing a comprehensive processing pipeline. This pipeline successfully transforms raw location records into high-resolution commuting mode classifications (distinguishing between railway, walking, cycling, and vehicle-based travel) \cite{sadeghian_stepwise_2022, sadeghian_review_2021} and specific route trajectories. Crucially, we overcame the computational challenge of route identification for large-scale GPS data \cite{sakamanee_methods_2020} by developing a novel spatial analysis method that accurately reconstructs chosen railway paths through transfer station identification.

Theoretically, our study provides strong empirical validation that the complex route choice system of Tokyo's railway network operates in a generally stable state. By confirming temporal consistency across periods, we demonstrate that the current transportation system maintains stable collective behavior even as individual users make personal trade-offs \cite{wilson_a_statistical_1967, zhang_analysis_2024, akamatsu_global_2023}. 

Notably, the finding that peak-hour commuters exhibit lower determinism is consistent with previous studies suggesting that these commuters adopt more strategic behaviors when choosing routes \cite{deng_heterogeneity_2025}, deviating from pure minimal-cost paths to secure seats or ensure punctual arrival under strict capacity constraints \cite{okubo_transportation_2022}. We contribute to this literature by providing a quantitative metric to measure this strategic shift.

Most importantly, the observed heterogeneity offers profound implications for future modeling studies. While the majority of route choice behaviors are predictable, we find that routes characterized by long travel times, long distances, and frequent transfers are a primary source of modeling difficulty, often being misclassified as unexplainable heterogeneity. Our results complement previous studies on vehicle travel which found that trip interruptions~\cite{ji_estimation_2015}, specifically traffic signals, induce bimodal distributions in commuting time when origin and destination locations are fixed. Our analysis reveals that this structural pattern is mirrored in long time and distance railway commuting: transfers act analogously to traffic signals as trip interruptions, inducing similar bimodal distributions in route choice behavior. This suggests that future route choice models should account for the bimodal nature of interruption-related uncertainty rather than assuming standard symmetric distributions.

\subsection{Connection to Maximum Entropy Inference}

The MNL model admits a well-established mathematical correspondence with the canonical ensemble of statistical mechanics: available routes correspond to microstates, generalized cost to energy, and the logit choice probability emerges as the maximum-entropy distribution subject to an observed expected-cost 
constraint~\cite{anas_discrete_1983,donoso_a_microeconomic_2010,helbing_the_physics_2004}. As detailed in Appendix~\ref{appendix:maxent} and Supplementary Section 16, the equivalence is algebraic: the MaxEnt approach maximizes Shannon entropy subject to the constraint that model-predicted expected route costs match their observed aggregate values, while the MLE approach maximizes the log-likelihood of observed choices under the MNL functional form. 

Despite their different starting points---one information-theoretic, the other statistical---both yield the same first-order condition and therefore the same parameter 
estimates~\cite{anas_discrete_1983,donoso_a_microeconomic_2010}. What differs is interpretation: in the econometric reading, 
$\boldsymbol{\omega}_{\mathrm{raw}}$ is the coefficient vector that 
best explains individual choices; in the MaxEnt reading, it is the 
Lagrange multiplier that enforces the macroscopic cost constraint. 
The logsum term likewise admits dual meaning---the log-partition 
function in statistical mechanics, and the expected maximum utility 
in discrete choice theory~\cite{train_discrete_2009}. Helbing \& Nagel~\cite{helbing_the_physics_2004} provide a representative review of this duality in the physics-of-cities literature. Building on this theoretical duality, the present manuscript establishes the empirical and methodological groundwork---a validated, large-scale route choice dataset with stable aggregate parameters, that enables future Maximum Entropy-based investigations.

\subsection{Limitations}

Despite its significant contributions, this study has several limitations. Fare data were not collected, which prevents direct estimation of monetary trade-offs in route choice. Furthermore, the multi-stage methodology, from identifying transportation modes and transfer stations to reconstructing final chosen routes from raw GPS data, introduces potential accuracy issues at each step, albeit minimal. A related limitation is the lack of semantic labels, meaning some brief interruptions could represent non-transit activities like buying breakfast. However, strict temporal and spatial filtering in our methodology minimizes this noise. Combined with the strong cultural norm of direct workplace travel during Tokyo's morning peak, these rare misclassifications are unlikely to skew the aggregate route choice models.

While travel time components such as access, egress, walking, and waiting times are known to influence commuter valuation, they were not explicitly calculated in this study. As illustrated in Supplementary Figure S7, the GPS logging frequency provided by Agoop is insufficiently dense to pinpoint precise state changes at this granular level. Specifically, there is typically a spatial and temporal gap between the final point of an \textit{access} walk and the first logged \textit{move} point on a train. According to the data provider’s handbook, the dataset most reliably captures trip initiations, movement state changes, or transitions when users emerge from underground for transfers. Consequently, our model focuses on the most robustly captured attributes: total door-to-door commuting time, high-speed movement points for network matching, and \textit{stroll} points to identify transfer behavior.

Furthermore, our reliance on door-to-door time means the model cannot explicitly differentiate between access and egress modes (e.g., walking, cycling, or driving). While these modes influence total travel time, treating these segments uniformly is a robust approximation for the Tokyo metropolitan context. According to the Ministry of Land, Infrastructure, Transport and Tourism (MLIT) \cite{mlit_transfer_survey_2017}, 87\% of regional commuters access stations via active transport (70\% walking, 17\% bicycles). Motorized access accounts for only 13\% due to parking constraints and high bus fares. However, because our model calculates a route's base time using the median duration of all commuters observed on that path, grouping these access modes introduces measurement bias. Specifically, a high proportion of cyclists on a given route will skew the median travel time lower, making distant stations appear more accessible than they actually are for pedestrians. Distinguishing access modes to calculate mode-specific travel times represents an important direction for future refinement.

Additionally, residual correlations may remain among route categories that share physical infrastructure. Future research should explicitly account for this overlap. Following Dixit et al. \cite{dixit_perception_2023}, who demonstrated that shared transfer nodes drive traveler perception in transit networks, integrating a Path Size Correction Logit (PSCL) represents a promising methodological extension.

Furthermore, to mitigate the inherent multicollinearity between travel time and distance, future model specifications should replace raw distance with distance circuity. As demonstrated by Dixit et al. \cite{dixit_perception_2023}, circuity effectively captures the penalty of indirect routing independent of travel time.

Additionally, train frequency was omitted due to unavailable unified timetables, though Tokyo's short headways minimize this impact. Finally, station capacity is treated as a static annual maximum in the current model. A more refined approach would assign time-specific capacity values matched to each commuter's actual transit time at each transfer station. We leave this as a direction for future work.

Nonetheless, the study makes important contributions: this is the first attempt to apply a mobility dataset of this unprecedented scale and spatial resolution to railway route choice modeling, yielding significant empirical findings. Crucially, the framework provides an interpretable, theory-driven approach, allowing us to directly interpret and quantify commuter behavior based on observable route attributes.



	\section*{Supplementary Materials} 
 The following supporting information can be downloaded at: \href{https://media.sciltp.com/articles/others/2606231539173202/J.-Soc.-Phys.-25100061-Author-Supplementary.pdf}{https://media.sciltp.com/articles/others/2606231\allowbreak539173202/J.-Soc.-Phys.-25100061-Author-Supplementary.pdf}. Reference \cite{Vittinghoff_2007} is cited in supplementary materials.

\section*{Author Contributions}
Y.Y.Z.: designed the research plan, developed methods of data analysis, performed the numerical calculations, and wrote the manuscript. H.T.: checked methods of data analysis and revised the manuscript. M.T.: led this project, and directed writing of the manuscript. All authors have read and agreed to the published version of the manuscript.

\section*{Funding}
This work was supported by the Japan Society for the Promotion of Science, Grant-in-Aid for Scientific Research (B) (GrantNumber 23K22980 to MT). The funder had no role in study design, data collection and analysis, decision to publish, or preparation of the manuscript.
 
\section*{Institutional Review Board Statement}
{Ethical approval was not required for the study involving humans in accordance with the local legislation and institutional requirements.}

\section*{Informed Consent Statement}
{Written informed consent to participate in this study was not required from the participants or the participants’ legal guardians/next of kin in accordance with the national legislation and the institutional requirements.}

\section*{Data Availability Statement}
The GPS data cannot be shared publicly because data is available only on request from a third party. Data are available from Agoop Corporation, a Japanese private company that provides location information big data acquired from smartphone applications. The specific product is “Pointo-gata ryoudou-jinkou data” (Point-type population data). Interested researchers who meet the criteria for access to confidential data can visit https://agoop.co.jp/service/dynamic-population-data/ for more information. Spatial data of the physical railway network were obtained from the National Land Numerical Information (N02, 2023) dataset provided by Japan’s Ministry of Land, Infrastructure, Transport and Tourism (MLIT). The data are freely available under the CC BY 4.0 license at https://nlftp.mlit.go.jp/ksj/gml/datalist/KsjTmplt-N02-2023.html.

\section*{Conflicts of Interest}
 {The authors declare no conflict of interest.}

\section*{Use of AI and AI-Assisted Technologies}
{During the preparation of this work, the authors used AI-assisted writing tools to check for grammatical errors and improve the English language style. After using this tool, the authors reviewed and edited the content as needed and take full responsibility for the content of the published article.}

\section*{Appendix A. Mathematical Derivations for Parameter Normalization}\label{app:derivations}
\addcontentsline{toc}{section}{Appendix A} 
\renewcommand{\thesubsection}{Appendix A.\arabic{subsection}}
\setcounter{subsection}{0}

\renewcommand{\thefigure}{A\arabic{figure}}
\setcounter{figure}{0}   

\renewcommand{\theequation}{A\arabic{equation}}
\setcounter{equation}{0}

This appendix provides detailed mathematical derivations supporting the Delta method calculations described in Section~\ref{subsect:model_estimation} of the main text. 

\begin{enumerate}[leftmargin=2.2em, labelsep=4mm, label=(\arabic*), itemsep=0pt, topsep=3pt, parsep=0pt]
    \item Estimate Raw Parameters: Use Maximum Likelihood Estimation (MLE) to find the unconstrained weight vector, $\hat{\boldsymbol{\omega}}_{\mathrm{raw}}$.
    
    \item Calculate Raw Covariance: Approximate the covariance matrix of these raw parameters, $\widehat{\Sigma}_{\boldsymbol{\omega}_{\mathrm{raw}}}$, by inverting the Hessian matrix of the negative log-likelihood function.
    
    \item Apply Delta Method: In order to get the variance and covariance of the non-linearly transformed parameters $(\beta, \boldsymbol{\omega})$, use the Delta method to calculate them so that it can directly get from the $\boldsymbol{\omega}_{\text{raw}}$, whose covariance $\widehat{\Sigma}_{\boldsymbol{\omega}_{\mathrm{raw}}}$ can be approximated MLE during optimization process. This gives:
  \begin{itemize}[leftmargin=*, labelsep=6.2mm, itemsep=0pt, topsep=3pt, partopsep=0pt, parsep=0pt]
        \item The variance of the scale parameter, $\operatorname{Var}(\hat{\beta})$.
        \item The full covariance matrix of the normalized weights, $\operatorname{Cov}(\hat{\boldsymbol{\omega}})$.
    \end{itemize}
    
    \item Test for Consistency: \textls[-15]{Using these variances and covariances, we perform statistical tests (like the Wald test) and construct confidence intervals to determine if the solutions for $\beta$ and $\boldsymbol{\omega}$ are consistent across the 12 months.}
\end{enumerate}

To maintain consistency, we use the same notation as the main text: $\boldsymbol{\omega}_{\text{raw}}$ for the unconstrained weight vector, $\beta$ for the scale parameter, and $\boldsymbol{\omega}$ for the normalized weight vector.

\subsection{Gradient and Jacobian Derivations}

\subsubsection{Gradient of the Norm $\beta = \|\boldsymbol{\omega}_{\text{raw}}\|_2$}

Let $f(\boldsymbol{\omega}_{\text{raw}}) = \|\boldsymbol{\omega}_{\text{raw}}\|_2 = \left( \sum_{j=1}^k \omega_{j, \text{raw}}^2 \right)^{1/2}$. The gradient is a vector of partial derivatives with respect to each element of $\boldsymbol{\omega}_{\text{raw}}$:
\begin{equation}
\nabla_{\boldsymbol{\omega}_{\text{raw}}} f = \begin{bmatrix} \frac{\partial f}{\partial \omega_{1, \text{raw}}} \\ \frac{\partial f}{\partial \omega_{2, \text{raw}}} \\ \vdots \\ \frac{\partial f}{\partial \omega_{k, \text{raw}}} \end{bmatrix}
\end{equation}

We compute the partial derivative for a single element $\omega_{i, \text{raw}}$ using the chain rule:
\begin{align*}
\frac{\partial f}{\partial \omega_{i, \text{raw}}} &= \frac{\partial}{\partial \omega_{i, \text{raw}}} \left( \sum_{j=1}^k \omega_{j, \text{raw}}^2 \right)^{1/2} \\
&= \frac{1}{2} \left( \sum_{j=1}^k \omega_{j, \text{raw}}^2 \right)^{-1/2} \cdot \frac{\partial}{\partial \omega_{i, \text{raw}}} \left( \sum_{j=1}^k \omega_{j, \text{raw}}^2 \right) \\
&= \frac{1}{2f} \cdot (2\omega_{i, \text{raw}}) = \frac{\omega_{i, \text{raw}}}{f}
\end{align*}

Since this result holds for every element, the full gradient vector is:
\begin{equation}
\nabla_{\boldsymbol{\omega}_{\text{raw}}} f = \frac{1}{f} \begin{bmatrix} \omega_{1, \text{raw}} \\ \omega_{2, \text{raw}} \\ \vdots \\ \omega_{k, \text{raw}} \end{bmatrix} = \frac{\boldsymbol{\omega}_{\text{raw}}}{f} = \frac{\boldsymbol{\omega}_{\text{raw}}}{\|\boldsymbol{\omega}_{\text{raw}}\|_2} = \boldsymbol{\omega}
\end{equation}

This demonstrates that the gradient of the norm with respect to the raw weight vector equals the normalized weight vector itself.

\subsubsection{Jacobian of the Normalization $\boldsymbol{\omega} = \boldsymbol{\omega}_{\text{raw}}/\|\boldsymbol{\omega}_{\text{raw}}\|_2$}
Applying the quotient rule for vector functions:
\begin{align}
\mathrm{d}\boldsymbol{\omega} &= \mathrm{d}\left(\frac{\boldsymbol{\omega}_{\text{raw}}}{f}\right) = \frac{1}{f}\mathrm{d}\boldsymbol{\omega}_{\text{raw}} + \boldsymbol{\omega}_{\text{raw}} \cdot \mathrm{d}\left(\frac{1}{f}\right)
\end{align}

The second term is derived from the chain rule:
\begin{align}
\mathrm{d}\left(\frac{1}{f}\right) &= -\frac{1}{f^2}\mathrm{d}f = -\frac{\boldsymbol{\omega}_{\text{raw}}^\top \mathrm{d}\boldsymbol{\omega}_{\text{raw}}}{f^3}
\end{align}

Substituting this back and simplifying, we get:
\begin{align}
\mathrm{d}\boldsymbol{\omega} &= \frac{1}{f}\mathrm{d}\boldsymbol{\omega}_{\text{raw}} - \frac{\boldsymbol{\omega}_{\text{raw}}(\boldsymbol{\omega}_{\text{raw}}^\top \mathrm{d}\boldsymbol{\omega}_{\text{raw}})}{f^3} \nonumber \\
&= \frac{1}{f}\left(I - \frac{\boldsymbol{\omega}_{\text{raw}}\boldsymbol{\omega}_{\text{raw}}^\top}{f^2}\right)\mathrm{d}\boldsymbol{\omega}_{\text{raw}} \\
&= \frac{1}{\beta}(I - \boldsymbol{\omega}\boldsymbol{\omega}^\top)\mathrm{d}\boldsymbol{\omega}_{\text{raw}}\nonumber 
\end{align}

Therefore, the Jacobian matrix is:
\begin{equation}
\mathbf{J}_{\omega} = \frac{\partial \boldsymbol{\omega}}{\partial \boldsymbol{\omega}_{\text{raw}}} = \frac{1}{\beta}(I - \boldsymbol{\omega}\boldsymbol{\omega}^\top)
\end{equation}

The matrix $(I - \boldsymbol{\omega}\boldsymbol{\omega}^\top)$ is a projection matrix that maps any vector onto the space orthogonal to $\boldsymbol{\omega}$, which is crucial for maintaining the unit length of the normalized vector under small perturbations.

\subsection{Delta Method for Standard Errors}
\label{app:delta}

\textls[-15]{The Delta method is a powerful statistical technique used to approximate the variance and covariance of a function of an asymptotically normal random variable. In simpler terms, if the variance of a parameter estimate is known, the Delta method helps to find the variance of a new parameter that is a nonlinear function of the original one.} 

The core of the Delta method is the first-order Taylor expansion, which approximates a nonlinear function with a linear one. We start with a vector of random variables, $\mathbf{X}$, with mean $\boldsymbol{\mu}$ and covariance matrix $\boldsymbol{\Sigma}$. We want to find the approximate covariance of a new vector, $g(\mathbf{X})$, which is a nonlinear function of $\mathbf{X}$.

The Taylor expansion of $g(\mathbf{X})$ around its mean $\boldsymbol{\mu}$ is:
\begin{equation}
g(\mathbf{X}) \approx g(\boldsymbol{\mu}) + \mathbf{J}(\mathbf{X} - \boldsymbol{\mu})
\end{equation}
where $\mathbf{J}$ is the Jacobian matrix of $g$ evaluated at the mean $\boldsymbol{\mu}$. It is a matrix of all the first-order partial derivatives of the function $g$. This Jacobian matrix is the linear part of the approximation, capturing how a small change in $\mathbf{X}$ affects the output of $g(\mathbf{X})$.

\subsubsection{General Framework}
Given that the raw MLE parameter estimates $\hat{\boldsymbol{\omega}}_{\text{raw}}$ are asymptotically normal with covariance matrix $\widehat{\Sigma}_{\boldsymbol{\omega}_{\text{raw}}}$, the Delta method provides the covariance matrix of a transformed vector $g(\hat{\boldsymbol{\omega}}_{\text{raw}})$ as:
\begin{equation}
\operatorname{Cov}(g(\hat{\boldsymbol{\omega}}_{\text{raw}})) \approx \mathbf{J} \widehat{\Sigma}_{\boldsymbol{\omega}_{\text{raw}}} \mathbf{J}^\top
\end{equation}
where $\mathbf{J}$ is the Jacobian of the transformation $g$ evaluated at $\hat{\boldsymbol{\omega}}_{\text{raw}}$.

\textls[-15]{Here, $g(\boldsymbol{\omega}_{\mathrm{raw}})$ represents the nonlinear transformation that normalizes $\boldsymbol{\omega}_{\mathrm{raw}}$ and extracts the scale parameter $\beta$.} 

\begin{itemize}[leftmargin=*, labelsep=6.2mm, itemsep=0pt, topsep=3pt, partopsep=0pt, parsep=0pt]
    \item $g(\boldsymbol{\omega}_{\mathrm{raw}})$ is the new vector $\boldsymbol{\theta}_{\mathrm{final}} = (\beta, \boldsymbol{\omega})^{T}$.
    \item This transformation is nonlinear because it involves a square root (in the norm calculation) and division by a variable quantity (in the normalization).
\end{itemize}

\paragraph{Variance of the Scale Parameter $\beta$}
The scale parameter is $\beta = g_1(\boldsymbol{\omega}_{\text{raw}}) = \|\boldsymbol{\omega}_{\text{raw}}\|_2$. Its gradient is the first row of the full Jacobian matrix.
\begin{equation}
\operatorname{Var}(\hat{\beta}) \approx (\nabla_{\boldsymbol{\omega}_{\text{raw}}} \beta)^\top \widehat{\Sigma}_{\boldsymbol{\omega}_{\text{raw}}} (\nabla_{\boldsymbol{\omega}_{\text{raw}}} \beta) = \boldsymbol{\omega}^\top \widehat{\Sigma}_{\boldsymbol{\omega}_{\text{raw}}} \boldsymbol{\omega}
\end{equation}

\paragraph{Covariance of $\boldsymbol{\omega}$}
The normalized weights are $\boldsymbol{\omega} = g_2(\boldsymbol{\omega}_{\text{raw}}) = \boldsymbol{\omega}_{\text{raw}}/\|\boldsymbol{\omega}_{\text{raw}}\|_2$.
\begin{equation}
\operatorname{Cov}(\hat{\boldsymbol{\omega}}) \approx \mathbf{J}_{\omega} \widehat{\Sigma}_{\boldsymbol{\omega}_{\text{raw}}} \mathbf{J}_{\omega}^\top = \frac{1}{\beta^2}(I - \boldsymbol{\omega}\boldsymbol{\omega}^\top) \widehat{\Sigma}_{\boldsymbol{\omega}_{\text{raw}}} (I - \boldsymbol{\omega}\boldsymbol{\omega}^\top)
\end{equation}

\subsection{Observed Information Matrix}
The asymptotic covariance matrix $\widehat{\Sigma}_{\boldsymbol{\omega}_{\text{raw}}}$ is estimated using the inverse of the observed information matrix. For the canonical ensemble model, the Hessian of the negative log-likelihood function is:
\begin{equation}
\mathbf{H} = \beta^2 \sum_{n} \left(\mathbf{X}_{n}^\top \operatorname{Diag}(P_n) \mathbf{X}_{n} - (\mathbf{X}_{n}^\top P_n)(\mathbf{X}_{n}^\top P_n)^\top\right)
\end{equation}
where $\mathbf{X}_{n}$ is the matrix of attributes for choice set $n$ and $P_n$ is the vector of choice probabilities. The asymptotic covariance matrix is then given by:
\begin{equation}
\widehat{\Sigma}_{\boldsymbol{\omega}_{\text{raw}}} = \mathbf{H}^{-1}
\end{equation}

\subsection{Testing Temporal Stability of $\beta$}
\label{ap:stability}

To assess whether the overall sensitivity to energy differences, represented by the scale parameter $\beta$, remains stable across the twelve-month study period, we conduct a formal hypothesis test. While we expect the normalized weights $\boldsymbol{\omega}$ to show some variation due to changing conditions, we hypothesize that the fundamental choice sensitivity $\beta$ is consistent.

\paragraph{Obtaining Standard Errors for $\beta$}

First, we estimate the scale parameter $\hat{\beta}^{(t)}$ and its associated standard error, $SE(\hat{\beta}^{(t)})$, for each month $t \in \{1, \ldots, 12\}$. Since $\beta$ is a nonlinear transformation of the unconstrained parameters $\boldsymbol{\omega}_{\mathrm{raw}}$, we use the Delta method to calculate its variance. The core idea of the Delta method is to use a first-order Taylor expansion to approximate a nonlinear function with a linear one, which allows the uncertainty of the original parameters to be translated into the uncertainty of the new, transformed parameters.

The variance of the scale parameter is calculated using the following formula:
\begin{equation}
\operatorname{Var}(\hat{\beta}) \approx 
(\nabla_{\boldsymbol{\omega}_{\mathrm{raw}}} \beta)^{\top}
\widehat{\Sigma}_{\boldsymbol{\omega}_{\mathrm{raw}}}
(\nabla_{\boldsymbol{\omega}_{\mathrm{raw}}} \beta)
\end{equation}
where $\nabla_{\boldsymbol{\omega}_{\mathrm{raw}}} \beta$ is the gradient of the scale parameter with respect to the raw parameter vector, and $\widehat{\Sigma}_{\boldsymbol{\omega}_{\mathrm{raw}}}$ is the covariance matrix of the raw parameters. We approximate $\widehat{\Sigma}_{\boldsymbol{\omega}_{\mathrm{raw}}}$ by inverting the Hessian matrix of the negative log-likelihood function,
\begin{equation}
\widehat{\operatorname{Cov}}(\hat{\boldsymbol{\omega}}_{\mathrm{raw}}) = [-\mathbf{H}(\hat{\boldsymbol{\omega}}_{\mathrm{raw}})]^{-1}.
\end{equation}

\section*{Appendix B. Maximum Entropy Framework}
\renewcommand{\thesection}{\Alph{section}}
\setcounter{section}{1}
\phantomsection
\refstepcounter{section}
\label{appendix:maxent}

\addcontentsline{toc}{section}{Appendix B}
\renewcommand{\thesubsection}{Appendix B.\arabic{subsection}}
\setcounter{subsection}{0}

\renewcommand{\thefigure}{B\arabic{figure}}
\setcounter{figure}{0}   

\renewcommand{\theequation}{B\arabic{equation}}
\setcounter{equation}{0}

The Maximum Entropy (MaxEnt) principle states that, given a set of known constraints, the least biased probability distribution is the one that maximizes Shannon entropy. We apply this principle to derive route choice probabilities under two different assumptions about spatial homogeneity. The complete step-by-step derivations, including the full MLE equivalence proof, are provided in Supplementary Material Section 16.

\subsection{Notation}
Let $O$ denote the set of all OD pairs, $N_o$ the set of commuters in OD pair $o$, and $R_o$ the number of route alternatives. For commuter $n \in N_o$ and route $j \in \{1, \dots, R_o\}$, let $P_{jn}$ denote the choice probability, $\mathbf{X}_{jn} \in \mathbb{R}^M$ the attribute vector, and $y_{jn} \in \{0,1\}$ the observed choice indicator.

\subsection{Objective Function}
We maximize the total Shannon entropy:
\begin{equation}
    S = -\sum_{o \in O} \sum_{n \in N_o} \sum_{j=1}^{R_o} P_{jn} \ln P_{jn}
    \label{eq:entropy}
\end{equation}

\subsection{Model 1 (City-Wide Pooled Estimation)}
\label{par:global}

This variant assumes spatial homogeneity: all commuters share a single parameter vector $\boldsymbol{\omega}_{\mathrm{raw}} \in \mathbb{R}^M$. Maximizing Equation~\eqref{eq:entropy} subject to (i)~per-commuter normalization $\sum_j P_{jn} = 1$ and (ii)~the global expected-value constraint
\begin{equation}
    \sum_{o,n,j} P_{jn}\, \mathbf{X}_{jn}
    = \sum_{o,n,j} y_{jn}\, \mathbf{X}_{jn}
    \label{eq:global_constraint}
\end{equation}
yields, via standard Lagrangian optimization (see Supplementary Section~S16 for derivation):
\begin{equation}
    P_{jn} = \frac{\exp\!\left( -\boldsymbol{\omega}_{\mathrm{raw}}^\top \mathbf{X}_{jn} \right)}
    {\displaystyle\sum_{j'=1}^{R_o} \exp\!\left( -\boldsymbol{\omega}_{\mathrm{raw}}^\top \mathbf{X}_{j'n} \right)}
    \label{eq:global_model}
\end{equation}
where $\boldsymbol{\omega}_{\mathrm{raw}}$ is the Lagrange multiplier enforcing Equation~\eqref{eq:global_constraint}.

\subsection{MLE--MaxEnt Equivalence}
\textls[5]{Maximizing the log-likelihood $\ell = \sum_{o,n,j} y_{jn} \ln P_{jn}$ with $P_{jn}$ given by Equation~\eqref{eq:global_model} produces the first-order condition}
\begin{equation}
    \sum_{n=1}^N \sum_{j \in C_n} \mathbf{X}_{nj} \bigl[ y_{nj} - P_{nj}(\boldsymbol{\omega}_{\mathrm{raw}}) \bigr] = 0
\end{equation}
which is algebraically identical to Equation~\eqref{eq:global_constraint}. The Lagrange multiplier and the MLE estimator therefore coincide, establishing the well-known duality between MaxEnt inference and discrete choice estimation \cite{anas_discrete_1983, donoso_a_microeconomic_2010}.

Decomposing $\boldsymbol{\omega}_{\mathrm{raw}} = \beta\,\boldsymbol{\omega}$ with $\beta = \|\boldsymbol{\omega}_{\mathrm{raw}}\|_2$ and $\|\boldsymbol{\omega}\|_2 = 1$ gives:
\begin{equation}
    P_{jn} = \frac{\exp\!\left( -\beta\,\boldsymbol{\omega}^\top \mathbf{X}_{jn} \right)}
    {\displaystyle\sum_{j'=1}^{R_o} \exp\!\left( -\beta\,\boldsymbol{\omega}^\top \mathbf{X}_{j'n} \right)}
    \label{eq:global_decomposed}
\end{equation}

Here $\boldsymbol{\omega}$ captures relative attribute preferences and $\beta$ controls overall cost sensitivity. These parameters are global: they represent city-wide average preferences and cost sensitivity.

\subsection{Model 2 (OD-Specific Estimation)}
\label{par:od_specific}

This variant assumes spatial heterogeneity: each OD pair $o$ constitutes an independent sub-system with its own parameter vector $(\boldsymbol{\omega}_{\mathrm{raw}})_o$. The normalization constraint is identical to Model~1. The expected-value constraint is imposed per OD pair:
\begin{equation}
    \sum_{n \in N_o} \sum_{j=1}^{R_o} P^o_{jn}\, \mathbf{X}_{jn}
    = \sum_{n \in N_o} \sum_{j=1}^{R_o} y_{jn}\, \mathbf{X}_{jn},
    \quad \forall\, o \in O
    \label{eq:od_constraint}
\end{equation}

Because the sub-systems are independent, the total entropy decomposes additively and each OD pair can be optimized separately. The resulting choice probability takes the same MNL form as Equation~\eqref{eq:global_model} with $(\boldsymbol{\omega}_{\mathrm{raw}})_o$ replacing $\boldsymbol{\omega}_{\mathrm{raw}}$, and the same MLE--MaxEnt equivalence holds within each OD pair. Decomposing $(\boldsymbol{\omega}_{\mathrm{raw}})_o = \beta^o\,\boldsymbol{\omega}^o$ yields OD-specific parameters characterizing the cost sensitivity and attribute preferences of each sub-population.

	\small
    \bibliographystyle{scilight}

\begin{thebibliography}{99}
\providecommand{\natexlab}[1]{#1}


\bibitem[Ardeshiri and Vij(2019)]{ardeshiri_lifestyles_2019}
Ardeshiri, A.; Vij, A.
\newblock Lifestyles, Residential Location, and Transport Mode Use: A Hierarchical Latent Class Choice Model.
\newblock {\em Transp. Res. Part A Policy Pract.} {\bf 2019}, {\em 126}, 342--359.

\bibitem[Wang \em{et~al.}(2021)Wang, Yan, and Wu]{wang_free_2021}
Wang, H.; Yan, X.Y.; Wu, J.
\newblock Free Utility Model for Explaining the Social Gravity Law.
\newblock {\em J. Stat. Mech.} {\bf 2021}, {\em 2021}, 033418.

\bibitem[Zhang \em{et~al.}(2024)Zhang, Chen, Liu, Yu, and Wang]{zhang_analysis_2024}
Zhang, L.; Chen, T.; Liu, Z.; et al.
\newblock Analysis of Multi-Modal Public Transportation System Performance Under Metro Disruptions: A Dynamic Resilience Assessment Framework.
\newblock {\em Transp. Res. Part A Policy Pract.} {\bf 2024}, {\em 183}, 104077.

\bibitem[Akamatsu \em{et~al.}(2023)Akamatsu, Satsukawa, and Oyama]{akamatsu_global_2023}
Akamatsu, T.; Satsukawa, K.; Oyama, Y.
\newblock Global Stability of Day-to-Day Dynamics for Schedule-Based Markovian Transit Assignment with Boarding Queues.
\newblock {\em arXiv} {\bf 2023}, arXiv:2304.02194.

\bibitem[Deng \em{et~al.}(2025{\natexlab{a}})Deng, Li, Yang, Yuan, and Chen]{deng_heterogeneity_2025}
Deng, J.; Li, T.; Yang, Z.; et al.
\newblock Heterogeneity in Route Choice During Peak Hours: Implications on Travel Demand Management.
\newblock {\em Travel Behav. Soc.} {\bf 2025}, {\em 38}, 100922.

\bibitem[Deng \em{et~al.}(2025{\natexlab{b}})Deng, Yang, Li, Gao, Bachmann, and Chen]{deng_unveiling_2025}
Deng, J.; Yang, Y.; Li, T.; et al.
\newblock Unveiling Route Choice Preferences and Classifying Travelers Based on Distinct Travel Patterns Using Trajectory Data.
\newblock {\em Case Stud. Transp. Policy} {\bf 2025}, {\em 19}, 101399.

\bibitem[Sakamanee \em{et~al.}(2020)Sakamanee, Phithakkitnukoon, Smoreda, and Ratti]{sakamanee_methods_2020}
Sakamanee, P.; Phithakkitnukoon, S.; Smoreda, Z.; et al.
\newblock Methods for Inferring Route Choice of Commuting Trip from Mobile Phone Network Data.
\newblock {\em ISPRS Int. J. Geo-Inf.} {\bf 2020}, {\em 9}, 306.

\bibitem[Kaneko \em{et~al.}(2018)Kaneko, Oka, Chikaraishi, Becker, and Fukuda]{kaneko_route_2018}
Kaneko, N.; Oka, H.; Chikaraishi, M.; et al.
\newblock Route Choice Analysis in the Tokyo Metropolitan Area Using a Link-Based Recursive Logit Model Featuring Link Awareness.
\newblock {\em Transp. Res. Procedia} {\bf 2018}, {\em 34}, 251--258.

\bibitem[Li \em{et~al.}(2016)Li, Miwa, Morikawa, and Liu]{li_incorporating_2016}
Li, D.; Miwa, T.; Morikawa, T.; et al.
\newblock Incorporating Observed and Unobserved Heterogeneity in Route Choice Analysis with Sampled Choice Sets.
\newblock {\em Transp. Res. Part C Emerg. Technol.} {\bf 2016}, {\em 67}, 31--46.

\bibitem[Arriagada \em{et~al.}(2025)Arriagada, Prato, and Munizaga]{arriagada_incorporating_2025}
Arriagada, J.; Prato, C.; Munizaga, M.
\newblock Incorporating the Inertia Effect into a Route Choice Model Using Fare Transaction Data from a Large-Scale Public Transport Network.
\newblock {\em Transp. Res. Part A Policy Pract.} {\bf 2025}, {\em 196}, 104467.

\bibitem[Fosgerau \em{et~al.}(2022)Fosgerau, Paulsen, and Rasmussen]{fosgerau_perturbed_2022}
Fosgerau, M.; Paulsen, M.; Rasmussen, T.K.
\newblock A Perturbed Utility Route Choice Model.
\newblock {\em Transp. Res. Part C Emerg. Technol.} {\bf 2022}, {\em 136}, 103514.

\bibitem[Cazor \em{et~al.}(2025)Cazor, Duncan, Watling, Nielsen, and Rasmussen]{cazor_closed-form_2025}
Cazor, L.; Duncan, L.C.; Watling, D.P.; et al.
\newblock A Closed-Form Bounded Route Choice Model Accounting for Heteroscedasticity, Overlap, and Choice Set Formation.
\newblock {\em Transp. Res. Part B Methodol.} {\bf 2025}, {\em 199}, 103275.

\bibitem[Wu \em{et~al.}(2019)Wu, Qu, Sun, Yin, Yan, and Zhao]{wu_data-driven_2019}
Wu, J.; Qu, Y.; Sun, H.; et al.
\newblock Data-Driven Model for Passenger Route Choice in Urban Metro Network.
\newblock {\em Physica A} {\bf 2019}, {\em 524}, 787--798.

\bibitem[McFadden(1972)]{mcfadden_conditional_1972}
McFadden, D.
\newblock Conditional Logit Analysis of Qualitative Choice Behavior.
\newblock In {\em Frontiers in Econometrics}; Zarembka, P., Ed.; Academic Press: New York, NY, USA, 1972; pp. 105--142.

\bibitem[Train(2009)]{train_discrete_2009}
Train, K.E.
\newblock {\em Discrete Choice Methods with Simulation}; Cambridge University Press: {Cambridge, UK}, 2009.

\bibitem[Ben-Akiva and Lerman(1985)]{ben_discrete_1985}
Ben-Akiva, M.E.; Lerman, S.R.
\newblock {\em Discrete Choice Analysis: Theory and Application to Travel Demand}; MIT Press: Cambridge, MA, USA, 1985.

\bibitem[Ma \em{et~al.}(2020)Ma, Yu, and Liu]{ma_nested_2020}
Ma, S.; Yu, Z.; Liu, C.
\newblock Nested Logit Joint Model of Travel Mode and Travel Time Choice for Urban Commuting Trips in Xi'an, China.
\newblock {\em J. Urban Plan. Dev.} {\bf 2020}, {\em 146}, 04020020.

\bibitem[Garcia-Martinez \em{et~al.}(2018)Garcia-Martinez, Cascajo, Jara-Diaz, Chowdhury, and Monzon]{garcia-martinez_transfer_2018}
Garcia-Martinez, A.; Cascajo, R.; Jara-Diaz, S.R.; et al.
\newblock Transfer Penalties in Multimodal Public Transport Networks.
\newblock {\em Transp. Res. Part A Policy Pract.} {\bf 2018}, {\em 114}, 52--66.

\bibitem[Okubo \em{et~al.}(2022)Okubo, Kitano, and Morimoto]{okubo_transportation_2022}
Okubo, T.; Kitano, N.; Morimoto, A.
\newblock A Transportation Choice Model on the Commuter Railroads Using Inverse Reinforcement Learning.
\newblock {\em Asian Transp. Stud.} {\bf 2022}, {\em 8}, 100072.

\bibitem[McFadden and Train(2000)]{mcfadden_mixed_2000}
McFadden, D.; Train, K.
\newblock Mixed MNL Models for Discrete Response.
\newblock {\em J. Appl. Econom.} {\bf 2000}, {\em 15}, 447--470.

\bibitem[Greene and Hensher(2003)]{greene_a_latent_2003}
Greene, W.H.; Hensher, D.A.
\newblock A Latent Class Model for Discrete Choice Analysis: Contrasts with Mixed Logit.
\newblock {\em Transp. Res. Part B Methodol.} {\bf 2003}, {\em 37}, 681--698.

\bibitem[Kim and Mokhtarian(2023)]{kim_finite_2023}
Kim, S.H.; Mokhtarian, P.L.
\newblock Finite Mixture (or Latent Class) Modeling in Transportation: Trends, Usage, Potential, and Future Directions.
\newblock {\em Transp. Res. Part B Methodol.} {\bf 2023}, {\em 172}, 134--173.

\bibitem[Ghorbani \em{et~al.}(2025)Ghorbani, Nassir, Lavieri, Beeramoole, and Paz]{ghorbani_enhanced_2025}
Ghorbani, A.; Nassir, N.; Lavieri, P.S.; et al.
\newblock Enhanced Utility Estimation Algorithm for Discrete Choice Models in Travel Demand Forecasting.
\newblock {\em Transportation} {\bf 2025}. \href{https://doi.org/10.1007/s11116-024-10579-1}{https://doi.org/10.1007/s11116-024-10579-1}.

\bibitem[Tang \em{et~al.}(2024)Tang, Jiang, Zou, Liu, Zhang, and Liao]{tang_mining_2024}
Tang, Y.; Jiang, Z.; Zou, X.; et al.
\newblock Mining Motif Periodic Frequent Travel Patterns of Individual Metro Passengers Considering Uncertain Disturbances.
\newblock {\em Int. J. Transp. Sci. Technol.} {\bf 2024}, {\em 15}, 102--121.

\bibitem[Mohammed and Oke(2023)]{mohammed_origin_2023}
Mohammed, M.; Oke, J.
\newblock Origin-Destination Inference in Public Transportation Systems: A Comprehensive Review.
\newblock {\em Int. J. Transp. Sci. Technol.} {\bf 2023}, {\em 12}, 315--328.

\bibitem[Yap \em{et~al.}(2025)Yap, Wong, and Cats]{yap_public_2025}
Yap, M.; Wong, H.; Cats, O.
\newblock Public Transport Crowding Valuation in a Post-Pandemic Era.
\newblock {\em Transportation} {\bf 2025}, {\em 52}, 287--306.

\bibitem[Cherchi and De~Dios~Ort\'{u}zar(2008)]{cherchi_empirical_2008}
Cherchi, E.; De Dios Ort\'{u}zar, J.
\newblock Empirical Identification in the Mixed Logit Model: Analysing the Effect of Data Richness.
\newblock {\em Netw. Spat. Econ.} {\bf 2008}, {\em 8}, 109--124.

\bibitem[Sadeghian \em{et~al.}(2022)Sadeghian, Zhao, Golshan, and Håkansson]{sadeghian_stepwise_2022}
Sadeghian, P.; Zhao, X.; Golshan, A.; et al.
\newblock A Stepwise Methodology for Transport Mode Detection in GPS Tracking Data.
\newblock {\em Travel Behav. Soc.} {\bf 2022}, {\em 26}, 159--167.

\bibitem[Montini \em{et~al.}(2017)Montini, Antoniou, and Axhausen]{montini_route_2017}
Montini, L.; Antoniou, C.; Axhausen, K.W.
\newblock Route and Mode Choice Models Using GPS Data. Available online: \href{https://ethz.ch/content/dam/ethz/special-interest/baug/ivt/ivt-dam/vpl/reports/1201-1300/ab1204.pdf}{https://ethz.ch/content/dam/ethz/special-interest/baug/ivt/ivt-dam/vpl/reports/1201-1300/ab1204.pdf} (accessed on {15 September 2025}).

\bibitem[Helbing and Nagel(2004)]{helbing_the_physics_2004}
Helbing, D.; Nagel, K.
\newblock The Physics of Traffic and Regional Development.
\newblock {\em Contemp. Phys.} {\bf 2004}, {\em 45}, 405--426.

\bibitem[Ozaki \em{et~al.}(2022)Ozaki, Shida, Takayasu, and Takayasu]{ozaki_direct_2022}
Ozaki, J.; Shida, Y.; Takayasu, H.; et al.
\newblock Direct Modelling from GPS Data Reveals Daily-Activity-Dependency of Effective Reproduction Number in COVID-19 Pandemic.
\newblock {\em Sci. Rep.} {\bf 2022}, {\em 12}, 17888.

\bibitem[Shida \em{et~al.}(2022)Shida, Ozaki, Takayasu, and Takayasu]{shida_potential_2022}
Shida, Y.; Ozaki, J.; Takayasu, H.; et al.
\newblock Potential Fields and Fluctuation-Dissipation Relations Derived from Human Flow in Urban Areas Modeled by a Network of Electric Circuits.
\newblock {\em Sci. Rep.} {\bf 2022}, {\em 12}, 9918.

\bibitem[{Ministry of Land, Infrastructure, Transport and Tourism (MLIT)}(2024)]{mlit_railway_data}
Ministry of Land, Infrastructure, Transport and Tourism (MLIT).
\newblock National Land Numerical Information: Railway Data (N02). Available online: \href{https://nlftp.mlit.go.jp/ksj/gml/datalist/KsjTmplt-N02-2023.html}{https://nlftp.mlit.go.jp/ksj/gml/datalist/KsjTmplt-N02-2023.html} (accessed on 14 October 2025). (In Japanese)

\bibitem[Su \em{et~al.}(2020)Su, McBride, and Goulias]{su_pattern_2020}
Su, R.; McBride, E.C.; Goulias, K.G.
\newblock Pattern Recognition of Daily Activity Patterns Using Human Mobility Motifs and Sequence Analysis.
\newblock {\em Transp. Res. Part C Emerg. Technol.} {\bf 2020}, {\em 120}, 102796.

\bibitem[Zhang and Li(2024)]{zhang_activity_2024}
Zhang, X.; Li, N.
\newblock An Activity Space-Based Gravity Model for Intracity Human Mobility Flows.
\newblock {\em Sustain. Cities Soc.} {\bf 2024}, {\em 101}, 105073.

\bibitem[{Ministry of Land, Infrastructure, Transport and Tourism (MLIT)}(2016)]{mlit_transfer_survey_2016}
Ministry of Land, Infrastructure, Transport and Tourism (MLIT).
\newblock Transfer Survey Report 2016. Available online: \href{https://www.mlit.go.jp/common/001179761.pdf}{https://www.mlit.go.jp/common/001179761.pdf} (accessed on {15 September 2025}). (In Japanese)

\bibitem[Takahashi(2019)]{takahashi_transportation_2019}
Takahashi, T.
\newblock Transportation Mode Choice and Spatial Structure of a City.
\newblock {\em {Technical Report, Center for Spatial Information Science}}; Working Paper; The University of Tokyo: Tokyo, Japan, 2019. Available online: \href{https://www.csis.u-tokyo.ac.jp/en/research/}{https://www.csis.u-tokyo.ac.jp/en/research/} (accessed on {15 September 2025}). (In Japanese)

\bibitem[Sadeghian \em{et~al.}(2021)Sadeghian, Håkansson, and Zhao]{sadeghian_review_2021}
Sadeghian, P.; Håkansson, J.; Zhao, X.
\newblock Review and Evaluation of Methods in Transport Mode Detection Based on GPS Tracking Data.
\newblock {\em J. Traffic Transp. Eng. (Engl. Ed.)} {\bf 2021}, {\em 8}, 467--482.

\bibitem[Markos and Yu(2020)]{markos_unsupervised_2020}
Markos, C.; Yu, J.J.
\newblock Unsupervised Deep Learning for GPS-Based Transportation Mode Identification.
\newblock In Proceedings of the 2020 IEEE 23rd International Conference on Intelligent Transportation Systems (ITSC), Rhodes, Greece, {20--23 September 2020}.

\bibitem[Montazeri-Gh and Fotouhi(2011)]{montazeri-gh_traffic_2011}
Montazeri-Gh, M.; Fotouhi, A.
\newblock Traffic Condition Recognition Using the K-Means Clustering Method.
\newblock {\em Sci. Iran.} {\bf 2011}, {\em 18}, 930--937.

\bibitem[Rasmussen \em{et~al.}(2015)Rasmussen, Ingvardson, Halld{\'o}rsd{\'o}ttir, and Nielsen]{rasmussen_improved_2015}
Rasmussen, T.K.; Ingvardson, J.B.; Halldórsdóttir, K.; et al.
\newblock Improved Methods to Deduct Trip Legs and Mode from Travel Surveys Using Wearable GPS Devices: A Case Study from the Greater Copenhagen Area.
\newblock {\em Comput. Environ. Urban Syst.} {\bf 2015}, {\em 54}, 301--313.

\bibitem[Zheng \em{et~al.}(2008)Zheng, Liu, Wang, and Xie]{zheng_learning_2008}
Zheng, Y.; Liu, L.; Wang, L.; et al.
\newblock Learning Transportation Mode from Raw GPS Data for Geographic Applications on the Web.
\newblock In Proceedings of the 17th International Conference on World Wide Web, {Beijing, China, 21--25 April 2008}; \mbox{pp. 247--256.}

\bibitem[Dabiri and Heaslip(2018)]{dabiri_inferring_2018}
Dabiri, S.; Heaslip, K.
\newblock Inferring Transportation Modes from GPS Trajectories Using a Convolutional Neural Network.
\newblock {\em Transp. Res. Part C Emerg. Technol.} {\bf 2018}, {\em 86}, 360--371.

\bibitem[Zhu \em{et~al.}(2016)Zhu, Zhu, Li, Fu, Huang, Gan, and Zhou]{che_identifying_2016}
Zhu, Q.; Zhu, M.; Li, M.; et al.
\newblock Identifying Transportation Modes from Raw GPS Data.
\newblock In {\em Social Computing}; Che, W., Han, Q., Wang, H., et al., Eds.; Springer: Singapore, 2016; Volume 623, pp. 395--409.

\bibitem[{Tokyo Metropolitan Government}(2019)]{tokyo_public_transport_status}
Tokyo Metropolitan Government.
\newblock {Current Status and Issues of Regional Public Transportation.} Available online: \href{https://www.toshiseibi.metro.tokyo.lg.jp/documents/d/toshiseibi/pdf_bunyabetsu_kotsu_butsuryu_pdf_chiiki_koutsu_kentoukai01_2-1}{https://www.toshiseibi.metro.tokyo.lg.jp/documents/d/toshiseibi/pdf\_bunyabetsu\_kotsu\allowbreak\_butsuryu\_pdf\_chiiki\_koutsu\_kentoukai01\_2-1} (accessed on {19 June 2026}). (In Japanese)

\bibitem[Schneider \em{et~al.}(2013)Schneider, Belik, Couronn{\'e}, Smoreda, and Gonz{\'a}lez]{schneider_unravelling_2013}
Schneider, C.M.; Belik, V.; Couronné, T.; et al.
\newblock Unravelling Daily Human Mobility Motifs.
\newblock {\em J. R. Soc. Interface} {\bf 2013}, {\em 10}, 20130246.

\bibitem[Jiang \em{et~al.}(2017)Jiang, Ferreira, and Gonzalez]{jiang_activity-based_2017}
Jiang, S.; Ferreira, J.; Gonzalez, M.C.
\newblock Activity-Based Human Mobility Patterns Inferred from Mobile Phone Data: A Case Study of Singapore.
\newblock {\em IEEE Trans. Big Data} {\bf 2017}, {\em 3}, 208--219.

\bibitem[Cao \em{et~al.}(2019)Cao, Li, Tu, and Wang]{cao_characterizing_2019}
Cao, J.; Li, Q.; Tu, W.; et al.
\newblock Characterizing Preferred Motif Choices and Distance Impacts.
\newblock {\em PLoS One} {\bf 2019}, {\em 14}, e0215242.

\bibitem[{At Home Co., Ltd.}(2018)]{at_home_trend_2018}
At Home Co., Ltd.
\newblock Trend Survey-Tokyo. Available online: \href{https://www.athome.co.jp/corporate/wp-content/themes/news/pdf/densha-tsuukin-201806/densha-tsuukin-201806.pdf}{https://www.athome.co.jp/corporate/wp-content/themes/news/pdf\allowbreak/densha-tsuukin-201806/densha-tsuukin-201806.pdf} (accessed on {3 March 2026}). (In Japanese)

\bibitem[Hibino \em{et~al.}(2001)Hibino, Morita, and Uchiyama]{hibino_alternative_2001}
Hibino, N.; Morita, Y.; Uchiyama, H.
\newblock A Study on Method of Setting Alternative Routes for Analysis on Route Choice Behavior.
\newblock {\em J. City Plan. Inst. JPN} {\bf 2001}, {\em 36}, 787--792. (In Japanese)

\bibitem[Dixit \em{et~al.}(2023)Dixit, Cats, Brands, van Oort, and Hoogendoorn]{dixit_perception_2023}
Dixit, M.; Cats, O.; Brands, T.; et al.
\newblock Perception of Overlap in Multi-Modal Urban Transit Route Choice.
\newblock {\em Transp. A Transp. Sci.} {\bf 2023}, {\em 19}, 2005180.

\bibitem[Anas(1983)]{anas_discrete_1983}
Anas, A.
\newblock Discrete Choice Theory, Information Theory and the Multinomial Logit and Gravity Models.
\newblock {\em Transp. Res. Part B Methodol.} {\bf 1983}, {\em 17}, 13--23.

\bibitem[Yoo \em{et~al.}(2025)Yoo, Kumagai, Aki, and Managi]{yoo_2025_railway}
Yoo, S.; Kumagai, J.; Aki, R.; et al.
\newblock Railway Network Expansion Reduces Air Pollution in Tokyo over 25 Years.
\newblock {\em Sustain. Cities Soc.} {\bf 2025}, {\em 127}, 106408.

\bibitem[Neter \em{et~al.}(1996)Neter, Kutner, Nachtsheim, Wasserman, et~al.]{neter_applied_1996}
Neter, J.; Kutner, M.H.; Nachtsheim, C.J.; et al.
\newblock {\em Applied Linear Statistical Models}; Irwin: Chicago, IL, USA, 1996.

\bibitem[Hair \em{et~al.}(2009)Hair, Black, Babin, and Anderson]{hair_multivariate_2009}
Hair, J.F.; Black, W.C.; Babin, B.J.; et al.
\newblock {\em Multivariate Data Analysis}, 7th ed.; Pearson Prentice Hall: Upper Saddle River, NJ, USA, 2009.

\bibitem[Casella and Berger(2002)]{casella_statistical_2002}
Casella, G.; Berger, R.L.
\newblock {\em Statistical Inference}, 2nd ed.; Duxbury Press: Pacific Grove, CA, USA, 2002.

\bibitem[Greene(2018)]{greene_econometric_2018}
Greene, W.H.
\newblock {\em Econometric Analysis}, 8th ed.; Pearson Education: New York, NY, USA, 2018.

\bibitem[Ozili(2023)]{ozili_acceptable_2023}
Ozili, P.K.
\newblock The Acceptable R-Square in Empirical Modelling for Social Science Research {\bf 2023}, 134--143. \href{https://doi.org/10.2139/ssrn.4128165}{https://doi.org/10.2139/ssrn.4128165}.

\bibitem[Ji \em{et~al.}(2015)Ji, Jiang, Du, and Zhang]{ji_estimation_2015}
Ji, Y.; Jiang, S.; Du, Y.; et al.
\newblock Estimation of Bimodal Urban Link Travel Time Distribution and Its Applications in Traffic Analysis.
\newblock {\em Math. Probl. Eng.} {\bf 2015}, {\em 2015}, 615468.

\bibitem[Kumagai(2025)]{kumagai_visualizing_2025}
Kumagai, J.
\newblock Visualizing Railway Transfer Penalties and Their Effects on Population Distribution in the Tokyo Metropolitan Area.
\newblock {\em Future Transp.} {\bf 2025}, {\em 5}, 114.

\bibitem[Wilson(1967)]{wilson_a_statistical_1967}
Wilson, A.
\newblock A Statistical Theory of Spatial Distribution Models.
\newblock {\em Transp. Res.} {\bf 1967}, {\em 1}, 253--269.

\bibitem[Donoso and de~Grange(2010)]{donoso_a_microeconomic_2010}
Donoso, P.; de Grange, L.
\newblock A Microeconomic Interpretation of the Maximum Entropy Estimator of Multinomial Logit Models and Its Equivalence to the Maximum Likelihood Estimator.
\newblock {\em Entropy} {\bf 2010}, {\em 12}, 2077--2084.

\bibitem[{Ministry of Land, Infrastructure, Transport and Tourism (MLIT)}(2017)]{mlit_transfer_survey_2017}
Ministry of Land, Infrastructure, Transport and Tourism (MLIT).
\newblock Transfer Survey Report 2017. Available online: \href{https://www.mlit.go.jp/common/001178977.pdf}{https://www.mlit.go.jp/common/001178977.pdf} (accessed on {5 April 2025}). (In Japanese)

\bibitem[Vittinghoff (2007)]{Vittinghoff_2007}
\newblock Vittinghoff, E.; McCulloch, C.E. 
\newblock Relaxing the Rule of Ten Events Per Variable in Logistic
and Cox Regression. 
\newblock {\em Am. J. Epidemiol.}  {\bf 2007}, {\em 165}, 710–-718.

\end{thebibliography}

\end{document}